\documentclass[aps,prb,showpacs,twocolumn,superscriptaddress]{revtex4-1}
\usepackage{amsmath}
\usepackage{microtype}
\usepackage{hyperref}
\usepackage{placeins}
\usepackage{graphicx}
\usepackage{siunitx}
\usepackage{bm}
\usepackage{adjustbox}
\usepackage{glossaries}
\usepackage[noabbrev, capitalise, nameinlink]{cleveref}
\usepackage[caption=false]{subfig}
\usepackage[dvipsnames]{xcolor}

\usepackage{physics}

\usepackage{ragged2e}
\edef\UrlBreaks{\do\-\UrlBreaks} 

\newcommand\hc{\mbox{h.\,c.}}

\DeclareMathOperator{\imag}{\mathrm{i}}

\renewcommand{\vec}[1]{\bm{#1}}
\newcommand{\mat}[1]{\bm{#1}}

\providecommand{\gls}[1]{\uppercase{#1}}
\providecommand{\glspl}[1]{\uppercase{#1}s}

\newcommand{\CC}{C\nolinebreak\hspace{-.05em}\raisebox{.4ex}{\tiny\bf +}\nolinebreak\hspace{-.10em}\raisebox{.4ex}{\tiny\bf +}}

\newcommand{\stfig}[1]{
  \begin{figure}
    \centering
    #1
  \end{figure}
}

\newcommand{\widefig}[1]{
  \begin{figure*}
    \renewcommand{\columnwidth}{\textwidth}
    \centering
    #1
  \end{figure*}
}

\newcommand{\homohead}{Homogeneous bilayer graphene system}
\newcommand{\hardhead}{Hard wall layer switching system}
\newcommand{\slswhead}{Sheared layer switching system}
\newcommand{\tlswhead}{Tensile layer switching system}

\providecommand{\somesize}{1cm}
\providecommand{\LDOSsize}{2cm}
\providecommand{\TRANSsize}{3cm}
\providecommand{\defaultFigWidth}{0.32\columnwidth}

\providecommand{\raisesize}{-0.5cm}

\providecommand{\RE}{\emph{R}}
\providecommand{\LE}{\emph{L}}
\providecommand{\BRE}{\emph{BR}}
\providecommand{\TRE}{\emph{TR}}
\providecommand{\BLE}{\emph{BL}}
\providecommand{\TLE}{\emph{TL}}

\providecommand{\SPass}{\sigma_{\mathrm{BL}\rightarrow\mathrm{BR}}}
\providecommand{\SLSW}{\sigma_{\mathrm{BL}\rightarrow\mathrm{TL}}}
\providecommand{\SRef}{\sigma_{\mathrm{R}\rightarrow\mathrm{TR}}}

\providecommand{\SkewWidthLong}[1]{The shear width parameter is $L_S=#1$}
\providecommand{\TensWidthLong}[1]{The tension length parameter is $L_T=#1$}
\providecommand{\TransNumberLong}[1]{The number of sites contained in the system used for the transmission calculation is $N_{\text{C}}\approx#1$}
\providecommand{\LDOSNumberLong}[1]{The number of sites contained in the system used for the LDOS calculation is $N_{\text{L}}\approx#1$}
\providecommand{\LowerCoupledLong}{The lower layer is fully coupled $t_{\mathrm{LSW}}^{(1)}=t$}

\providecommand{\UpperDecoupledLong}{The upper layer is decoupled $t_{\mathrm{LSW}}^{(2)}=0$}
\providecommand{\NearCutoffLong}{The cutoff between the two layers in the \gls{lsw} is short, $D_{\text{C}}=1.3a_0$}
\providecommand{\FarCutoffLong}{The cutoff between the two layers in the \gls{lsw} is long, $D_{\text{C}}=4a_0$}
\providecommand{\LDOSphiLong}[1]{The \gls{ldos} calculation is performed at an external magnetic field $\phi^{-1}_{\mathrm{LDOS}}=#1\phi_0^{-1}$}

\providecommand{\TransNumbersLong}[1]{The number of sites in the system used for the transmission calculations is $N_{\text{C}}\approx#1$}

\providecommand{\SkewWidth}[1]{\SkewWidthLong{#1}}
\providecommand{\TensWidth}[1]{\TensWidthLong{#1}}
\providecommand{\TransNumber}[1]{\TransNumberLong{#1}}
\providecommand{\LDOSNumber}[1]{\LDOSNumberLong{#1}}
\providecommand{\LowerCoupled}{\LowerCoupledLong}

\providecommand{\UpperDecoupled}{\UpperDecoupledLong}
\providecommand{\NearCutoff}{\NearCutoffLong}
\providecommand{\FarCutoff}{\FarCutoffLong}
\providecommand{\LDOSphi}[1]{\LDOSphiLong{#1}}

\providecommand{\HallbarPlot}[1]{Hall bar plot for #1. On the left is a logarithmic plot of the \gls{ldos} due to electron injection from the lower left and right lead with a cutoff at $\num{1e-4}$. On the right are the elements of the conductance tensor. The different conductances have different colors and marker styles. The indices for the conductances are as indicated in \ref{figure:hallbar}}

\providecommand{\HallbarPlotRef}[1]{Hall bar plot, see \ref{figure:homogenous_hallbar}, for #1}

\providecommand{\TransPlotRef}[1]{Transmission calculation like in \textbf{Hall bar plot}, see \ref{figure:homogenous_hallbar}, for #1}

\providecommand{\parametercap}[3]{Parameter study of the #1 for different external magnetic fields $\phi^{-1}$ for #2. On the left is $\SPass{}$ and on the right is $\SLSW{}$. The two plots show the conductance in color with the #1 on the y-axis and the magnetic field on the x-axis. \TransNumbersLong{#3}}

\newcommand{\finiteSizeCap}[1]{#1 Hall bar system finite-size analysis for transmission diagrams, see \ref{figure:homogenous_hallbar} for a description of transmission diagrams}

\providecommand{\ShearLargeWidth}{\num{2.3e5}}
\providecommand{\ShearMiddleWidth}{\num{1.8e5}}
\providecommand{\ShearSmallWidth}{\num{1.4e5}}
\providecommand{\TensVLargeWidth}{\num{3.0e5}}
\providecommand{\TensLargeWidth}{\num{2.0e5}}
\providecommand{\TensMiddleWidth}{\num{1.6e5}}

\providecommand{\tlsw}{t_{\mathrm{LSW}}^{(U)}}
\providecommand{\maybefbox}[1]{#1}

\newdimen\hspacing
\newcommand{\widewidth}{0.9\columnwidth}
\newcommand{\narrowwidth}{0.75\columnwidth} 
\DeclareCaptionJustification{justified}{\justifying}
\bibpunct{[}{]}{;}{n}{}{,~}

\AtBeginDocument{\renewcommand{\ref}[1]{\cref{#1}}}
\glsdisablehyper{}
 \newcommand{\lswLong}{layer switching wall}
\newcommand{\hlswLong}{hard layer switching wall}
\newcommand{\slswLong}{shear layer switching wall}
\newcommand{\tlswLong}{tension layer switching wall}
\newcommand{\blgLong}{bilayer graphene}
\newcommand{\slgLong}{single-layer graphene}
\newcommand{\tblgLong}{twisted bilayer graphene}
\newcommand{\ldosLong}{lcoal density of states}
\newcommand{\iqheLong}{integer quantum Hall effect}

\newacronym{lsw}{LSW}{\lswLong}
\newacronym{hlsw}{HLSW}{\hlswLong}
\newacronym{slsw}{SLSW}{\slswLong}
\newacronym{tlsw}{TLSW}{\tlswLong}
\newacronym{blg}{BLG}{\blgLong}
\newacronym{slg}{SLG}{\slgLong}
\newacronym{tblg}{TBLG}{\tblgLong}
\newacronym{ldos}{LDOS}{\ldosLong}
\newacronym{iqhe}{IQHE}{\iqheLong}
\newacronym{ssh}{SSH}{Su Schrieffer Heeger}
\newacronym{tem}{TEM}{transmission electron microscopy}
 \newcommand{\authorinfo}{
\title{Effects of domain walls in bilayer graphene in an external magnetic field}
\author{Nico S. Ba\ss ler}
\affiliation{Lehrstuhl f\"ur Theoretische Physik I, Staudtstra{\ss}e 7, FAU Erlangen-N\"urnberg, D-91058 Erlangen, Germany}
\author{Kai Phillip Schmidt}
\affiliation{Lehrstuhl f\"ur Theoretische Physik I, Staudtstra{\ss}e 7, FAU Erlangen-N\"urnberg, D-91058 Erlangen, Germany}}
 \graphicspath{{./img/}}

\crefname{section}{Sect.}{Sects.}
\crefname{figure}{Fig.}{Figs.}
\crefname{equation}{Eq.}{Eqs.}

\begin{document}
\begin{abstract}
  We investigate bilayer graphene systems with layer switching domain walls separating the two energetically equivalent Bernal stackings in the presence of an external magnetic field. To this end we calculate quantum transport and local densities of three microscopic models for a single domain wall: a hard wall, a defect due to shear, and a defect due to tension. The quantum transport calculations are performed with a recursive Green's function method. Technically, we discuss an explicit algorithm for the separation of a system into subsystems for the recursion and we present an optimization of the well known iteration scheme for lead self-energies for sparse chain couplings.
 We find strong physical differences for the three different types of domain walls in the integer quantum Hall regime. For a domain wall due to shearing of the upper graphene layer there is a plateau formation in the magnetoconductance for sufficiently wide defect regions. For wide domain walls due to tension in the upper graphene layer there is only an approximate plateau formation with fluctuations of the order of the elementary conuctance quantum $\sigma_0$. A direct transition between stacking regions like for the hard wall domain wall shows no plateau formation and is therefore not a good model for either of the previously mentioned extended domain walls.
\end{abstract}
 \authorinfo{}\maketitle{}
\section{Introduction}\label{section:introduction}
Two dimensional quantum materials have risen in popularity quite dramatically in recent years. One aspect of this is the rapid progress in manufacturing processes \cite{BONACCORSO2012564} which allows the creation of a multitude of materials such as Van der Waals heterostructures \cite{Geim2013}. Another aspect is the discovery of topological quantum phases like the topological insulator \cite{Xiao2011, Hasan2010}, topological superconductors \cite{schnyder2009,Kitaev2001,kitaev2009} as well strongly correlated intrinsic topological phases, which are relevant for topological quantum computing \cite{kitaev2003,freedman2003} and potentially realized in fractional quantum Hall systems \cite{laughlin1983, tsui1982} as well as in certain frustrated quantum magnets \cite{Balents2010}. The first material proposed to be a topological insulator was \gls{slg} \cite{kane2005quantum}, but the spin-orbit coupling required for graphene to exhibit a stable quantum spin Hall effect turned out to be too weak. Nevertheless, \gls{slg} has attracted an enormous scientific interest due its extraordinary mechanical \cite{Frank2007} and electronic \cite{Schwierz2010, Xia2009} properties.

A variant of graphene which came into focus recently is \gls{blg} \cite{Ohta951, Zhang2009, McCann_2013}, where most fascinatingly a correlated superconducting state has been identified experimentally in so-called twisted \gls{blg} \cite{Suarez2010, Huang2018, Yankowitz1059, Cao2018Insulator, Cao2018}. Conventional \gls{blg} has many interesting physical properties which are often fundamentally different from its single-layer counterpart \cite{Castro2009}. As an example, although \gls{blg} and \gls{slg} display an anomalous integer quantum Hall effect \cite{Gusynin2005}, the anomaly in \gls{blg} is quite different due to the quadratic low-energy band \cite{novoselov2006unconventional}. Indeed, the linear low-energy dispersion of \gls{slg} can be modeled by a Dirac Hamiltonian so that chiral quasiparticles induce a Berry phase of $\pi$ leading naturally to conductance jumps of $4\mathrm{e}^2/\hbar$ in the magnetoconductance at zero energy. In contrast, the quadratic low-energy band of \gls{blg} gives rise to chiral quasiparticles inducing a Berry phase of $2\pi$ resulting in two-times larger conductance jumps of $8\mathrm{e}^2/\hbar$ in the magnetoconductance at charge neutrality.

Microscopically, \gls{blg} realizes a Bernal stacking \cite{Yan2011} so that two neighboring lattice sites from opposite layers correspond to different sublattices A and B. This leads to an exact two-fold degeneracy of the electronic structure with respect to AB and BA stacking which can be obtained from each other through inversion. As a consequence, \gls{blg} samples are typically not homogeneous but consist of a network of domains with AB or BA stacking. These domains are separated by defects which are created in the manufacturing process of \gls{blg}, e.g.~, with epitaxy \cite{riedl2009quasi, speck2010quasi} or are separated due to the natural structure of slightly twisted \gls{blg} \cite{Huang2018}. These extended defects, which cause a registry shift in \gls{blg} over \SIrange{6}{11}{\nano\meter} \cite{Alden11256} are called \glspl{lsw} (or alternatively either partial dislocations or strain solitons). There have been extensive studies on the precise nature of \glspl{lsw} in real materials. The fact, that \gls{blg} is a two-dimensional material allows for a release of strain with out-of-plane buckling \cite{Butz2014}. There are also restrictions on the physical stacking textures due to energetic considerations \cite{Gong2014} and stacking boundary conditions.
    
In general, it is an interesting and important question to understand the impact of such \glspl{lsw} on the physical properties of \gls{blg}. In an external electric field, topological modes protected from scattering are known to exist along these \glspl{lsw} \cite{Li2016, Ju2015}. Since samples usually contain many of these \glspl{lsw}, applying an external electric field leads to the formation of networks of topological channels \cite{rickhaus2018transport} due to the gap introduced into the electronic structure \cite{Yin2016}. 

Another important aspect is the influence and role of \glspl{lsw} on \gls{blg} in an external magnetic field or, more concretely, on the anomalous integer as well as fractional quantum Hall effect in this material. Recently, the physical properties of \gls{lsw} networks in the presence of an external magnetic field have been studied \cite{kisslinger2015linear, Yin2016}. It is found that the presence of \glspl{lsw} can lead to rich conductance features even in the single particle transport regime. It is argued that transport energy gaps in the \si{\milli\electronvolt} regime are not caused by an electronic instability due to electronic interactions as typically assumed \cite{SanJose2014, Shallcross2017}, but simply due to the structure of transport across \glspl{lsw} and, in particular, due to ``hot'' charge carrying \glspl{lsw} \cite{Weckbecker2019}. A similar argument for the presence of plateaus at fractional fillings \cite{Maher61, Kou2014, Barlas2008} in \gls{blg} is made in \cite{Kisslinger2019}, where arbitrary fractional plateaus are engineered in artificial \gls{slg} mosaics interconnected with metallic strips. It is proposed that there is an ambiguity in two-terminal transport experiments between the purely single-particle plateaus at fractional fillings due to \glspl{lsw} and fractional plateaus originating from electron-electron interactions. However, this conclusion relies on a classical approximation of the couplings along linear \glspl{lsw}. It is one purpose of our work to validate these findings by approaching the problem from a quantum perpespective using an explicit real-space lattice model.

To this end we will use recursive Green's function methods \cite{Thouless_1981, Lewenkopf2013, petersen2008block, khomyakov2004real, khomyakov2005conductance, rammer1986} to calculate magnetoconductance and the \gls{ldos} \cite{Cresti2003} for various \gls{blg} systems with \glspl{lsw}. The systems used to model \glspl{lsw} are two fairly realistic domain wall models, where shear and tension lead to the formation of an AB-BA domain wall \cite{Alden11256}, as well as an unrealistic hard wall model. Here we do not explicitly model out-of-plane buckling and simply describe these systems with a shear or stretch transformation of the upper graphene layer. We find relevant physical differences for these three different types of \gls{lsw} in the integer quantum Hall regime. First, the hard wall model does not capture the essential physical properties for either of the other two extended domain walls due to shear or tension. Second, a domain wall due to shearing (tension) of the upper graphene layer yields an unexpected (approximate) plateau formation in the magnetoconductance for sufficiently wide defect regions. 

This article is organized as follows. In \ref{section:model} we introduce microscopic models for \gls{blg} with \glspl{lsw} in the presence of an external magnetic field and we detail the relevant physical observables in \ref{section:obs_geom}. Technical details on the implementation are given in \ref{section:implementation} while all obtained results are contained in \ref{section:results}. Finally, we conclude in \ref{section:conclusion} and present remaining open questions. 
\section{Models}\label{section:model}
We begin by describing a tight binding model of homogeneous \gls{blg} and three models of \glspl{lsw}. The three models are \gls{blg} with \gls{hlsw}, \gls{slsw} and \gls{tlsw} respectively. Afterwards, we discuss the transport setup used to determine the properties of the particular \gls{lsw} models and the relevant observables. 

\stfig{
  \includegraphics[width=0.5\columnwidth]{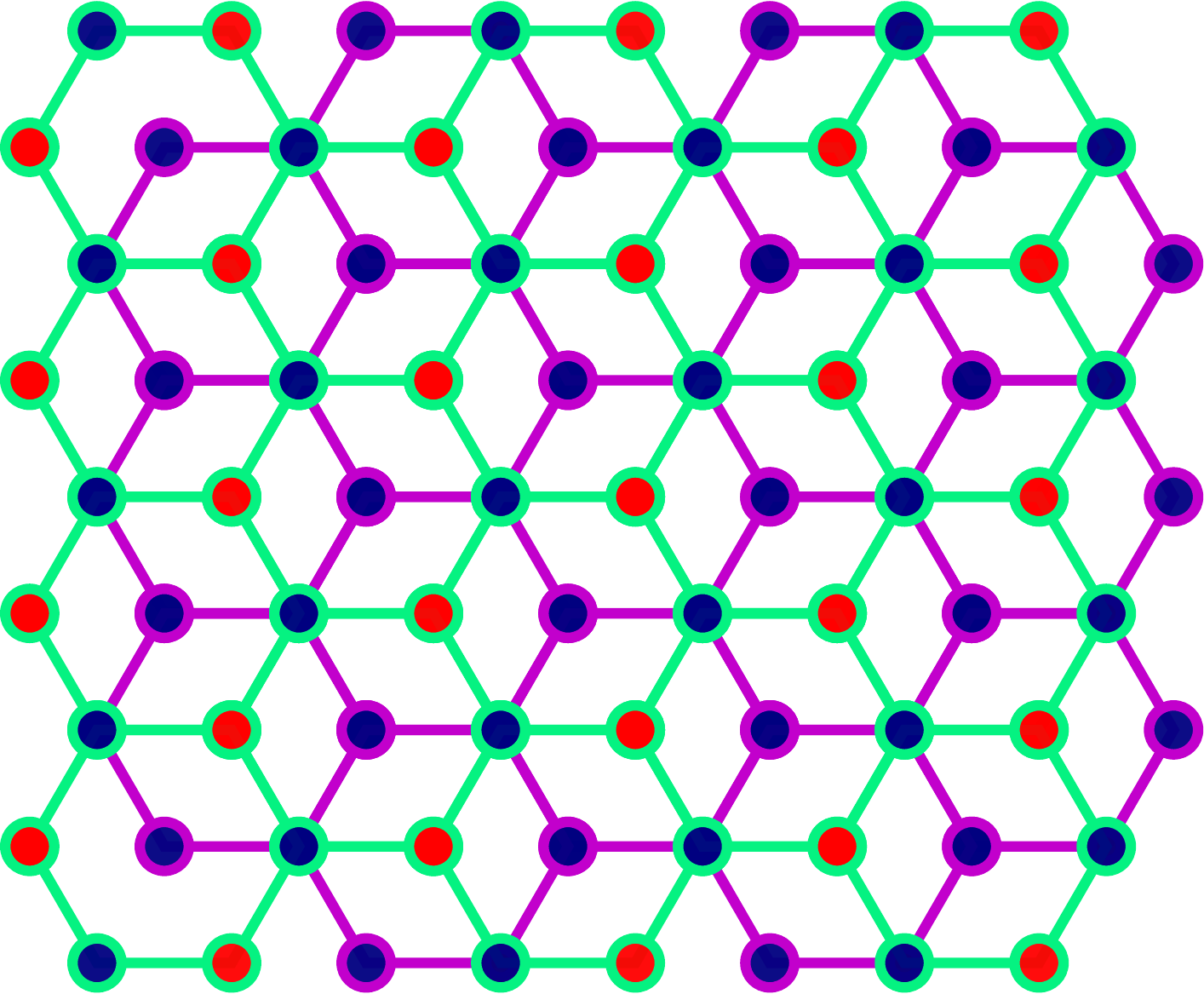}
  \caption{Illustration of Bernal stacked \gls{blg} for a nearest-neighbour model. The green sheet is on top and the magenta sheet is in the bottom. The blue and red circles indicate the A and B sublattice respectively for both sheets.}
  \label{fig:bilayer_graphene}
}  
\subsubsection{Homogeneous blayer graphene}
\label{subsubsection:homoblg}
\gls{blg} is described generically by the simple tight binding model given in \cite{tb_param_2012}. In the following we denote by $a_0=a/\sqrt{3}\approx\SI{0.142}{\nano\meter}$ the nearest-neighbor distance between carbon atoms of \gls{slg} and by $d_0\approx\SI{0.335}{\nano\meter}$ the interlayer distance between vertically located atoms. In actual graphene, $a\approx\SI{0.246}{\nano\meter}$, but all calculations are performed in units, such that $a=1$. In addition, we omit the spin degree of freedom which gives only rise to a trivial degeneracy of 2. Consequently, we describe \gls{blg} with the following microscopic Hamiltonian
\begin{equation}
 \label{eq::H_BLG}
  \mathcal{H}_{\mathrm{BLG}}=-\sum_{\norm{\vec{r}_i-\vec{r}_j}<D_{\rm c}}t(\vec{r}_i, \vec{r}_j)\;\hat{c}^\dagger_i\hat{c}^{\phantom{\dagger}}_j+\sum\limits_i\mu_i\hat{c}^\dagger_i\hat{c}^{\phantom{\dagger}}_i\,,
\end{equation}
where $\hat{c}^\dagger_i$ and $\hat{c}_i$ are the usual fermionic creation and annihilation operators on the site with index $i$. The form of the onsite chemical potential $\mu_i$ will be discussed alongside the system parameters. The sum runs in principal over all pairs of sites, but we included a numerical cutoff $D_{\rm c}$ for practical calculations. The nearest-neighbor model is obtained for $D_{\rm c}=a_0$. The hopping amplitudes are given by
\begin{equation}
  -t(\vec{r}_i,\vec{r}_j) =  V_{pp\pi}\left[1-\left(\frac{\vec{d}_z}{\norm{\vec{d}}}\right)^2\right]+ V_{pp\sigma}\left(\frac{\vec{d}_z}{\norm{\vec{d}}}\right)^2\; ,
\end{equation}
where $\vec{d} =\vec{r}_i-\vec{r}_j$ is the atom-atom distance, \mbox{$V_{pp\pi} = V_{pp\pi}^0\exp \left[-(\norm{\vec{d}}-a_0)/\delta\right]$} with $V_{pp\pi}^0$ the intralayer overlap integral between the nearest-neighbor atoms at a distance $a_0$, and \mbox{$V_{pp\sigma} = V_{pp\sigma}^0\exp \left[-(\norm{\vec{d}}-d_0)/\delta\right]$} with $V_{pp\sigma}^0$ the interlayer overlap integral at a distance $d_0$. The explicit values $V_{pp\pi}^0=\SI{-2.7}{\electronvolt}$ and $ V_{pp\sigma}^0=\SI{0.48}{\electronvolt}$ for the overlap integrals can be obtained by fitting the low-energy dispersion of bulk graphite. Further, $\delta=0.184a$ is the decay length of the overlap integral.

The Hamiltonian in \ref{eq::H_BLG} is quite general and can in principle model \gls{blg} at any twist angle, in particular the for \gls{blg} most relevant Bernal stacking which is illustrated in \ref{fig:bilayer_graphene}. Note that when not mentioned otherwise, we use a nearest-neighbor model with $D_{\rm c}=a_0$ to describe the system.

For a constant external magnetic field perpendicular to the sheet plains of \gls{blg}, the Peierl's substitution
\begin{equation}
  t(\vec{r}_i, \vec{r}_j)\rightarrow{}t(\vec{r}_i, \vec{r}_j)\exp\left(-\imag \frac{e}{\hbar}\int_{\vec{r}_i}^{\vec{r}_j}\vec{A}(\vec{r})\,\mathrm{d}\vec{r}\right)
\end{equation}
is chosen with a gauge that respects the translational symmetry of one of the electrodes which is usually a Landau gauge. The solution of this model for \gls{blg} in a perpendicular magnetic field yields then an anomalous \gls{iqhe}, which is calculated and discussed in \ref{subsection:homoblg_calc}.

\subsubsection{Hard layer switching wall}

The simplest model of a \gls{blg} system with an \gls{lsw} is with a so-called hard wall. This established simplification of the \gls{lsw} in \gls{blg} corresponds to an abrupt change between AB and BA Bernal stacking either parallel to the armchair or the zigzag nanoribbon. An illustration for the latter one is shown in comparison to a homogeneous \gls{blg} system in \ref{figure:normal_hard_wall}. Study of \glspl{lsw} parallel to the zigzag nanoribbon is preferred since they are simpler to model and also encapsulate the topological properties of the stacking transition \cite{zhang2013valley}.

\stfig{
  \subfloat{{\includegraphics[width=0.45\columnwidth]{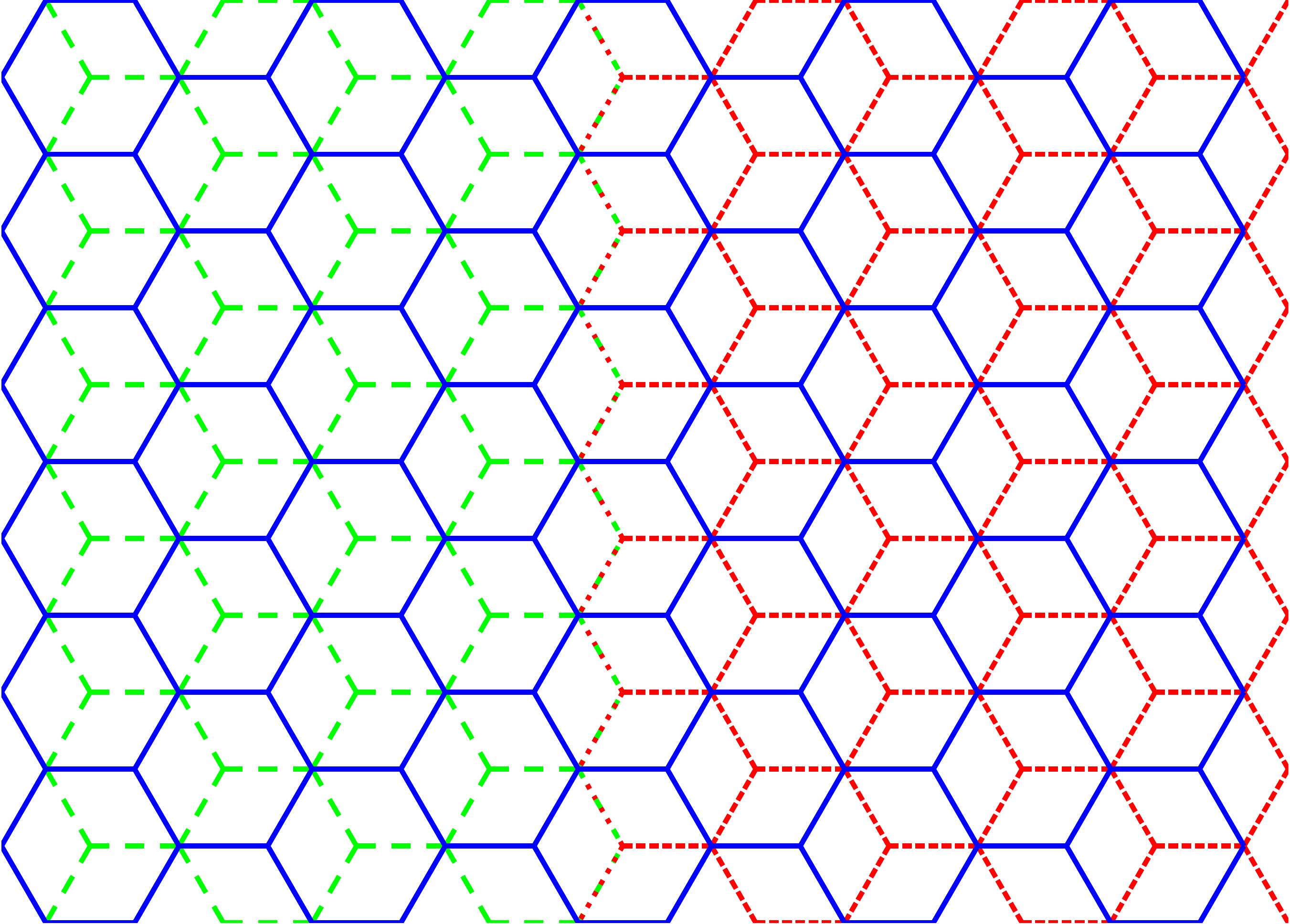} }}%
  \hfill
  \subfloat{{\includegraphics[width=0.45\columnwidth]{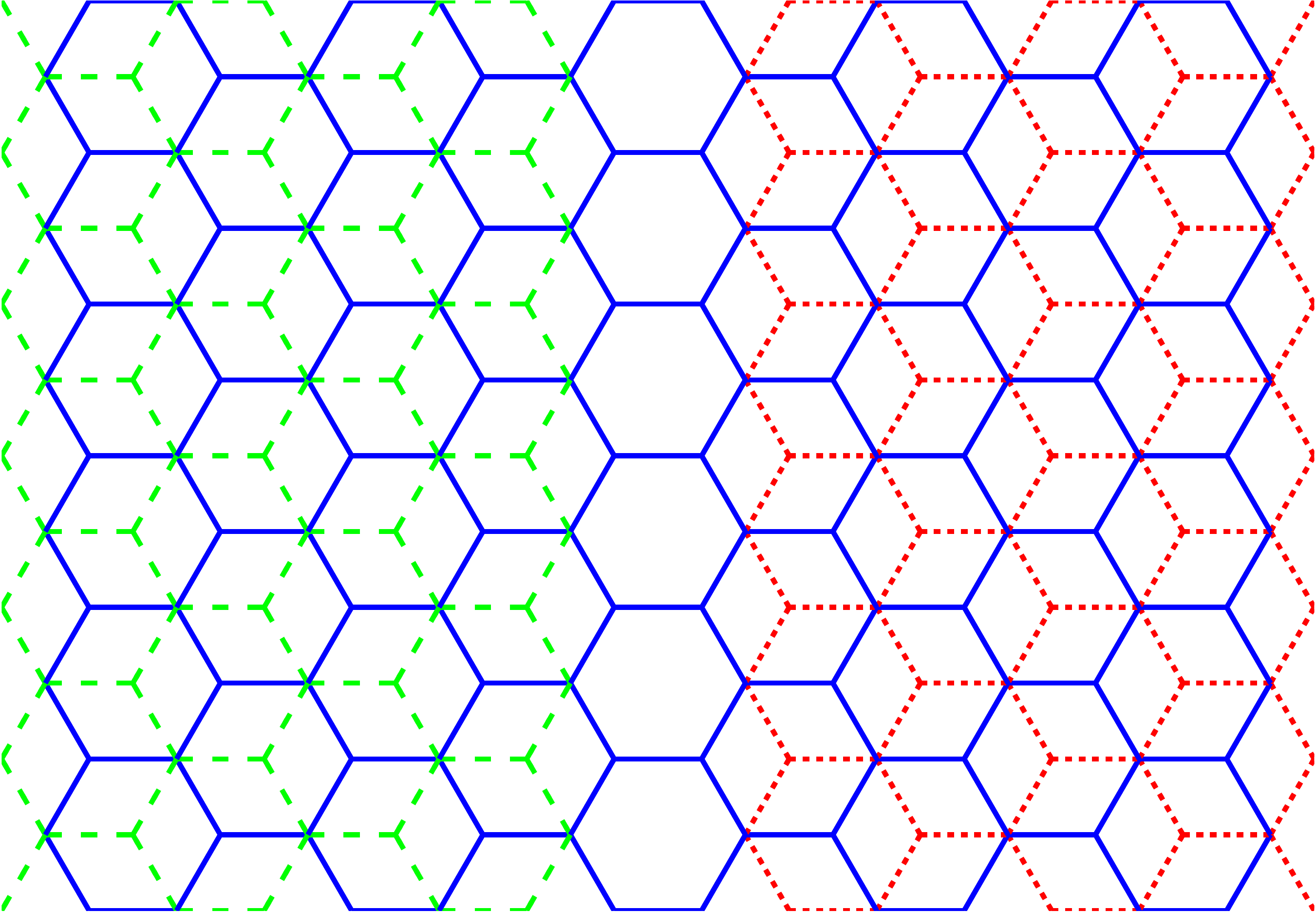} }}%
  \caption{Homogeneous and \gls{hlsw} system. The lower layer is in blue and the upper layer is striped or dotted in red and green. On the left is a homogeneous AB stacked \gls{blg} system, which will be used as a reference system. On the right is the hard wall model for a \gls{lsw} at the zigzag nanoribbon. The upper layer is decoupled, but any effective interaction may be of interest, in particular a fully coupled upper layer.}%
  \label{figure:normal_hard_wall}%
}
  
To define the Hamiltonian for a system with \gls{hlsw} we split the system into a left side $\mathcal{L}$ and a right side $\mathcal{R}$ like in \ref{figure:normal_hard_wall}. The full Hamiltonian of a \gls{hlsw} system is then given by
\begin{equation}
  \mathcal{H}_{\mathrm{HLSW}}=\mathcal{H}_{\mathrm{BLG}}^{\mathrm{(AB,\mathcal{L})}}+\mathcal{H}_{\mathrm{BLG}}^{\mathrm{(BA,\mathcal{R})}}+\mathcal{H}_{\mathrm{HLSW}}^{(L)}+\mathcal{H}_{\mathrm{HLSW}}^{(U)},
\end{equation}
in which $H_{\mathrm{BLG}}^{\mathrm{(AB,\mathcal{L})}}$ is the Hamiltonian for the \gls{blg} AB stacked bulk to the left and $H_{\mathrm{BLG}}^{\mathrm{(BA,\mathcal{R})}}$ for the BA stacked \gls{blg} bulk to the right as introduced in \ref{eq::H_BLG}. The remaining two terms $\mathcal{H}_{\mathrm{HLSW}}^{(L)}$ and $\mathcal{H}_{\mathrm{HLSW}}^{(U)}$ describe the hopping elements between the left and right part and therefore represent the hard wall given by

\begin{align}
  H_{\mathrm{HLSW}}^{(L)}&=-t_{\mathrm{LSW}}^{(L)}\sum_{\langle i,j\rangle}\left(a_i^{(L,\mathcal{L})}\right)^\dagger b_{j}^{(L,\mathcal{R})}+\hc{}\\
  H_{\mathrm{HLSW}}^{(U)}&=-t_{\mathrm{LSW}}^{(U)}\sum_{\langle i,j \rangle}\left(a_i^{(U,\mathcal{L})}\right)^\dagger b_{j}^{(U,\mathcal{R})}+\hc{}\, .
\end{align}

Here $\left(\hat{a}_i^{(n,\kappa )}\right)^{(\dagger )}$ $\left[\left(\hat{b}_i^{(n,\kappa )}\right)^{(\dagger )}\right]$ are the creation and annihilation operators for the $A$-sublattice [$B$-sublattice ] of the layer $n\in\{L,U \}$ and side $\kappa\in\{\mathcal{L},\mathcal{R} \}$ of the system.

In this work we always set the hopping integrals of the lower layer to be equal to the nearest-neighbor value $t_{\mathrm{LSW}}^{(L)}\equiv t=\SI{2.8}{\electronvolt}$. In contrast, the one of the upper layer is a model parameter which we will tune. For $t_{\mathrm{LSW}}^{(U)}=0$, we call the upper layer \emph{decoupled} while for $t_{\mathrm{LSW}}^{(U)}=t$ the upper layer is called \emph{fully coupled}.

A similar model can also be introduced as a simplification of an \gls{lsw} parallel to the armchair nanoribbon. The armchair hard wall model is, however, not nearly as widely used in the literature and we will not study it in detail.

\subsubsection{Shear and tensile layer switching wall}

The hard wall model is adequate if properties are investigated which are related only to the topology of the material such as the formation of conducting channels in an external electric field. However, from an ab initio perspective, this is not clear for the case of an external magnetic field. A more realistic microscopic description of the domain walls is therefore important. The first step to model the \gls{lsw} in \gls{blg} more realistically is to introduce an explicit finite lattice model for realistic \gls{lsw} geometries. The natural deformation of the underlying lattice, that allows a smooth transition between domains is obtained by tensing or shearing a single layer of graphene. The existence of the layer transition due to such transformations has been experimentally confirmed in \cite{Alden11256}. Let us stress that such a modeling of \gls{blg}, that is fixed tightly in two dimensions, is not expected to adequately describe a \gls{blg} system in which buckling occurs \cite{Butz2014}.

In order to explicitly define the shear and stretching transformation for \gls{blg} let us define first a general transformation. Given a lattice generated by primitive vectors $\left(\vec{p}_n\right)_{n=1,\ldots,N}$ and basis vectors $\left(\vec{b}_m\right)_{m=1,\ldots,M}$, denoted by

\begin{equation}
\mathcal{L}:=\left\{\left(\vec{p}_n\right)_{n=1,\ldots,N}, \left(\vec{b}_m\right)_{m=1,\ldots,M}\right\},
\end{equation}

we define its transformation due to a matrix $\mat{M}$ by

\begin{equation}
\mat{M}\mathcal{L}:=\left\{\left(\mat{M}\vec{p}_n\right)_{n=1,\ldots,N}, \left(\mat{M}\vec{b}_m\right)_{m=1,\ldots,M}\right\}.
\end{equation}

AB stacked \gls{blg} can then be described symbolically by the union of two lattices by $\mathcal{G}^{(AB)}:=\mathcal{G}_L^{(AB)}\cup\mathcal{G}_U^{(AB)}$, where $\mathcal{G}_L^{(AB)}$ and $\mathcal{G}_U^{(AB)}$ correspond to the lower and upper layer of an AB stacked \gls{blg} respectively.

We are now in a position to define a sheared \gls{blg} lattice by $\mathcal{G}_s^{(AB)}(L_s)=\mathcal{G}_L^{(AB)}\cup\mat{S}(L_s)\mathcal{G}_U^{(AB)}$, where

\begin{equation}
  \mat{S}(L_s)=\begin{pmatrix}
    1 & 1/L_s \\
    0 & 1
  \end{pmatrix}
\end{equation}

is a shear transformation of the upper layer and $L_s$ is the shear strength (or width). Similarly, tensed \gls{blg} is described by $\mathcal{G}^{(AB)}_t(L_t)=\mathcal{G}_L^{(AB)}\cup\mat{T}(L_t)\mathcal{G}_U^{(AB)}$, where

\begin{equation}
  \mat{T}(L_t)=\begin{pmatrix}
    1 & 0 \\
    0 & 1+1/L_t
  \end{pmatrix}
\end{equation}

is a stretching transformation and $L_t$ is the tensile strength (or width). Given the proper extent of the tensed region ($2aL_t/\sqrt{3}$) and of the sheared region ($aL_{s}/\sqrt{3}$), such a section of \gls{blg} transforms AB stacked \gls{blg} to BA stacked \gls{blg}. Equivalent lattices $\mathcal{G}^{(BA)}_t(L_t)$ and $\mathcal{G}^{(BA)}_s(L_s)$ can also be defined.

\stfig{
  \subfloat{{\includegraphics[width=0.45\columnwidth]{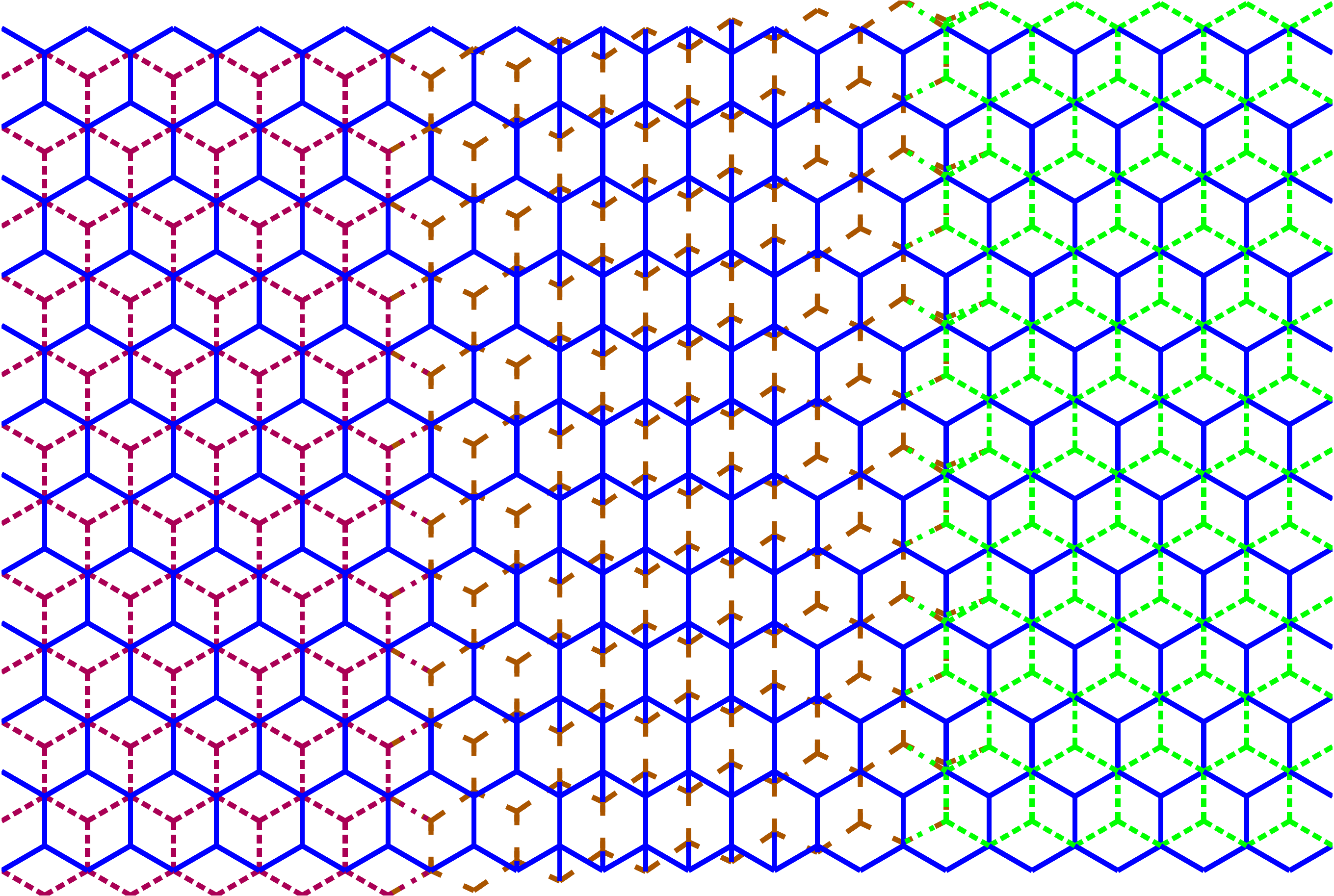} }}%
  \hfill
  \subfloat{{\includegraphics[width=0.45\columnwidth]{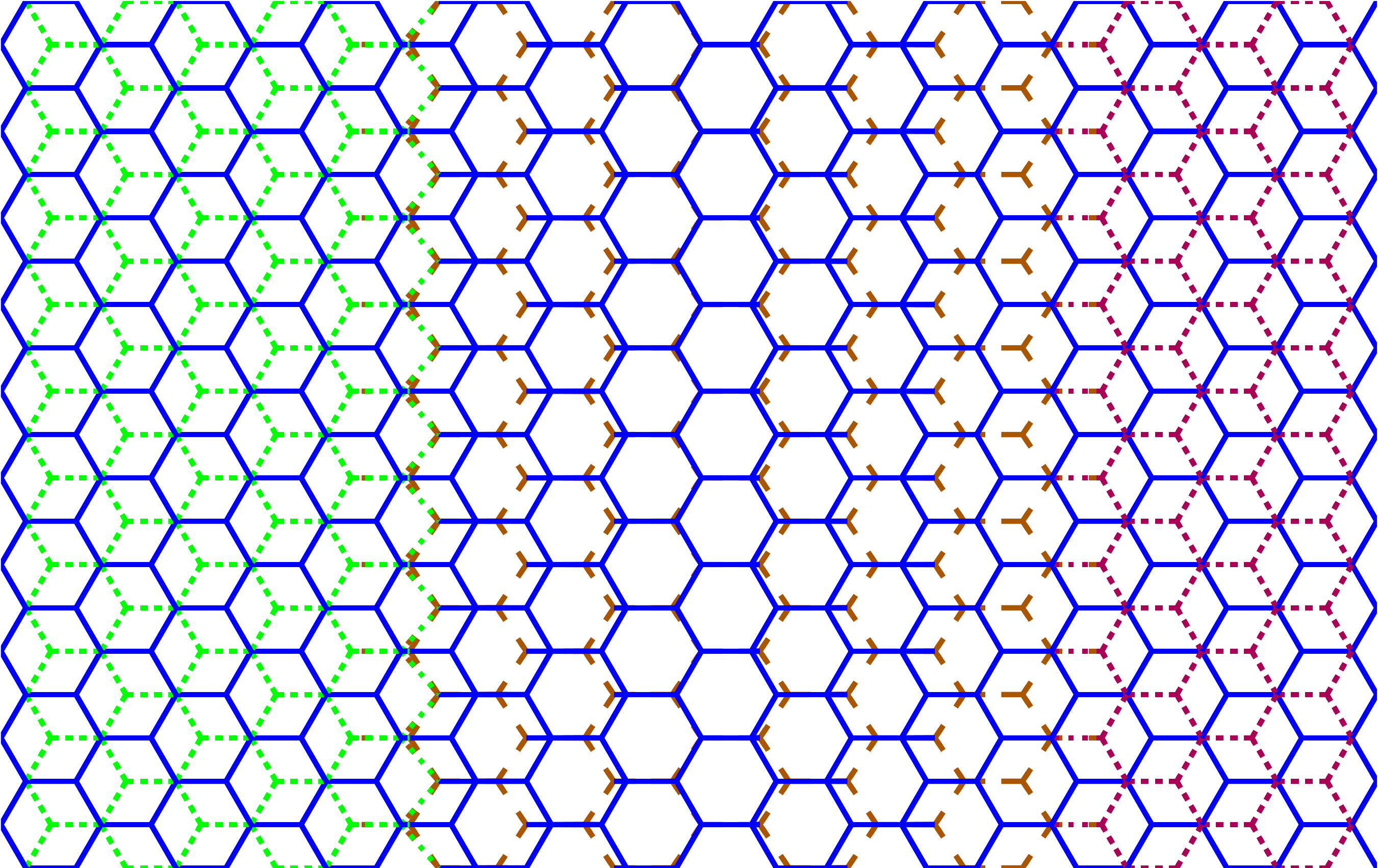} }}%
  \caption{Heterogeneous systems with \glspl{lsw} containing \gls{blg} under shear (left) and tension (right). The lower layer is the solid blue lattice. The upper layer is in both cases either dotted or striped in green, orange and red respectively. Only the nearest-neighbour hoppings are shown.}%
  \label{figure:skew_tension}%
}
  
The previously defined lattices are shown in \ref{figure:skew_tension}, where the \gls{slsw} and \gls{tlsw} regions were chosen such that they exactly transform from AB stacked \gls{blg} bulk (left) to BA \gls{blg} bulk (right). It is clear by examining \ref{figure:skew_tension} that shear causes an \gls{lsw} parallel to the armchair nanoribbon and tension an \gls{lsw} parallel to the zigzag nanoribbon.  
\section{Observables and system geometry}\label{section:obs_geom}
We start by defining the observables of interest. Then we describe the transport system setup and which system parameters are important for the calculation of the relevant observables.

These observables can be calculated using the usual nonequilibrium Green's function formalism. The retarted, advanced, and lesser Green's function are defined as
\begin{equation}
  \begin{split}
    G_{\rm r}(\vec{r},t,\vec{r}',t')&=-\imag\theta(t-t')\left\langle\anticommutator{\Psi(\vec{r},t)}{\Psi^\dagger(\vec{r}',t')}\right\rangle\\
    G_{\rm a}(\vec{r},t,\vec{r}',t')&=\imag\theta(t'-t)\left\langle\anticommutator{\Psi(\vec{r},t)}{\Psi^\dagger(\vec{r}',t')}\right\rangle\\
    G^<(\vec{r},t,\vec{r}',t')&=\imag\left\langle\Psi^\dagger(\vec{r}',t')\Psi(\vec{r},t)\right\rangle
  \end{split}
\end{equation}
and have interpretations as particle and hole propagators as well as correlators, respectively. $\Psi(\vec{r},t)$ and $\Psi^\dagger(\vec{r},t)$ are the usual field creation and destruction operators, that represent a particle field localized in space and will be replaced by their discrete counterparts for the lattice model. If the temperature of the system is zero, $\langle\ldots\rangle$ is evaluated with respect to the ground state and if the temperature is nonzero, it is evaluated with respect to the density matrix of the system. In this work, we will fully focus on the zero-temperature case. Given time homogeneity, the evaluation of the retarded and advanced quantities $G_{\rm r}$ and $G_{\rm a}$ simplify in the Fourier domain to $G_{r/a}(E)=\lim_{\eta\to 0}(E\pm\imag\eta-\mathcal{H}_S)^{-1}$ so that their evaluation corresponds to an inversion of the system Hamiltonian. The effect of the electrodes (leads) is modeled with an appropriate retarded self energy $\Sigma_{r,\mathrm{lead}}=V_{S,\mathrm{lead}}G_{r,\mathrm{lead}}V_{S,\mathrm{lead}}^\dagger$ obtained from the retarded surface Green's function of a semi-infinite chain $G_{r,\mathrm{lead}}$ and coupling $V_{S,\mathrm{lead}}$ of the chain to the system without electrodes (scattering region). For all systems under consideration in this article $G_a=G_r^{\dagger}$. The lesser Green's function can be obtained from the appropriate quantum kinetic equations, which are the fluctuation dissipation theorem and the Keldysh equation. These read

\begin{equation}
  \begin{split}
    \Sigma^< (E)&=-\imag\sum\displaylimits_nf(E, \mu_n)\left(G_a^{(n)}(E)-G_r^{(n)}(E)\right),\\
    G^<(E)&=G_r(E)\Sigma^<(E)G_a(E),
  \end{split}
\end{equation}

where the index $(n)$ enumerates the electrodes at fixed chemical potentials $\mu_n$ and energy $E$, and $f(E,\mu_n)$ corresponds to the Fermi distribution describing the electron filling of the $n$th lead.

Given an algorithm for the evaluation of these Green's functions, most relevant observables may be evaluated. Two observables of particular interest are the conductance from electode $A$ to $B$

  \begin{equation}
    \sigma_{A\rightarrow{}B}(E)=\frac{2\mathrm{e}}{\hbar}\trace\left(\Gamma^{(A)}G_r\Gamma^{(B)}G_a\right)
  \end{equation}

  with the coupling matrices $\Gamma^{(A/B)}=\imag\left(\Sigma^{(A/B)}_r-\Sigma^{(A/B)}_a\right)$ for the repspective leads and the \gls{ldos} at site $i$

  \begin{equation}
    \rho_i(E)=\frac{1}{2\pi}\Im\left(G^<_{ii}(E)\right)
  \end{equation}

  in the zero temperature, zero bias limit at energy $E$. All equations presented above become matrix expressions in the case of a discrete lattice model and a calculation for large systems is possible with sophisticated algorithms as outlined in \ref{section:implementation}.

\subsubsection{Hall bar geometry}

The Hall bar geometry is an appropriate setup to examine systems in the \gls{iqhe} regime and is particularly useful for the analysis of systems with \gls{lsw}s. A sketch of the generic geometry of this measurement is shown in \ref{figure:hallbar}. It is a six-terminal device, that can resolve the transport physics at the \gls{lsw}, since transmitted and reflected modes can easily be distinguished by the structure of $\SPass{}$ and $\SLSW{}$. The system displayed in \ref{figure:hallbar} has an \gls{lsw} due to tension in the upper layer in the center of the region. For other \glspl{lsw}, the central region is replaced by the appropriate lattices as defined in \ref{section:obs_geom}.

\stfig{
  \centering
  \includegraphics[width=\columnwidth]{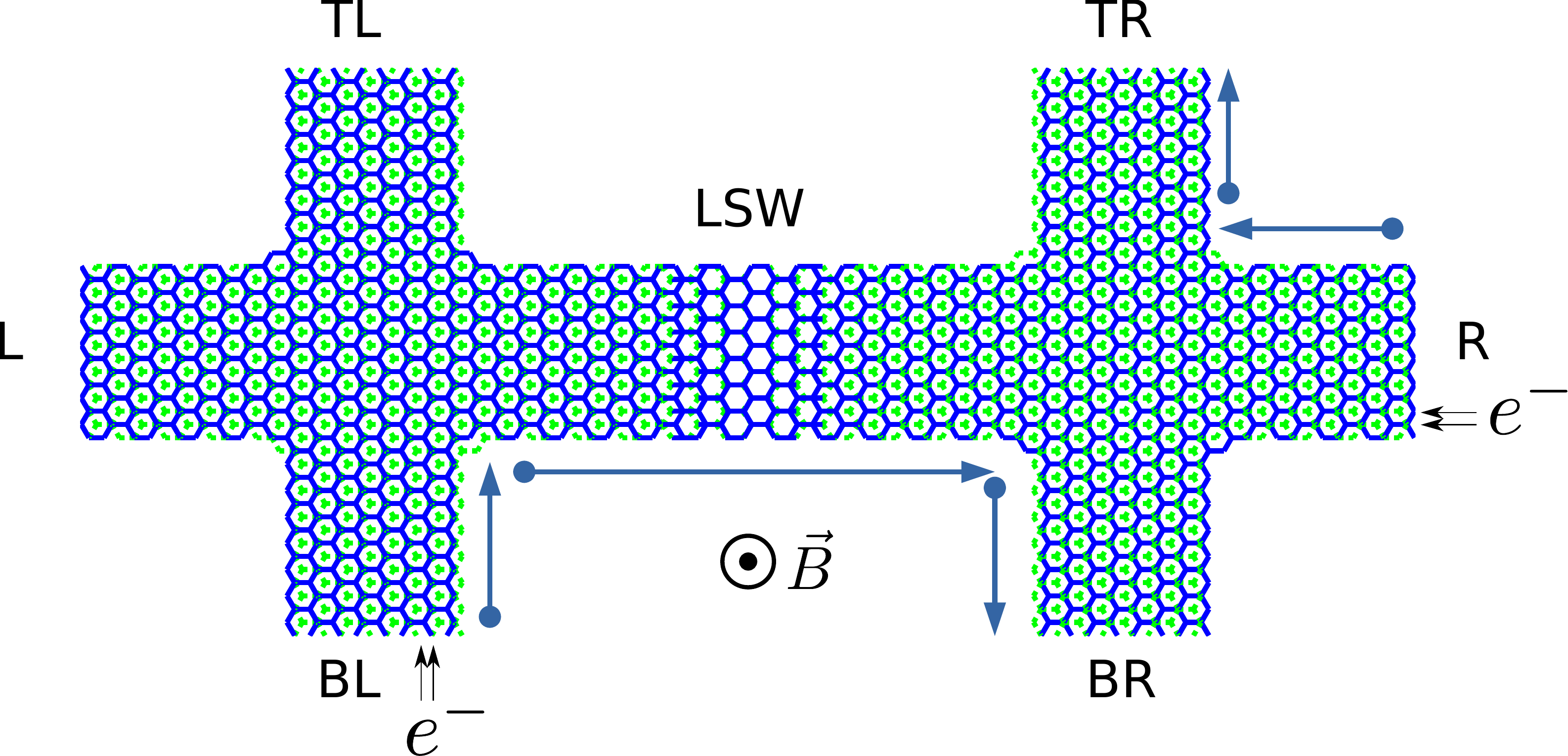}
  \caption{Hall bar system with \gls{tlsw} represeting a general \gls{lsw}. The lower graphene layer is solid blue and the upper layer is striped green. This is the generic system which will be used to discuss the \gls{iqhe} in inhomogenous or homogeneous \gls{blg} systems. It is a six-terminal structure with a lead connected to the system at each of the labels \LE{} (left), \RE{} (right), \BLE{} (bottom left), \BRE{} (bottom right), \TLE{} (top left) and \TRE{} (top right). The electron propagation in a homogeneous medium is indicated by blue arrows. The electron injection electrodes are also indicated with arrows.}
  \label{figure:hallbar}
}
  
\subsubsection{System parameters}

We will use the following definitions and nomenclature for discussing the physical properties of \gls{blg} with layer switching domain walls:

\begin{itemize}
  \item $N_{\text{C}}$/$N_{\text{L}}$: Number of sites in the conductance/\gls{ldos} calculation.
\item $D_{\text{C}}$: Cutoff distance for hopping integrals in the upper layer and between layers in the \gls{lsw}, where the graphene layers are thought of as being at the same distance as two nearest carbon atoms within a layer, $d_0 = a_{0}$. The couplings within the lower layer are always a nearest neighbor hopping.
\item $\phi_{\mathrm{LDOS}}^{-1}$: Magnetic field value for the \gls{ldos} calculation. This value represents a physical magnetic field, but the Fermi energies for the transport calculations were chosen to be very large to suppress finite-size effects. For this reason the resulting magnetic fields would be unrealistically large. Thus this is kept as a parameter $\phi_{\mathrm{LDOS}}^{-1}$ and not converted into absolute Tesla units.
\item $E_{\text{F}}$: Fermi energy at which the calculations take place. All calculations are performed at zero temperature and zero bias, such that only the definition of the Fermi energy and no other chemical potential is required. In all calculations in the main part of the paper we fixed the Fermi energy as $E_{\text{F}}=0.35t\approx\SI{0.95}{\electronvolt}$.
\end{itemize}

Finally, we have added a random onsite disorder for all calculations corresponding to Gaussian noise with a width of $\Delta=0.05t\approx\SI{0.14}{\electronvolt}$ around zero. This disorder is added to model the disorder required in real systems to observe the \gls{iqhe} and to break unrealistic symmetries of the Hall bar geometry. The results however shall not depend vitally on the presence of such a disorder (calculations not shown). 
\section{Implementaion}\label{section:implementation}
The transport experiment simulation is essentially a recursive Green's function method implemented in \CC{}. From the various particular implementation strategies of the bulk \cite{Groth_2014, khomyakov2004real, khomyakov2005conductance, Kazymyrenko2008} and lead self-energy calculation \cite{Lee1981, Ando1991, sorensen2008krylov}, a block Gaussian elimination solver for tridiagonal matrices \cite{petersen2008block} was chosen for the bulk recursion and a modified iteration \cite{sancho1985highly} strategy was chosen for the lead self-energy calculation. Two particular additions to the implementation used for the results presented here are not discussed in these respective publications. They are presented including a summary of the algorithms from \cite{petersen2008block} and \cite{sancho1985highly}.

\subsection{Calculation of lead self-energies}

As mentioned previously, the electrodes (leads) are modeled by a semi-infinite chain. Calculation of the surface Green's function and thus the self-energy of such a chain can be efficiently performed by a chain decimation algorithm described in \cite{sancho1985highly}. A short summary will be given here. To this end, define a periodically coupled chain with chain vertex (onsite) term $\mat{H}_{\mathrm{chain}}$ and chain hopping (interaction) term $\mat{V}_{\mathrm{chain}}$. Then recursive equations for a chain decimation with starting conditions \mbox{$\mat{\epsilon}_0=\mat{\epsilon}_0^s=\mat{H}_{\mathrm{chain}}$} equal to the chain onsite Hamiltonian and \mbox{$\mat{\alpha}_0=\mat{\beta}_0^\dagger=\mat{V}_{\mathrm{chain}}$} equal to the chain hopping Hamiltonian read

  \begin{equation}\begin{split}
    \mat{\alpha}_i&=\mat{\alpha}_{i-1}(\omega\mat{I}-\mat{\epsilon}_{i-1})^{-1}\mat{\alpha}_{i-1}, \\
    \mat{\beta}_i&=\mat{\beta}_{i-1}(\omega\mat{I}-\mat{\epsilon}_{i-1})^{-1}\mat{\beta}_{i-1},\\
    \mat{\epsilon}_i&=\mat{\epsilon}_{i-1}+\mat{\alpha}_{i-1}(\omega\mat{I}-\mat{\epsilon}_{i-1})^{-1}\mat{\beta}_{i-1}\\
    &\hphantom{=\mat{\epsilon}_{i-1}\,\,}+\mat{\beta}_{i-1}(\omega\mat{I}-\mat{\epsilon}_{i-1})^{-1}\mat{\alpha}_{i-1},\\
    \mat{\epsilon}^s_i&=\mat{\epsilon}^s_{i-1}+\mat{\alpha}_{i-1}(\omega\mat{I}-\mat{\epsilon}_{i-1})^{-1}\mat{\beta}_{i-1},
  \end{split}
  \end{equation}

with $\omega=E+\imag\eta$ for the retarded solution which encodes a decimation of $2^n$ chain links after $n$ iterations and the relevant chain surface Green's function becomes \mbox{$G(\omega)=(\omega\mat{I}-\mat{\epsilon}_n^s)^{-1}$} for a sufficiently large $n$.

An optimization to this algorithm which has not been discussed in the literature so far to the best of our knowledge is a reduction in matrix dimension of the Hamiltonian of the semi-infinite chain describing the lead. This reduction in matrix dimension is effective for chains, where the periodic coupling is sufficiently sparse. Such an optimization is effective since matrix inversion of sparse matrices does not produce sparse matrices in general and the expressions thus become dense after a few iteration steps. The size reduction is obtained by integrating the parts of the chain Hamiltonian, that do not participate in the periodic coupling into an effective Hamiltonian of reduced size including the interior parts of the chain.

For this purpose, we define the following quantities: The \emph{left} section of the chain, which is periodically coupled to the previous chain link and the interior section, which is not coupled to the previous chain link. From these definitions, we obtain $\mat{H}_L$ and $\mat{H}_I$, the respective subsystem matrices and $\mat{V}_{IL}$ and $\mat{V}_{LI}$, the couplings between them ($\mat{V}_{IL}$ is the coupling to the next period's \emph{left} section). We also define the retarded Green's function $\mat{G}_I$ of the interior subsystem. Using the notation from \cite{sancho1985highly}, the following starting parameters can be used:

\begin{equation}\begin{split}
  \mat{\alpha}_0&= \mat{\beta}_0^\dagger=\mat{V}_{LI} \mat{G}_I \mat{V}_{IL}\\
  \mat{\epsilon}_0&= \mat{H}_L + \mat{V}_{LI} \mat{G}_I \mat{V}_{LI}^\dagger + \mat{V}_{IL}^\dagger \mat{G}_I \mat{V}_{IL}\\
  \mat{\epsilon}_0^s&= \mat{V}_{LI} \mat{G}_I \mat{V}_{LI}^\dagger.
\end{split}\end{equation}

This effectively reduces the system size in the self-energy iteration from the dimension of an entire chain link $d_{\mathrm{full}}$ to the dimension of its \emph{left} section $d_{\mathrm{left}}$ implying a reduction in complexity for all matrix operations performed from $\mathcal{O}(d_{\mathrm{full}}^3)$ to $\mathcal{O}(d_{\mathrm{left}}^3)$.

\subsection{Calculation of select Green's function matrix elements}

 Given the definition $\mat{A}=\mat{E}-\mat{H}-\mat{\Sigma}_\mathrm{lead}$ of the transport problem we want to solve for matrix elements of $\mat{G}=\mat{A}^{-1}$. The reference \cite{petersen2008block} defines a modified Gaussian elimination, that uses a partition of $\mat{A}$ into blocks $\mat{a}_{nm}$ such that $\mat{A}$ is block tridiagonal. To this end, a forward and backward elimination (left and right sweep in the more physics-oriented discussions of the recursive Green's function technique) are performed to obtain the matrices

  \begin{align}
      \mat{d}^{\mathrm{L}}_{ii}&=\mat{a}_{ii}+\mat{c}^{\mathrm{L}}_{i-1}\mat{a}_{i-1,i}\quad &&i=2,3,\ldots,n\nonumber\\
      \mat{c}^{\mathrm{L}}_i&=-\mat{a}_{i+1,i}(\mat{d}^{\mathrm{L}}_{ii})^{-1}\quad &&i=1,2,\ldots,n-1\\
      \mat{d}^{\mathrm{R}}_{ii}&=\mat{a}_{ii}+\mat{c}^{\mathrm{R}}_{i+1}\mat{a}_{i+1,i}\quad &&i=n-1,\ldots,2,1\nonumber\\
      \mat{c}^{\mathrm{R}}_i&=-\mat{a}_{i-1,i}(\mat{d}^{\mathrm{R}}_{ii})^{-1}\quad &&i=n,\ldots,3,2\nonumber
  \end{align}

  with $\mat{d}^{\mathrm{L}}_{11}=\mat{a}^{\mathrm{L}}_{11}$ and $\mat{d}^{\mathrm{R}}_{nn}=\mat{a}^{\mathrm{R}}_{nn}$, which can be combined to compute select  matrix blocks of the retarded Green's function

  \begin{align}
    \mat{G}_{ii}&=(-\mat{a}_{ii}+\mat{d}_{ii}^{\mathrm{L}}+\mat{d}_{ii}^{\mathrm{R}})\nonumber\\
    \mat{G}_{ij}&=\mat{G}_{ii}\mat{c}^{\mathrm{R}}_{i+1}\mat{c}^{\mathrm{R}}_{i}\ldots\mat{c}_{j}^{\mathrm{R}}\quad &&\text{for }i<j\\
    \mat{G}_{ij}&=\mat{G}_{ii}\mat{c}^{\mathrm{L}}_{i-1}\mat{c}^{\mathrm{L}}_{i-2}\ldots\mat{c}_{j}^{\mathrm{R}}\quad &&\text{for }i>j\nonumber.
  \end{align}

  Depending on the choice of blocks for a given system, these matrix elements may be used to obtain the desired observable.

\stfig{
  \includegraphics[width=0.8\columnwidth]{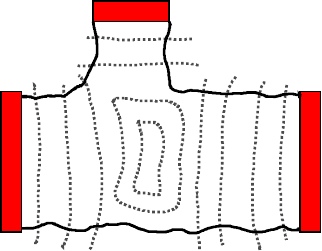}
  \caption{Schematic depiction of a partition of a quite complicated geometry into appropriate subsystems such that the resulting Hamiltonian matrix is block tridiagonal with small blocks. Leads are shown in red and the dotted lines in grey show the lines that define the subsystems.}
  \label{fig:split}
}

To apply a solver for tridiagonal matrices to arbitrary systems, a reduction of fairly arbitrary geometries into blocks shown schematically in \ref{fig:split} must be defined. This is in principle a difficult task, closely related to the graph partitioning problem. Thus a simple and versatile graph partitioning method was defined. Though non-optimal, the simplicity of the algorithm makes it suitable for the application to any lattice geometry.

\begin{itemize}
\item Define a starting set of system sites $\mathcal{S}_0$. For a transmission calculation, this contains the source and drain lead. For a density calculation, only the source lead sites are included (source and drain leads are abstractions for multiple electrodes, where electrons are injected and not injected respectively).
\item Given $\mathcal{S}_i$, define $\mathcal{S}_{i+1}$ as the collection of all sites $s$, such that some site in $\mathcal{S}_i$ has a nonzero hopping element to $s$ and none in $\mathcal{S}_j$ for $j<i$ do. Formally
\begin{align*}
\mathcal{S}_{i+1}:=&\{\,s\mid{}\exists{}g\in\mathcal{S}_i:t(s,g)\neq 0\\
&\land{}k<i\implies\forall{}g\in\mathcal{S}_k:t(s, g)=0\,\}.
\end{align*}
\item Stop when $\mathcal{S}_i=\emptyset$.
\end{itemize}

\widefig{
  \adjustbox{width=\widewidth}{%
    \includegraphics[height=\LDOSsize]{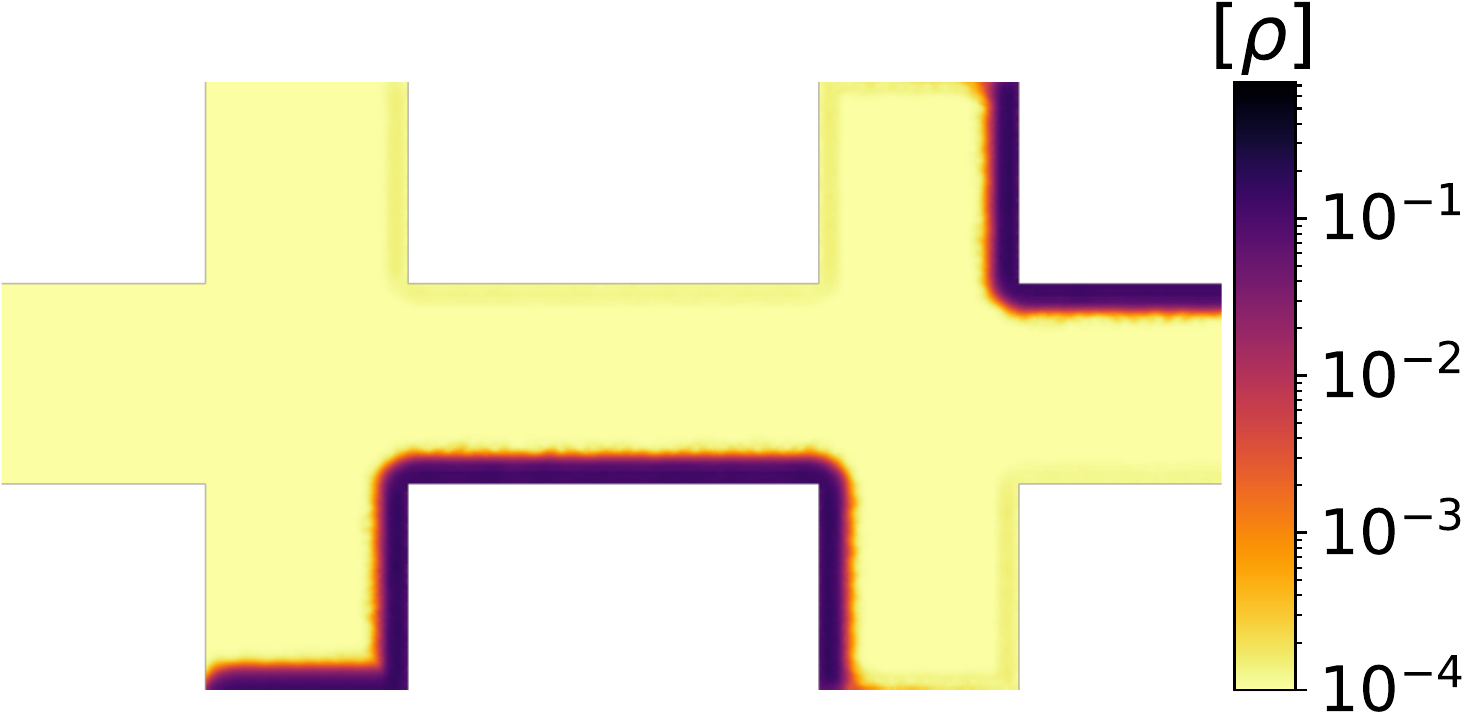}%
    \hspace{\hspacing}%
    \raisebox{\raisesize}{%
      \includegraphics[height=\TRANSsize]{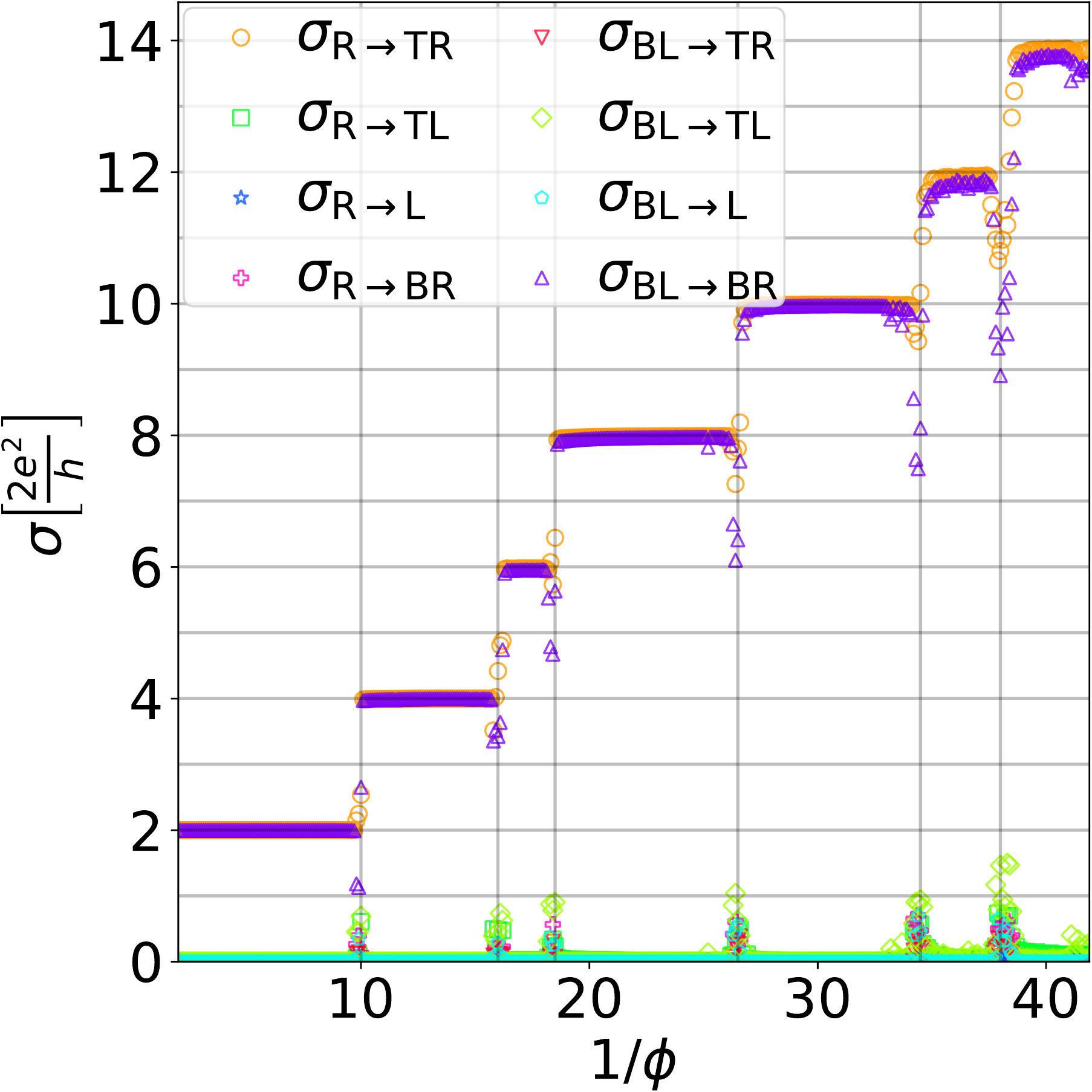}
    }%
  }%
  \caption{\HallbarPlot{a homogeneous AB stacked Hall bar system}. \LDOSphi{10.5}. \LDOSNumber{\num{5e5}}. \TransNumber{\num{1.5e5}}.}%
  \label{figure:homogenous_hallbar}%
}
 
These $\mathcal{S}_i$ define a natural block structure for any system geometry with reasonably small matrix dimensions producing a block tridiagonal Hamilton matrix. For the transmission calculations only the matrix elements contained in $\mathcal{S}_0$ are calculated and for density calculations matrix elements from $\mathcal{S}_0$ to all $\mathcal{S}_i$ are calculated. This has the formal difference to \cite{petersen2008block}, that for transmission calculations the source and drain electrodes are considered to both be elements of the first matrix block leading to more favourable block structures for particularly complicated geometries.

The methods described here will all be applied to the Hallbar geometry discussed earlier for which each lead self-energy calculation can be performed separately and for which the system blocking should be similar to the one shown in \ref{fig:split}. In order to perform parameter studies in a reasonable time frame, systems of the order $10^5$ sites were chosen implying a maximal matrix block size of $\approx 10^3$. The self-energy calculation without optimization would involve matrices of the order $\approx 4 \cross 10^3$, but is reduced to $\approx 10^3$ reducing the runtime considerably due to the cubic complexity of all matrix algorithms involved. 
\section{Results}\label{section:results}
In this section we present and discuss our magnetoconductance calculations for the introduced \gls{blg} systems. First, results for the the homogeneous system will be described as an important reference. In a next step the hard wall layer switching system is investigated as a very simple model for an \gls{lsw}. Then the more realistic sheared Hall bar with an \gls{lsw} along the armchair nanoribbon and the tensile Hall bar with an \gls{lsw} along the zigzag nanoribbon are studied. For each system an \gls{ldos} calculation will be shown. In addition, since the latter only show the properties at a single magnetic field value, conductance calculations are also shown for more in-depth analysis of system properties with respect to the external magnetic field.

\subsection{\homohead}\label{subsection:homoblg_calc}

\widefig{
  \adjustbox{width=\widewidth}{%
    \includegraphics[height=\LDOSsize]{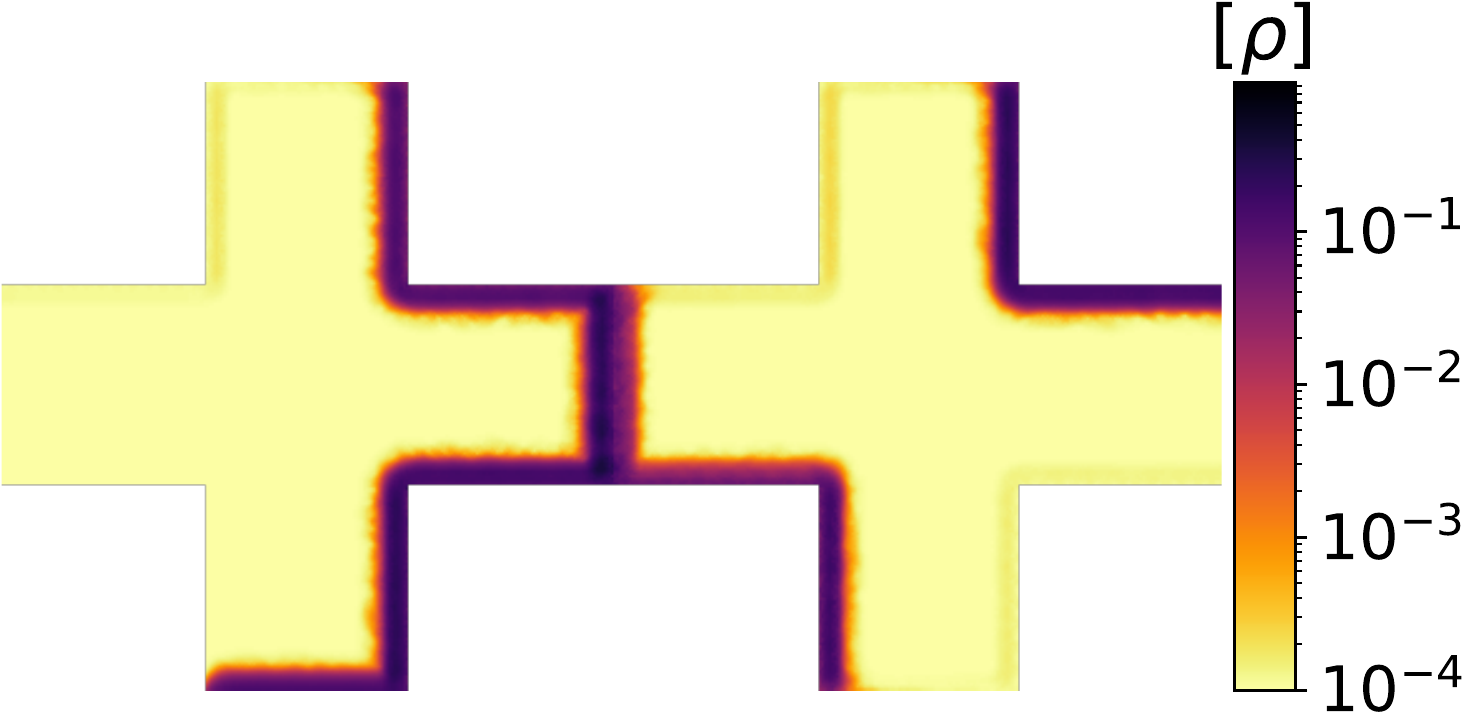}%
    \hspace{\hspacing}%
    \raisebox{\raisesize}{%
      \includegraphics[height=\TRANSsize]{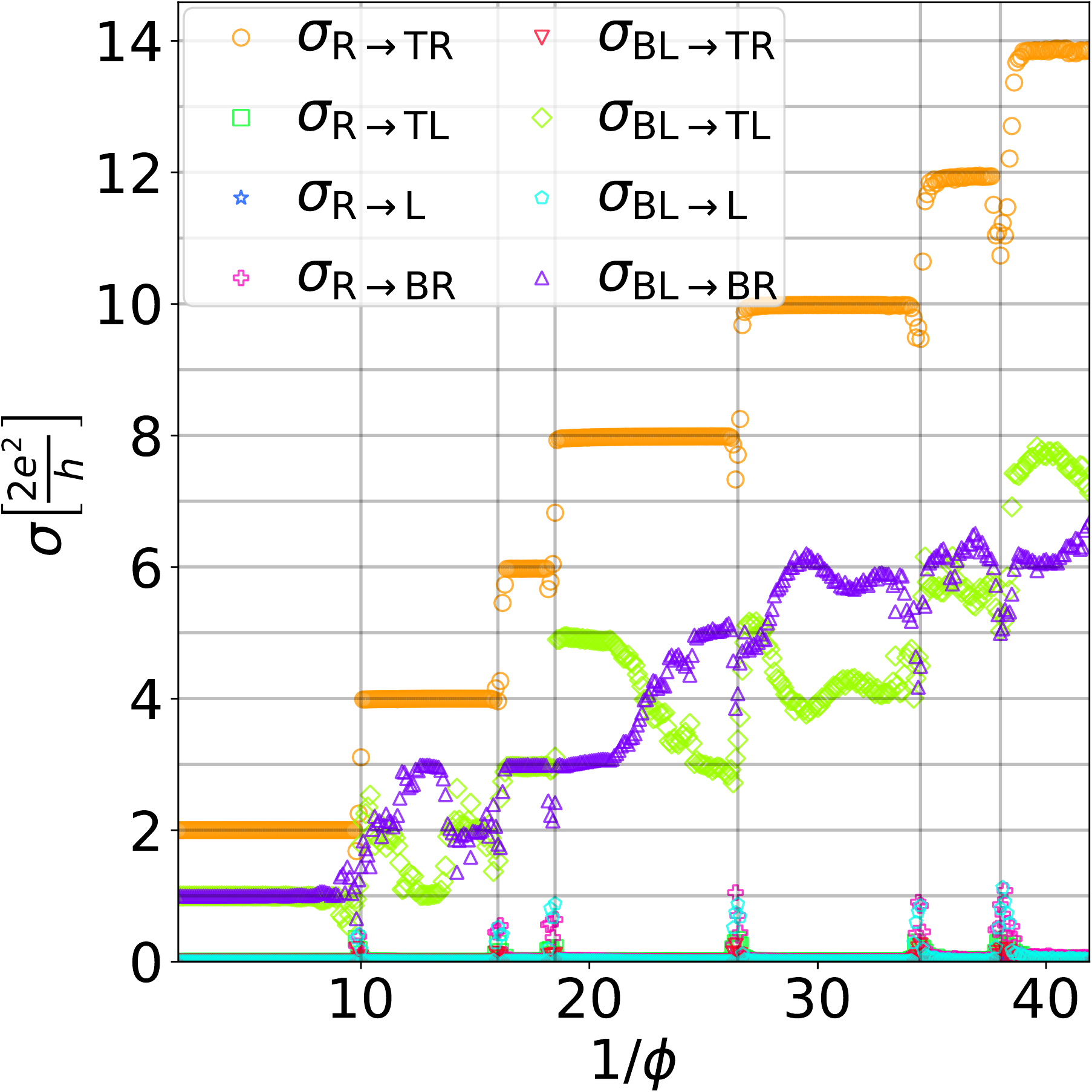}%
    }%
  }%
  \caption{\HallbarPlotRef{a Hall bar system with \gls{hlsw}}. \LDOSphi{10.5}. \LowerCoupled. \UpperDecoupled. \LDOSNumber{\num{5e5}}. \TransNumber{\num{1.5e5}}.}%
  \label{figure:hardwall_hallbar}%
}
 
As a reference and in order to check the validity of our numerical implementation, we first discuss a homogeneous AB stacked \gls{blg} system with a Hall bar geometry. The \gls{ldos} and conductance calculations for the homogeneous system are displayed in \ref{figure:homogenous_hallbar}. Both the \gls{ldos} and conductance calculation indeed show the expected behaviour for a homogeneous \gls{blg} system. The \gls{ldos} calculation in \ref{figure:homogenous_hallbar} clearly displays modes for a magnetic field $\phi^{-1}_{\rm LDOS}=10.5\phi_0^{-1}$ located within the second conductance plateau, that are localized at the system edge. The propagation direction due to the magnetic field defined in \ref{figure:hallbar} is also in agreement with the expected anti-clockwise cyclotron orbits.

The conductance calculation displayed on the right of \ref{figure:homogenous_hallbar} already has a lot of structure, even for the homogeneous system. Since spin is not accounted for, the plateau structure is $2(n+1)$ for $n=0,1,2,\ldots$ affirming an anomalous integer quantum Hall effect due to the degeneracy of the first Landau level. The weakening of the localization argument at the edges of the Landau levels can also be clearly seen in the nonzero contributions to the conductance e.g. in $\sigma_{\text{BL}\rightarrow{}\text{TL}}$, which would be characterized by nonzero \gls{ldos} inside the material bulk implying the contribution of previously gapped bulk states. Another striking aspect of the conductance plateau is their irregular spacing caused by the underlying electronic structure of \gls{blg}, that modifies the equidistant free (purely parabolic dispersion) electron solution \cite{novoselov2006unconventional}. At larger values of $1/\phi$ (smaller magnetic fields), in particular for the sixth plateau, the larger magnetic length scale implied by the external magnetic field already leads to some finite-size effects, causing the occupation of bulk modes which do not appear in a macroscopically large material. This is particularly obvious, when comparing $\SRef{}$ for which the gauge was chosen to be commensurable with the injection lead periodicity and $\SPass{}$ for which the gauge was thus not compatible with the injection lead periodicity, such that localization of bulk modes in the smaller material section in the bottom left fails for higher Landau levels.

\subsection{\hardhead}

Representative results for the \gls{hlsw} model as illustrated in \ref{figure:normal_hard_wall} with a decoupled upper layer and a fully coupled lower layer are displayed in \ref{figure:hardwall_hallbar}. The conductance curves of particular interest are $\SPass{}$, which corresponds to the conductance across the \gls{lsw}, and $\SLSW{}$, which corresponds to the conductance parallel to the \gls{lsw}. As long as the transport states are localized at the system edge, implying little finite-size effects and a magnetic field value not located at a plateau edge, $\SRef{}$ is a reference to the integer quantum Hall conductance in a homogeneous \gls{blg} sample as discussed in the last subsection. Indeed, as expected, the conductance $\SRef{}$ far away from the \gls{hlsw} is the same as the one for the homogeneous system in \ref{figure:homogenous_hallbar}, exhibiting a typical anomalous \gls{iqhe} in \gls{blg}.

The \gls{ldos} calculation in \ref{figure:hardwall_hallbar} shows localized edge modes just as in \ref{figure:homogenous_hallbar} when far away from the \gls{hlsw}. There is however also an increase in electron density in the region of the \gls{hlsw}, indicating transport parallel to the \gls{hlsw} from the electrode \BLE{} towards the electrode \TLE{}. The quantities $\SLSW{}$ and $\SRef{}$ most related to the structure of the \gls{lsw} have an entirely different structure from the homogeneous \gls{blg} system. First of all, $\SPass{}+\SLSW{}=\SRef{}$ for all values of $\phi^{-1}$ not close to plateau transitions in $\SRef{}$. This strengthens the intuition, that the \gls{tlsw} either causes conductance parallel or transverse to it and no scattering to any other electrode occurs when the edge modes are sufficiently localized in the homogeneous system.

For large magnetic fields (small $\phi^{-1}$), the homogeneous system exhibits a conductance of $\SPass{}=2\sigma_o$, whereas the decoupled \gls{hlsw} system has the value $\SPass{}=\sigma_o$. This reflects the fact, that only the lower layer is coupled and the mode localized at the edge of the upper layer thus propagates parallel to the \gls{hlsw}, which acts as a system edge of the upper sheet. This can also be confirmed by closer examination of an \gls{ldos} calculation for large magnetic fields (not shown). With a similar intuition, the magnitude of $\SPass{}$ and $\SLSW{}$ seems to be similar when averaged over an entire plateau. $\SLSW{}$ is larger at the edges of plateaus and $\SPass{}$ at the centers. Physically, an explanation similar to the reasoning for the existence of larger longitudinal conductance at plateau transitions is reasonable corresponding to available delocalized modes to scatter into.

\stfig{
\includegraphics[width=\columnwidth]{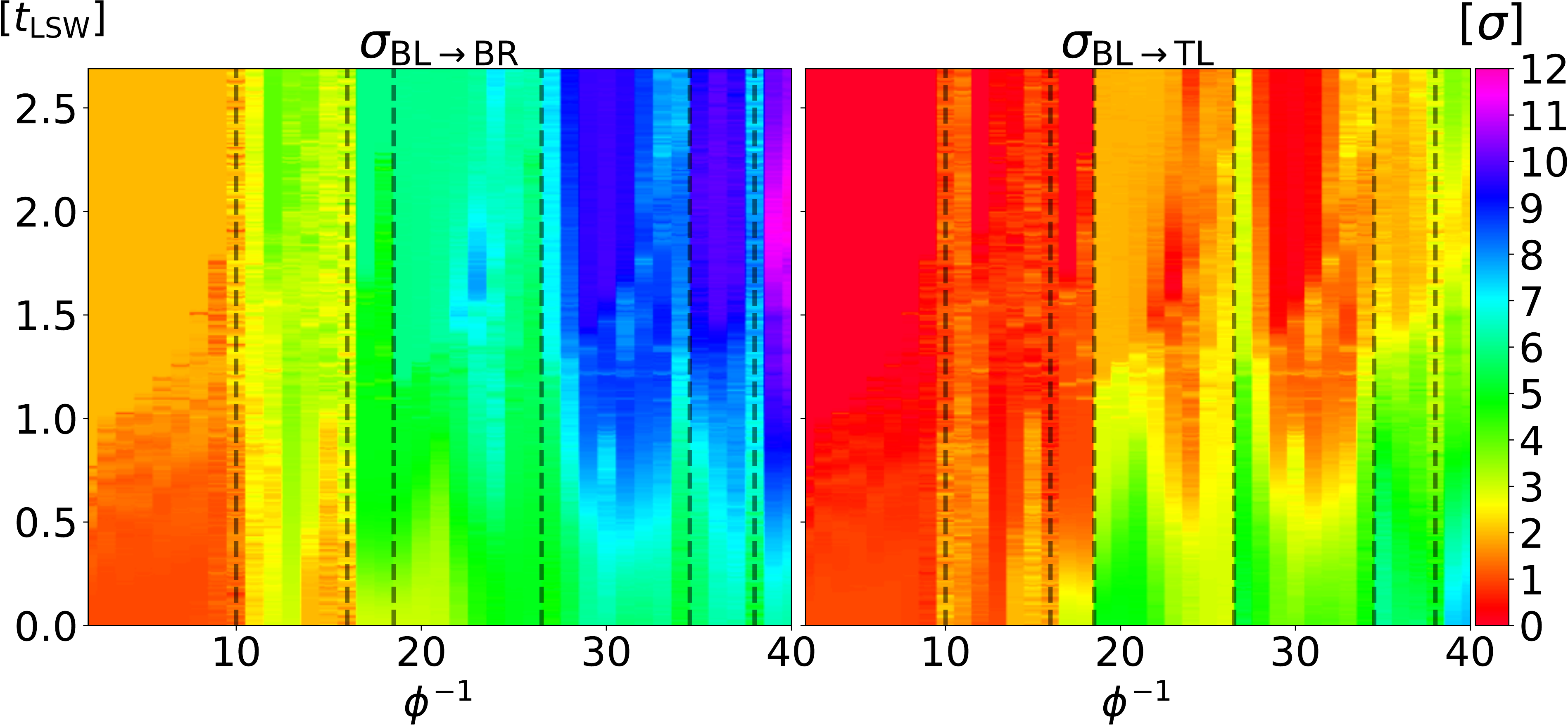}
\caption{\parametercap{\gls{hlsw} interaction strength $\tlsw$}{a \gls{hlsw} system}{\num{1.6e5}}.}
\label{figure:hlsw_study}
}
  
A study of the full parameter dependence on \mbox{$t_{\mathrm{LSW}}^{(U)}\in[0, t]$} is shown in \ref{figure:hlsw_study}. Only $\SPass{}$ and $\SLSW{}$ are shown, since the behaviour of the other conductance are essentially the same as for the homogeneous system for all parameter values. The limiting cases $\tlsw{}=0$ and \mbox{$\tlsw{}=t$} have no special properties. As expected, $\SPass{}$ is larger for larger $\tlsw{}$ and $\SLSW{}$ is smaller. There is no formation of a plateau structure in $\SPass{}$ or $\SLSW{}$ for any value of $\tlsw{}$. The tendency of a larger magnitude in $\SPass{}$ close to the centers of plateaus persists for all $\tlsw\in[0,t]$.

\subsection{\slswhead}
\widefig{
  \adjustbox{width=\widewidth}{
    \includegraphics[height=\LDOSsize]{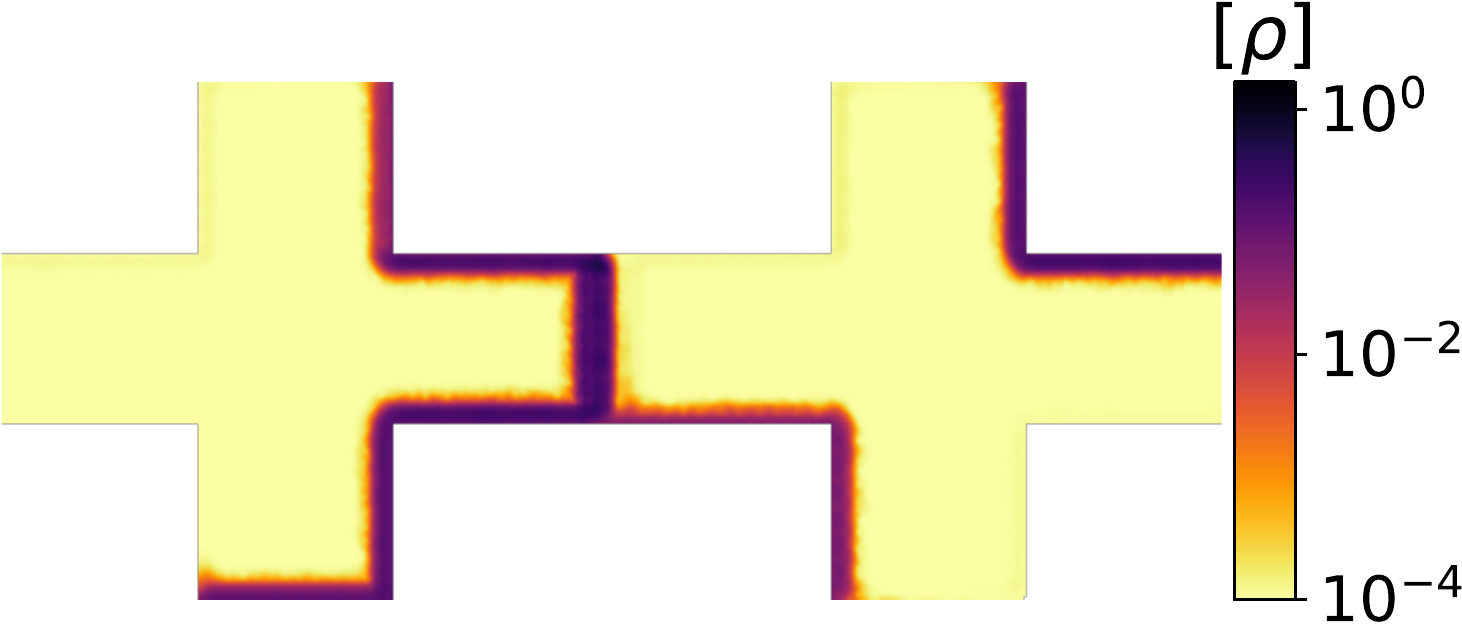}
    \hspace{\hspacing}%
    \raisebox{\raisesize}{
      \includegraphics[height=\TRANSsize]{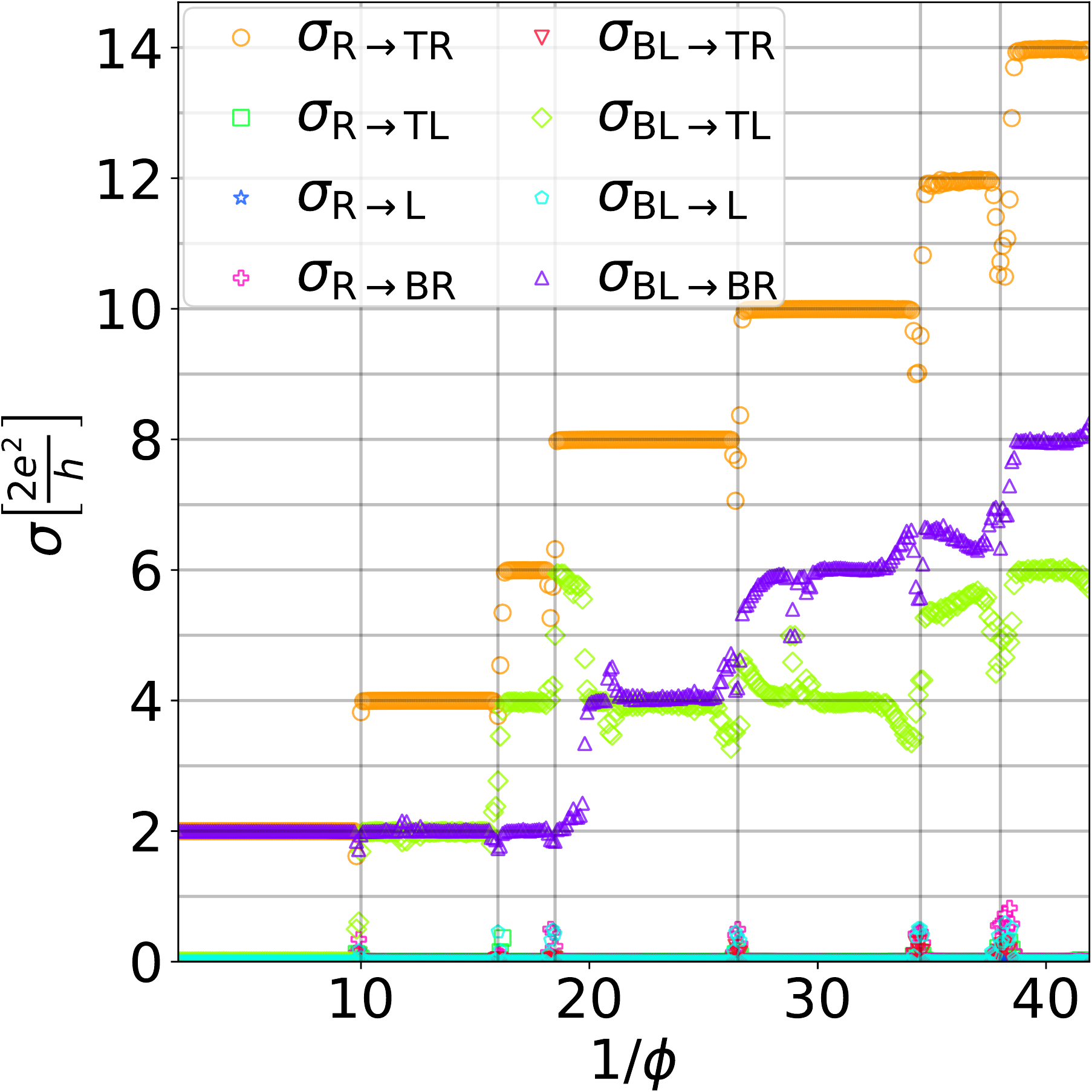}
    }
  }
  \caption{\HallbarPlotRef{a Hall bar system with \gls{slsw}}. \LDOSphi{10.5}. \FarCutoff{}. \SkewWidth{44}. \LDOSNumber{\num{1.2e6}}. \TransNumber{\ShearLargeWidth}.}
  \label{figure:skew_hallbar}
}
 
Next we turn to our findings for a more realistic model of an \gls{lsw}, which is formed due to shear in the upper graphene layer as illustrated in \ref{figure:skew_tension}. Representative results are shown in \ref{figure:skew_hallbar}. Just as for the \gls{hlsw} system in \ref{figure:hardwall_hallbar}, the \gls{ldos} calculation shows localized edge modes away from the \gls{slsw} and an increased density in the region of the \gls{slsw} itself. The region of increased density however is larger than the region for the \gls{hlsw} system in \ref{figure:hardwall_hallbar}. This is due to the finite extent of the \gls{lsw}, which allows for propagation parallel to the \gls{lsw} over the whole range of the \gls{lsw}.

The particular choice of $L_s=44$ in \ref{figure:skew_hallbar} is such that the system is neither in the limit of a very small nor a very large \gls{slsw}. The corresponding conductance calculations in \ref{figure:skew_hallbar} confirm that there is conductance parallel to the \gls{slsw}. Both conductance functions $\SPass{}$ and $\SLSW{}$ are almost monotonous and form approximate plateaus, albeit not for the same $\phi$-range as $\SRef{}$. In particular around the center of the plateaus, the conductances $\SPass{}$ and $\SLSW{}$ are almost constant.

\stfig{
  \includegraphics[width=\columnwidth]{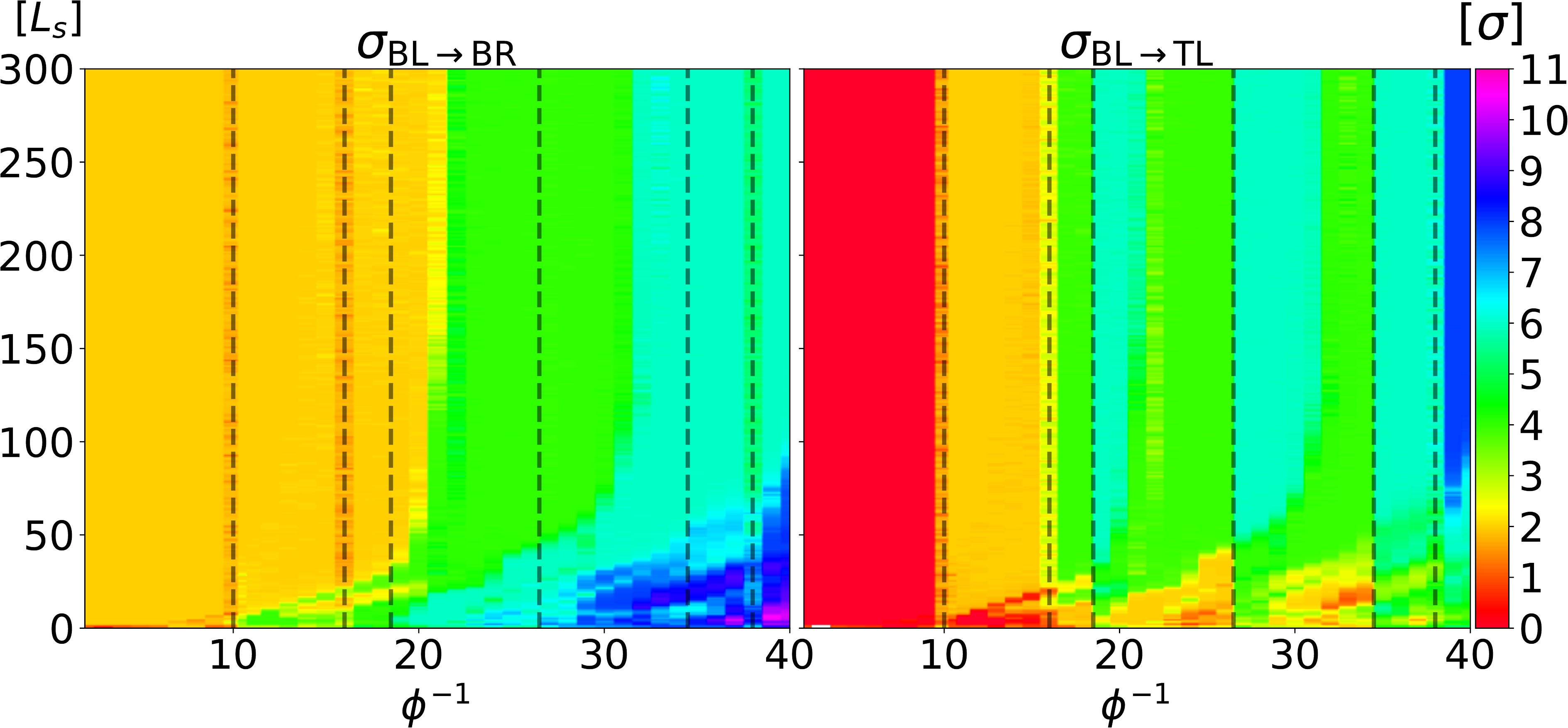}
  \caption{\parametercap{\gls{slsw} width $L_s$}{a \gls{slsw} system}{\ShearLargeWidth}. \FarCutoffLong{}.}
  \label{figure:slsw_study}
}
 
A parameter study of the shear strength parameter $L_s$ is displayed in \ref{figure:slsw_study}. There is a well defined limit of the conductances $\SPass{}$ and $\SLSW{}$ for $L_s\rightarrow\infty$, which is reached in the interval $\phi^{-1}\in[0,40]$ for any values $L_s>100$. The convergence clearly depends on the value of the external magnetic field. This is reasonable, since a larger value of $\phi^{-1}$ implies a larger magnetic length $l_B=\sqrt{\frac{\hbar\mathrm{c}}{\mathrm{e}B}}$. Thus the limit $L_s\rightarrow\infty$ is reached for larger $L_s$ when $\phi^{-1}$ is chosen larger. The resolution of the conductances along the magnetic field axis is however not good enough to confirm a quadratic dependence on $L_s$, which would be expected. It is clear that $\SLSW{}$ being almost monotonous for $L_s=44$ in \ref{figure:skew_hallbar} is no longer true for the infinitely smooth shear transition $L_s\rightarrow\infty$. However, both $\SLSW{}$ and $\SPass{}$ exhibit a plateau structure in the limit $L_s\rightarrow\infty$. Further, the limit for $\SPass{}$ is a monotonous plateau structure. The non-converged values in the small shear width limit are generically smaller than the values in the opposite limit of large shear widths. This is reasonable, if the \gls{slsw} were to be modeled by some potential wall that is thicker for a wider \gls{slsw}.

\stfig{
  \adjustbox{width=\narrowwidth}{%
    \includegraphics[height=\somesize]{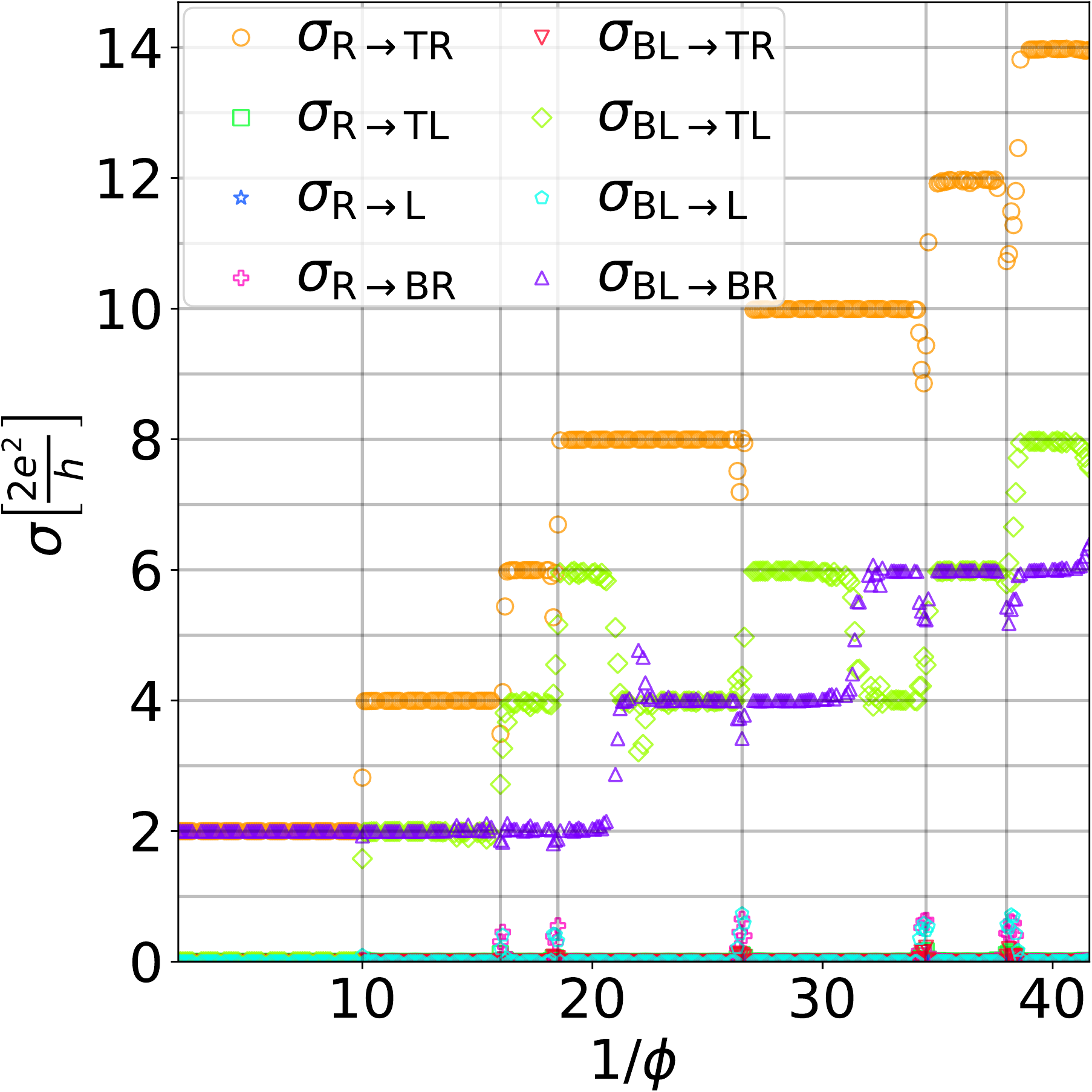}%
  }%
  \caption{\TransPlotRef{a Hall bar system with \gls{slsw}}. \FarCutoff{}. \SkewWidth{150}. For both calculations \TransNumber{\ShearLargeWidth}.}%
  \label{figure:skew_other_skews}%
}
  
The limiting case $L_s\rightarrow\infty$ is investigated in \ref{figure:skew_other_skews} for the choice $L_s=150$. The monotonous plateau structure of $\SPass{}$ mentioned in the discussion of \ref{figure:slsw_study} is now readily apparent. It is also confirmed, that $\SLSW{}$ exhibits a non-monotonous plateau structure since the plateaus of $\SPass{}$ and $\SRef{}$ are not aligned and $\SPass{}+\SLSW{}=\SRef{}$. The plateau structure of $\SPass{}$ is:

\begin{itemize}
\item $\phi^{-1}\in[0, 21]\Rightarrow\SPass{}=2\sigma_0$.
\item $\phi^{-1}\in[21, 32]\Rightarrow\SPass{}=4\sigma_0$.
\item $\phi^{-1}\in[32, 40]\Rightarrow\SPass{}=6\sigma_0$.
\end{itemize}

\subsection{\tlswhead}\label{subsection:tlswhead}

\widefig{
  \centering
  \adjustbox{width=\widewidth}{
    \includegraphics[height=\LDOSsize]{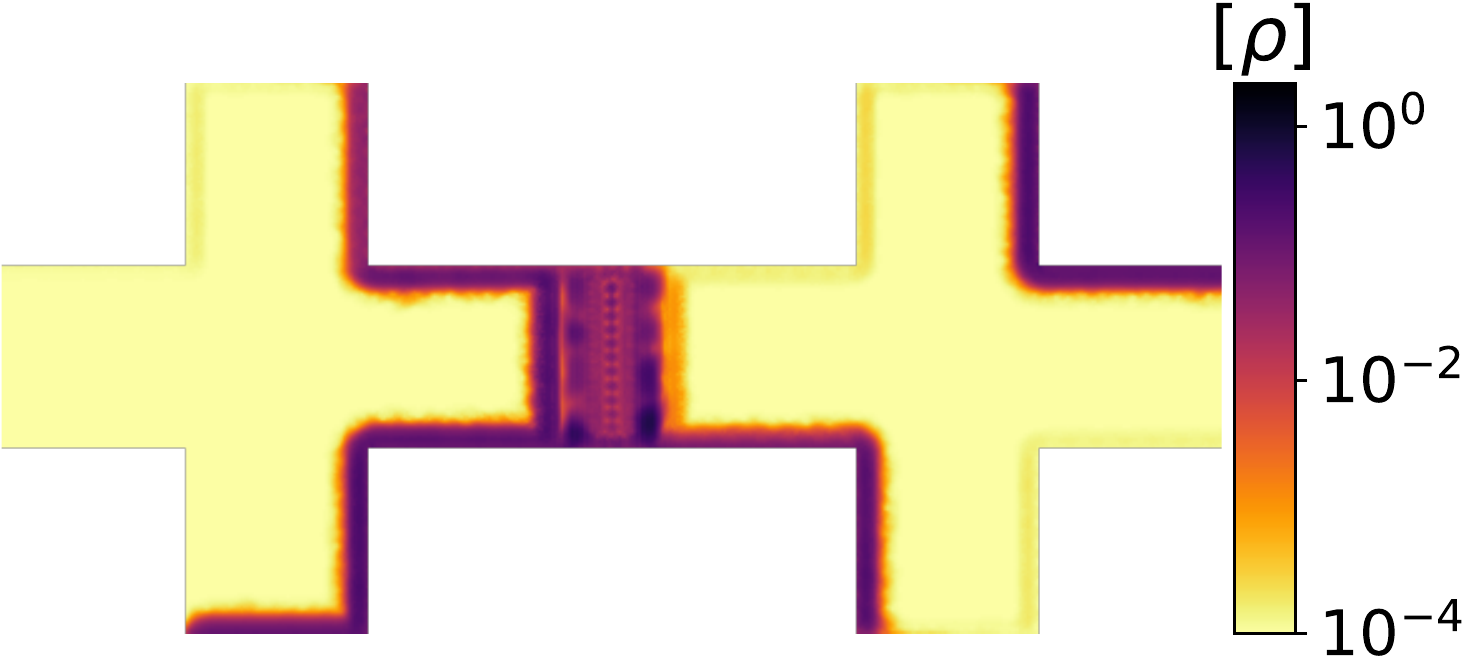}
    \hspace{\hspacing}%
    \raisebox{\raisesize}{
      \includegraphics[height=\TRANSsize]{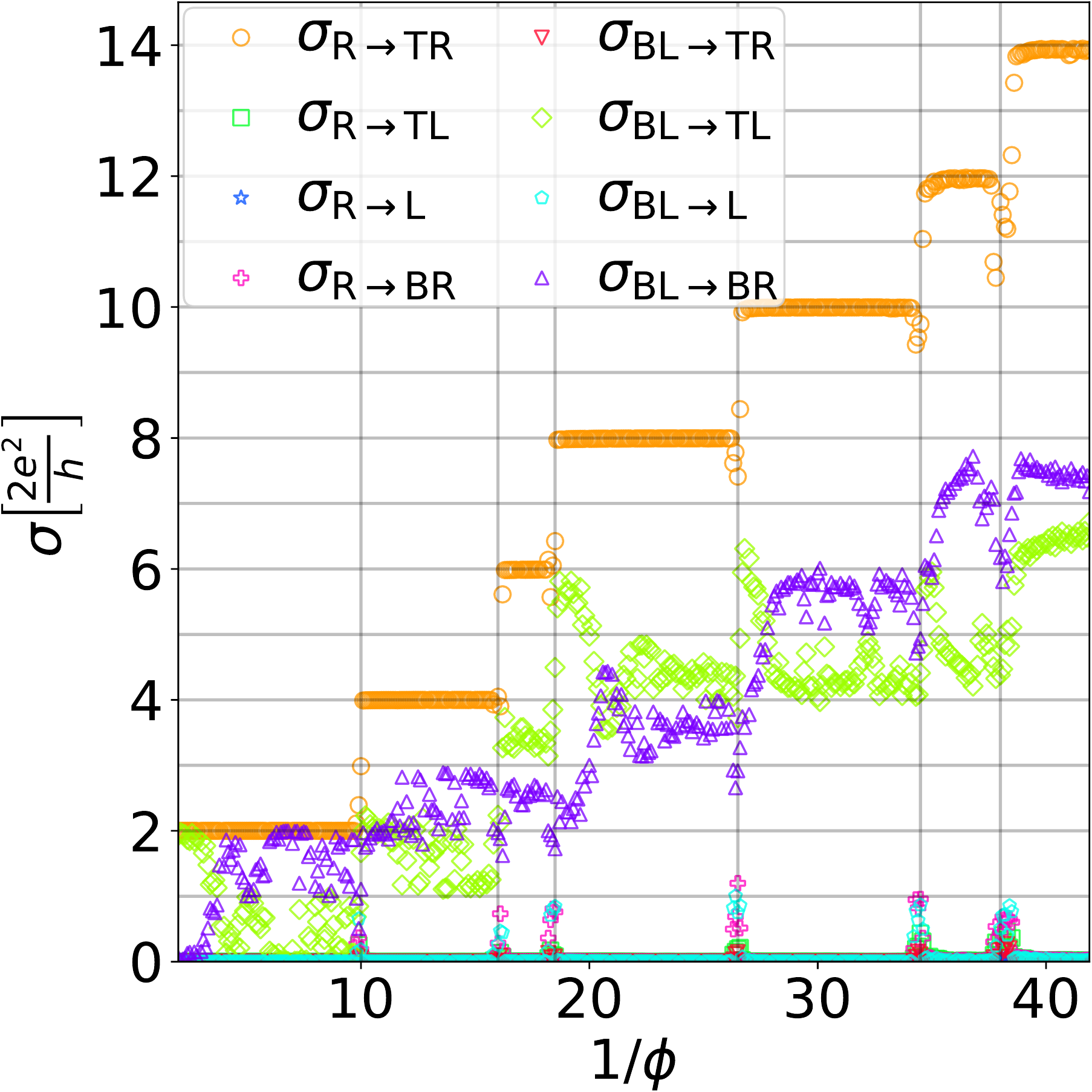}
    }
  }
  \caption{\HallbarPlotRef{a tensed system}. \LDOSphi{10.5}. \NearCutoff{}. \TensWidth{60}. \LDOSNumber{\num{8.1e5}}. \TransNumber{\TensLargeWidth}.}
  \label{figure:tensile_hallbar}
}
 
Representative results for a system where the \gls{lsw} is formed due to tension, see \ref{figure:skew_tension}, are shown in \ref{figure:tensile_hallbar}. Just as for the \gls{slsw} system the choice of $L_t=60$ is an intermediate value, neither in the $L_t\rightarrow 0$ nor the $L_t\rightarrow\infty$ limit. The \gls{ldos} calculation yields similar results as calculations for the \gls{hlsw} and \gls{slsw} systems, an increased density near edges away from the \gls{tlsw} and an increased density in the region of the \gls{tlsw}. The density increase in the region of the \gls{tlsw} is however much more extended than for the cases of the \gls{slsw} and the \gls{hlsw}. This is due to the fact, that the choice of \gls{lsw} here is wider than in any of the other systems. Note that a similar distribution of the electron density may be also observed for systems of the SLSW type with wider LSW. Thus, although the \gls{ldos} in the transition regime of the \gls{lsw} is the largest, there are also relevant contributions in its interior.

$\SRef{}$ is once again not affected by the presence of an \gls{lsw}. The conductance functions $\SLSW{}$ and $\SPass{}$ in \ref{figure:tensile_hallbar} are not as simple as in \ref{figure:hardwall_hallbar} or \ref{figure:skew_hallbar} due to the \gls{tlsw} geometry. The conductances, in particular $\SPass{}$, clearly have regions where they are approximately constant. This indicates a tendency for a plateau formation, but there are still relevant fluctuations of the order $\sigma_0$ in regions of almost constant conductance. These areas of plateau formation are also interrupted by plateau transitions in $\SRef{}$, but this is to be expected with a similar explanation as for the \gls{slsw} and \gls{hlsw} systems. 

\stfig{
  \includegraphics[width=\columnwidth{}]{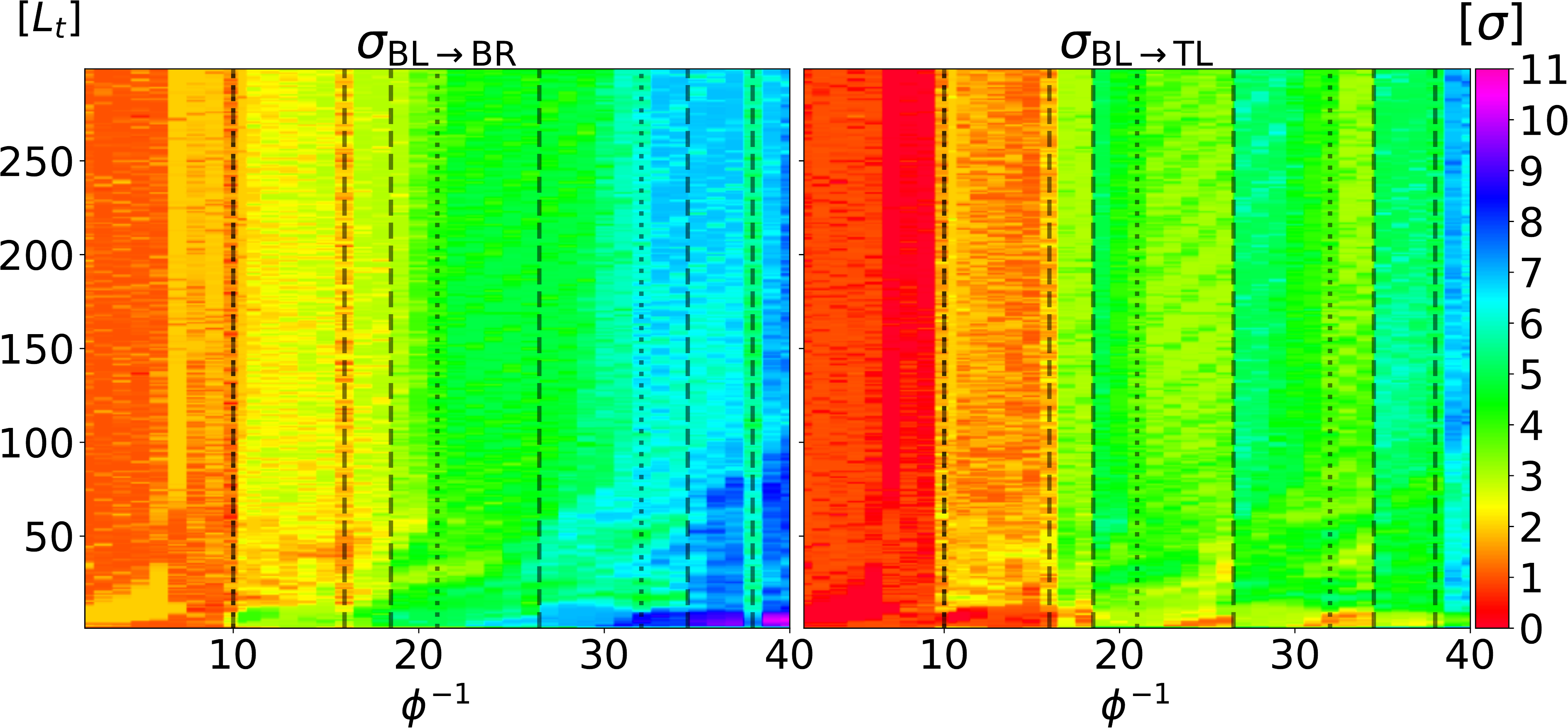}
  \caption{\parametercap{\gls{tlsw} width $L_t$}{a \gls{tlsw} system}{\TensLargeWidth}. \FarCutoffLong{}.}
  \label{figure:tlsw_study}
}
  
Since the plateau formation might be more obvious in the limit $L_t\rightarrow\infty$ as for the \gls{slsw} system, a parameter study of the \gls{tlsw} system with respect to $L_t$ is shown in \ref{figure:tlsw_study}. However, the fluctuations seen in \ref{figure:tensile_hallbar} are persistent for all studied values of $L_t$, including the $L_t\rightarrow\infty$ limit. There is a fixed limit $L_t\rightarrow\infty$ up to persistent fluctuations. Our conclusion to the persistence of the fluctuations is that the \gls{tlsw} system simply reacts more strongly to small changes in external parameters than the \gls{slsw} or \gls{hlsw} systems. This is also confirmed by further discussions in \ref{section:finite_size,section:numerical_cutoff}. When examining the whole range of $L_t$ it becomes apparent, that the $L_t\rightarrow\infty$ limit exhibits a very similar plateau formation as the $L_s\rightarrow\infty$ limit for the \gls{slsw} system, but with fluctuations of order $\sigma_0$. This means, there is a monotonous plateau structure in $\SPass{}$ and a non-monotonous plateau structure in $\SLSW{}$. Similar to the \gls{slsw} system, the convergence with respect to system size is faster for smaller $\phi^{-1}$ and the magnitude of $\SPass{}$ is smaller for smaller $L_t$. The physical interpretations for these phenomena are therefore the same as the \gls{slsw} system.

\stfig{
  \adjustbox{width=\narrowwidth}{
    \centering
    \includegraphics[height=\somesize]{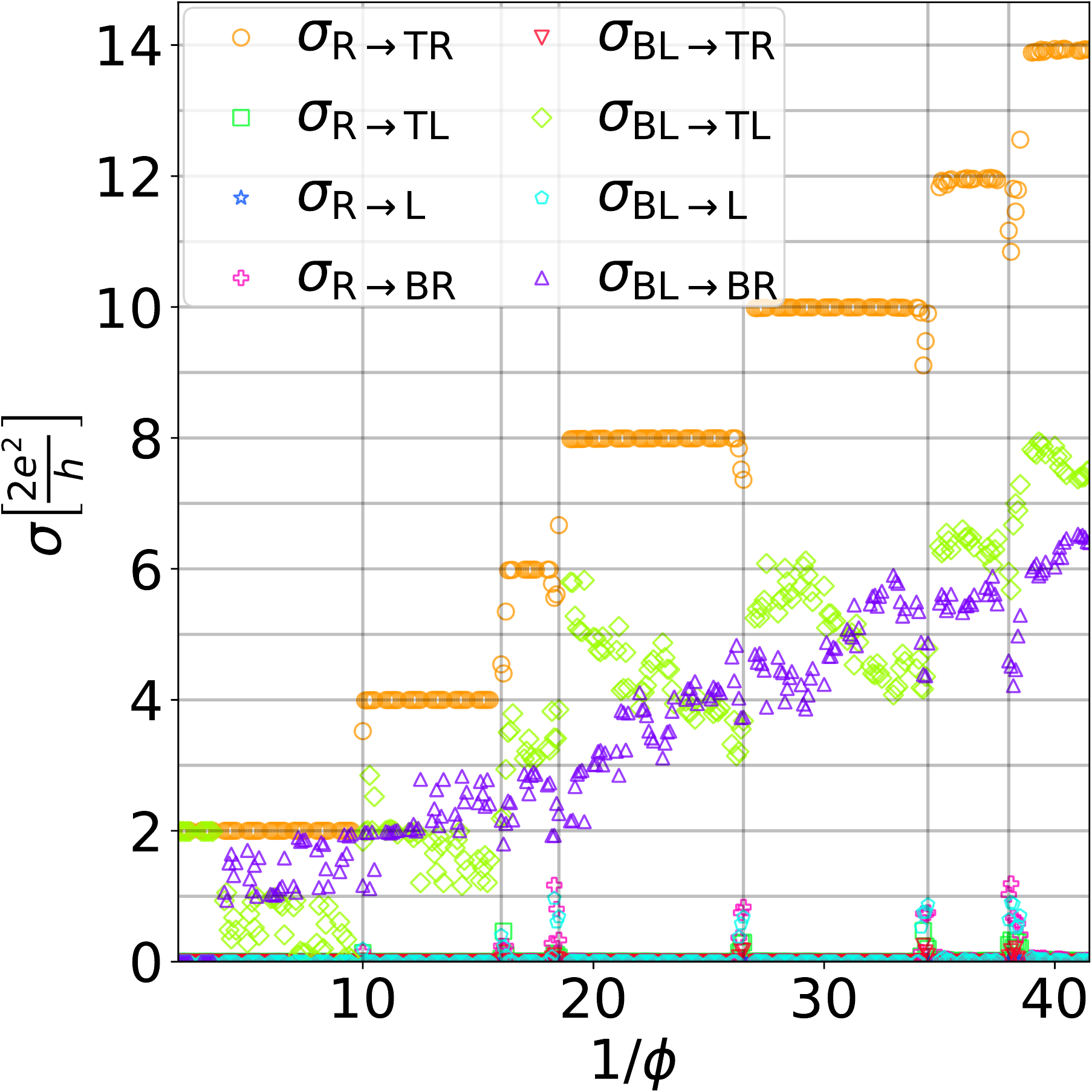}%
  }%
  \caption{\TransPlotRef{a tensed Hall bar system}. \FarCutoff{}. \TensWidth{150}. For both calculations \TransNumber{\TensLargeWidth}.}%
  \label{figure:tension_other_tensions}%
}
  
As an example for the $L_t\rightarrow\infty$ limit, a \gls{tlsw} system with $L_t=150$ is shown in \ref{figure:tension_other_tensions}. When examining a single magnetoconductance slice of the $L_t$ parameter study, the plateau formation is not as obvious. In the context of \ref{figure:tlsw_study}, we can identify the regions of almost constant conductance. Close to the plateau centers of $\SRef{}$, the following implications hold:

\begin{itemize}
\item $\phi^{-1}\in[2,10]\Rightarrow\SPass{}\in[\sigma_0,2\sigma_0]$.
\item $\phi^{-1}\in[10,21]\Rightarrow\SPass{}\in[2\sigma_0,3\sigma_0]$.
\item $\phi^{-1}\in[21,32]\Rightarrow\SPass{}\in[4\sigma_0,5\sigma_0]$.
\item $\phi^{-1}\in[32,40]\Rightarrow\SPass{}\in[5\sigma_0,6\sigma_0]$.
\end{itemize}

These plateau definitions are consistent with \ref{figure:tlsw_study}. 
\section{Conclusions}\label{section:conclusion}
We have investigated the effects of different \glspl{lsw} in \gls{blg} in the \gls{iqhe} regime on the conductance and the \gls{ldos} given a Hall bar geometry. The particular types of \glspl{lsw}  investigated were a hard wall \gls{lsw} parallel to the zigzag nanoribbon, a shear \gls{lsw} parallel to the armchair nanoribbon, and a tensile \gls{lsw} parallel to the zigzag nanoribbon. Additionally, the homogeneous \gls{blg} system was discussed as a reference and to establish the basic properties of AB stacked \gls{blg} in the \gls{iqhe} regime.

Aside from the investigation of the different \gls{blg} systems we have expanded upon the usual methods employed in the calculation of such transport properties. For the calculation of the self energy due to a semi-ifinite lead, an optimization for sparse chain couplings with a significant dimensional reduction was introduced. To perform the simulation of transport properites in a Hall bar geometry a general and stable scheme for the partition of a \gls{blg} lattice on arbitrary geometries was formulated. Although this solution is not optimal for all geometries, the performance in all apllications presented in this paper was satisfactory.

The results for the homogeneous \gls{blg} without \glspl{lsw} were as expected, with all features in the \gls{ldos} and conductance accounted for like the sequence of non-equidistant conductance plateaus with the expected sequence of conductance quantum multiples and the appropriate edge localization of the corresponding modes. We found that the hard wall system behaves significantly different from the tensile and shear systems. It is therefore not a particularly good model for any choice of hopping elements. Further, there are no obvious limiting cases for $t_{\mathrm{LSW}}^{(U)}=0$ or $t_{\mathrm{LSW}}^{(U)}=t$. In contrast, the tensile system shows persistent fluctuations of order $\sigma_0$ in $\SPass{}$ and $\SLSW{}$ even for large values of $L_t$. Both the tensile and the shear system have indeed a well defined limit $L_{s/t}\to\infty$ (up to fluctuations in case of the tensile system). Convergence for both limits $L_s$ and $L_t$ depends on the external magnetic field where larger $\phi^{-1}$ imply slower convergence. Most importantly, both limits $L_s\to\infty$ and $L_t\to\infty$ show monotonous plateau structure in $\SPass$ and non-monotonous plateau structure in $\SLSW{}$. The values of these plateaus are always integer multiples of the elementary conductance quantum $\sigma_0$.

One motivation for our study was the perceived ambiguity between fractionalization due to geometry and due to electron-electron interaction effects. For all types of \glspl{lsw} investigated, conductance plateaus that do not appear for the ordinary \gls{iqhe} can certainly be achieved simply due to geometrical effects and the effect of multiple \gls{lsw} geometries on the conductance is still unclear from the results presented here. What can be said is that the effect of \glspl{lsw} is highly nontrivial and very geometry-dependant. Thus, it certainly requires a fully quantum mechanical description in general.

There are several additional issues, for which further investigations would be desirable. The most obvious one stems from the motivation of the research and is the calculation of conductance functions and densities for systems with multiple \glspl{lsw} and ultimately to model entire defect networks like in real materials. In particular, a potential fractionalization of the conductance for multiple \glspl{lsw} in different geometrical configurations would be very interesting. With more computing time larger systems and thus a larger range of magnetic field values could be reliably studied even in a full parameter study. Another aspect, that has not been fully investigated is the different system geometries that are possible. On one hand the proper modeling of buckling would extend the discussion to more experimentally available systems. On the other hand, the hard wall model parallel to the armchair nanoribbon should at least be investigated for completeness. Another \gls{lsw} geometry, that has not been investigated in our work is an \gls{lsw} due to a change of interaction cutoff as discussed in the appendix to differentiate between effects due to the choice of a different value in the \gls{lsw} region and effects of the actual \gls{lsw} geometry. Since the \glspl{lsw} due to shear and tension show some similarities to twisted bilayer graphene, for which interactions are indeed important for small twist angles, electron-electron interactions might also be relevant and should thus be modeled for the tensile and shear \gls{lsw} models. For a microscopic lattice model this is computationally not feasible for reasonably large systems. Thus, a different approach to the system modeling or some simplification to the interaction would have to be employed.

\section{Acknowledgments}
We thank Sam Shallcross and Heiko Weber for fruitful discussions.
\FloatBarrier{}
%
\FloatBarrier{}

\appendix{}
\section{Finite-size effects}\label{section:finite_size}
The magnitude of finite-size effects is magnetic field dependent and it is of great importance to check the previous calculations for convergence with respect to system size. We will discuss three system sizes for \gls{hlsw} with decoupled upper sheet, \gls{slsw} and \gls{tlsw} each.

\subsection{\hardhead}
\widefig{
\captionsetup[subfigure]{justification=centering}
  \adjustbox{width=\columnwidth}{
  \centering
  \subfloat[$N\approx\num{1.6e5}$]{{ \includegraphics[width=\defaultFigWidth]{FIG_7_B.pdf}\label{figure:finite_size_HLSW_nocoup_left} }}%
  \subfloat[$N\approx\num{2.0e5}$]{{ \includegraphics[width=\defaultFigWidth]{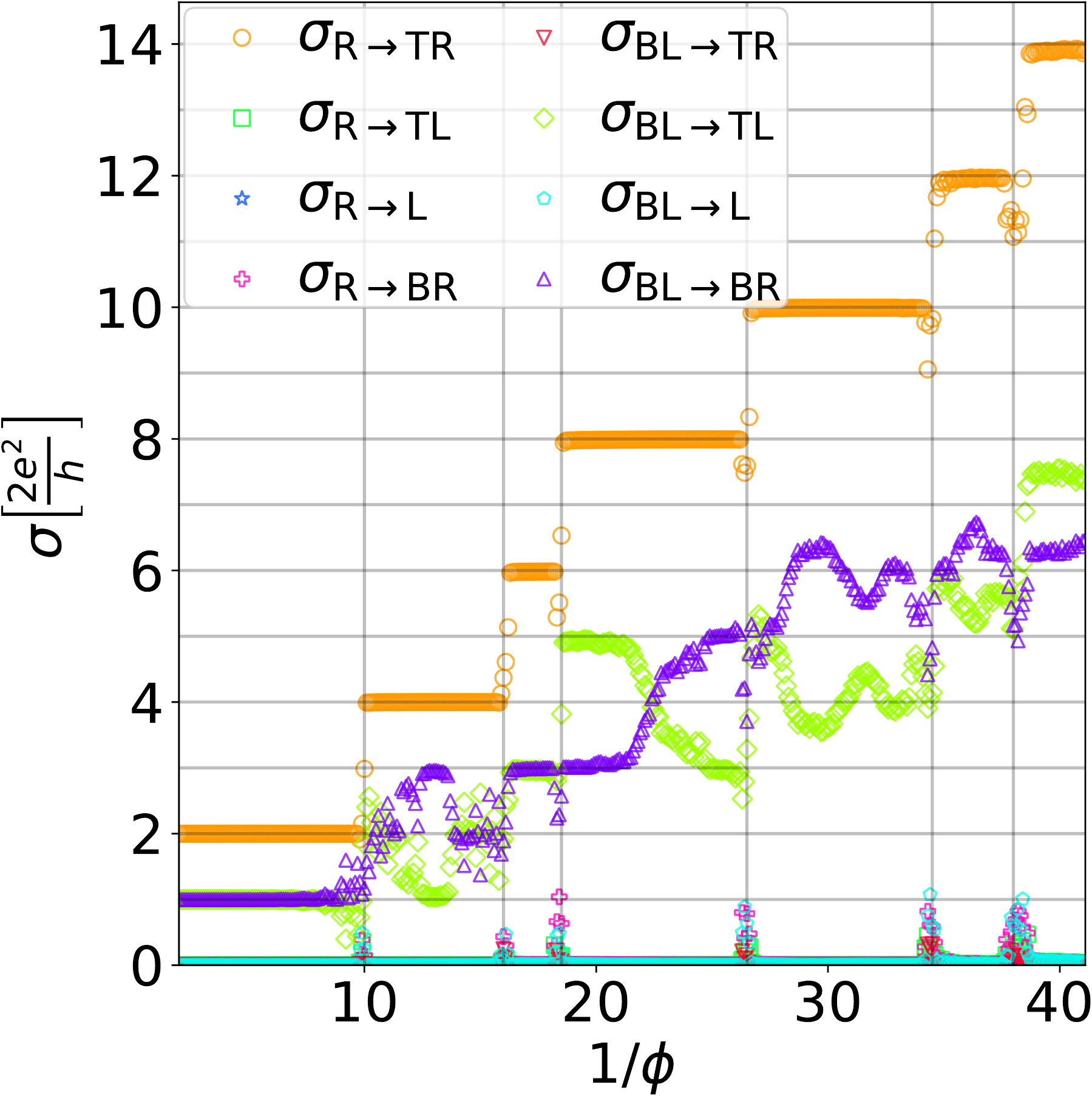}\label{figure:finite_size_HLSW_nocoup_middle} }}%
  \subfloat[$N\approx\num{2.7e5}$]{{ \maybefbox{\includegraphics[width=\defaultFigWidth]{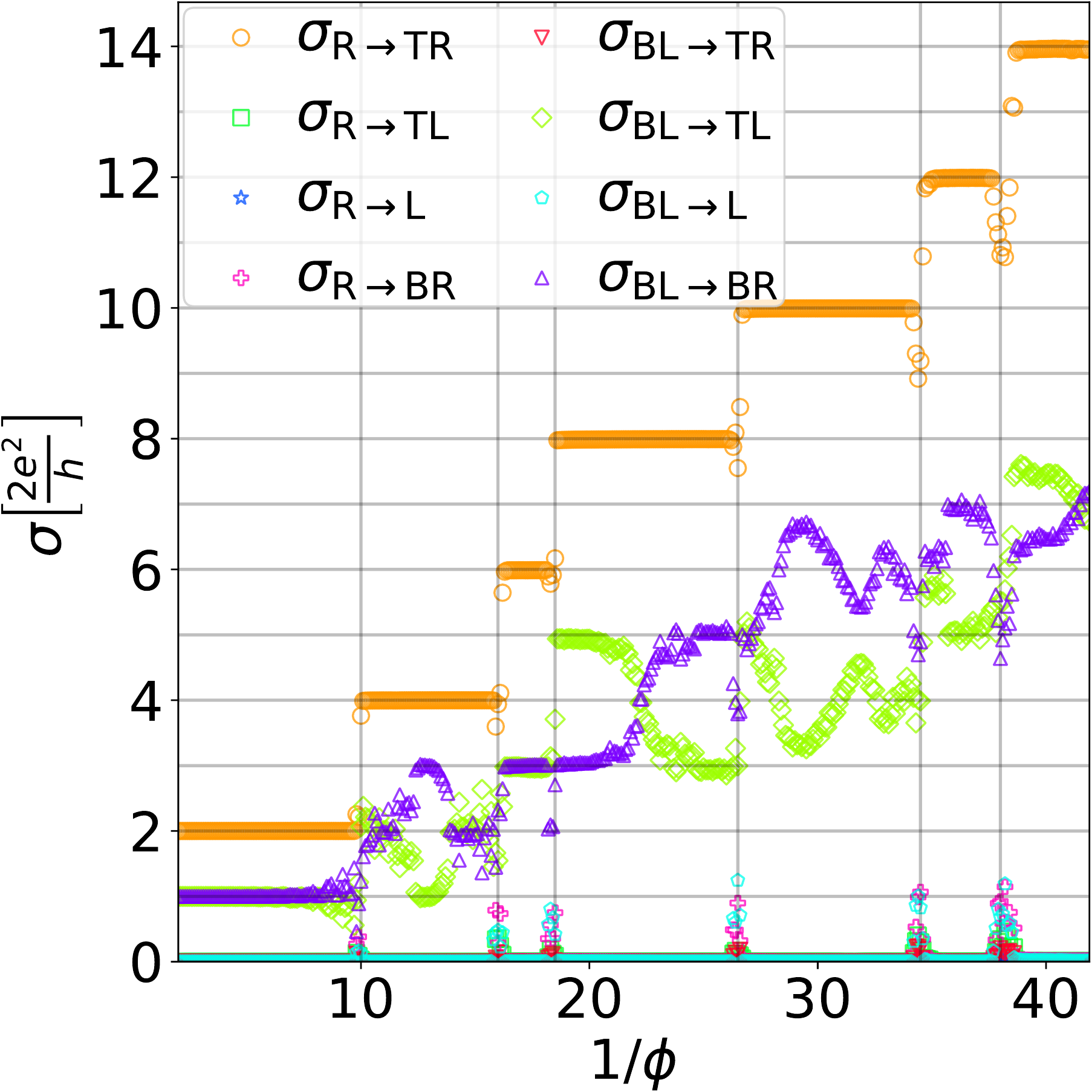}}\label{figure:finite_size_HLSW_nocoup_right} }}%
  }%
  \caption{\finiteSizeCap{\gls{hlsw}}. \LowerCoupled{}. \UpperDecoupled{}.}
  \label{figure:finite_size_HLSW_nocoup}
}
 
A finite-size study for \gls{hlsw} systems with decoupled upper layer is shown in \ref{figure:finite_size_HLSW_nocoup}. Comparing \ref{figure:finite_size_HLSW_nocoup_left}, \ref{figure:finite_size_HLSW_nocoup_middle} and \ref{figure:finite_size_HLSW_nocoup_right}, it is clear that there are some finite-size effects in the calculations of smaller systems. Particularly for the transmissions $\SPass$ and $\SLSW$ in the regime of smaller magnetic field (the last three plateaus), features are still changing with system size. Increases and decreases in the conductivity are sharper for larger system sizes in this region. For smaller plateaus however, the system has already converged to a satisfactory degree for the smallest system size. The general shape of the conductivities does also not change with system size. 

\subsection{\slswhead}
\widefig{
\captionsetup[subfigure]{justification=centering}
  \adjustbox{width=\columnwidth}{
  \centering
  \subfloat[$N\approx\ShearSmallWidth$]{{ \includegraphics[width=\defaultFigWidth]{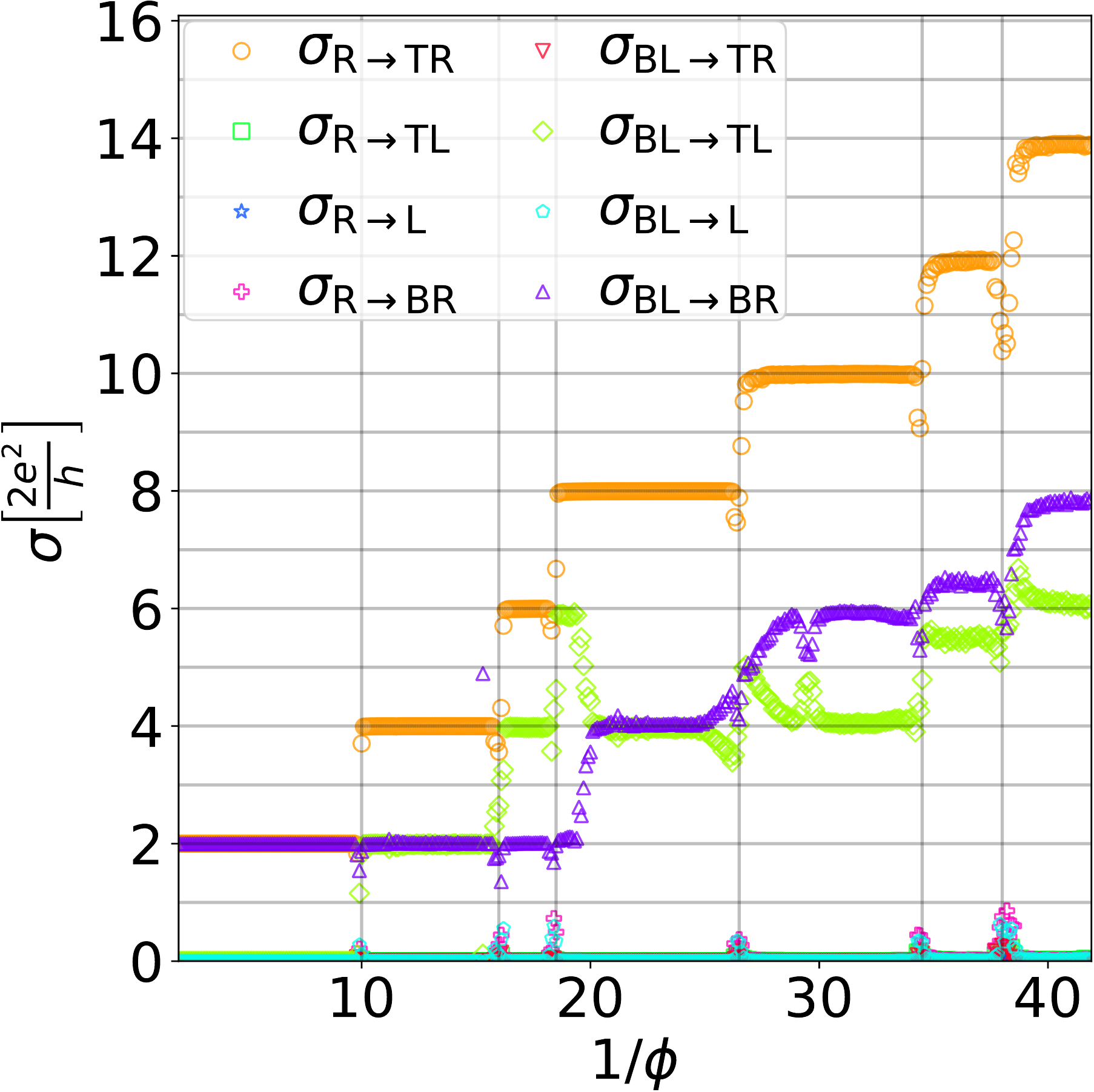} \label{figure:finite_size_SLSW_left} }}%
  \subfloat[$N\approx\ShearMiddleWidth$]{{ \includegraphics[width=\defaultFigWidth]{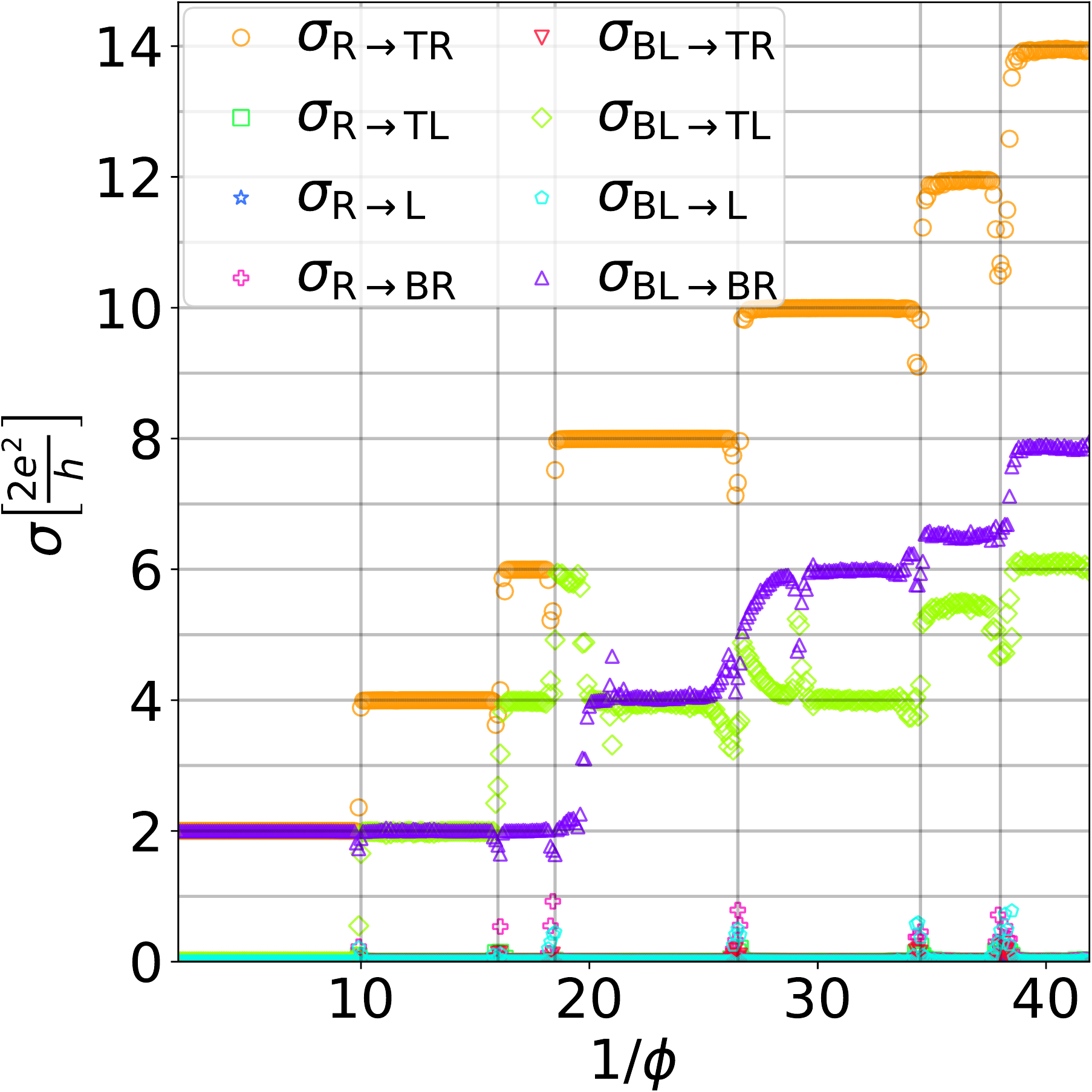} \label{figure:finite_size_SLSW_middle} }}%
  \subfloat[$N\approx\ShearLargeWidth$]{{ \maybefbox{\includegraphics[width=\defaultFigWidth]{FIG_9_B.pdf}} \label{figure:finite_size_SLSW_right} }}%
  }%
  \caption{\finiteSizeCap{\gls{slsw}}. \FarCutoff{}. \SkewWidth{44}.}
  \label{figure:finite_size_SLSW}
}
 
A finite-size study for \gls{slsw} systems is shown in \ref{figure:finite_size_SLSW}. The conductance functions for all three system sizes are almost identical, particularly for the first four plateaus. For larger $1/\phi$ some features show minor change with system size at plateau transitions. The increases and decreases due to plateau transitions become more localized with larger system size.

\subsection{\tlswhead}
\widefig{
\captionsetup[subfigure]{justification=centering}
  \adjustbox{width=\columnwidth}{
  \centering
  \subfloat[$N\approx\TensMiddleWidth$]{{ \includegraphics[width=\defaultFigWidth]{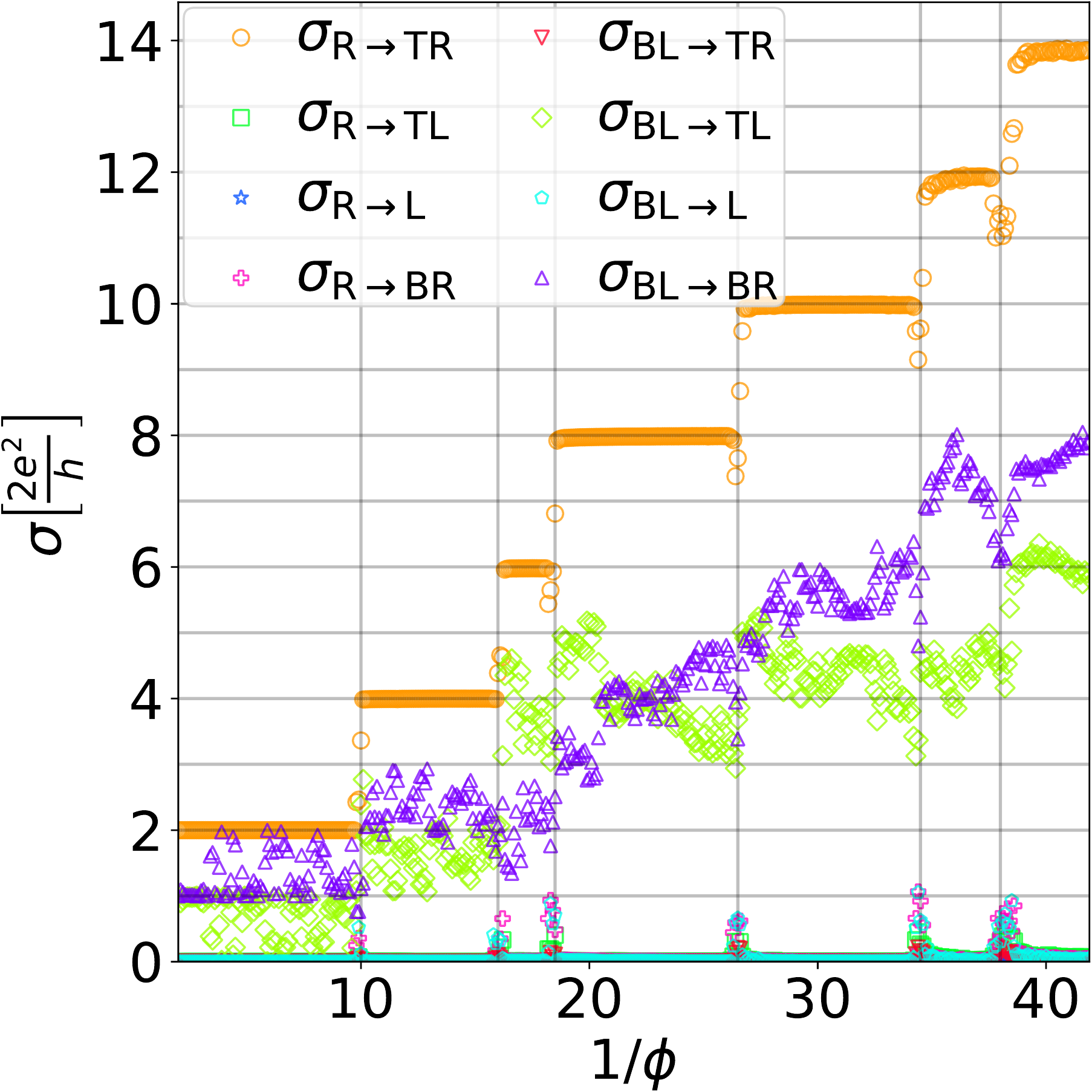}\label{figure:finite_size_HLSW_left} }}%
  \subfloat[$N\approx\TensLargeWidth$]{{ \includegraphics[width=\defaultFigWidth]{FIG_12_B.pdf}\label{figure:finite_size_HLSW_middle} }}%
  \subfloat[$N\approx\TensVLargeWidth$]{{ \maybefbox{\includegraphics[width=\defaultFigWidth]{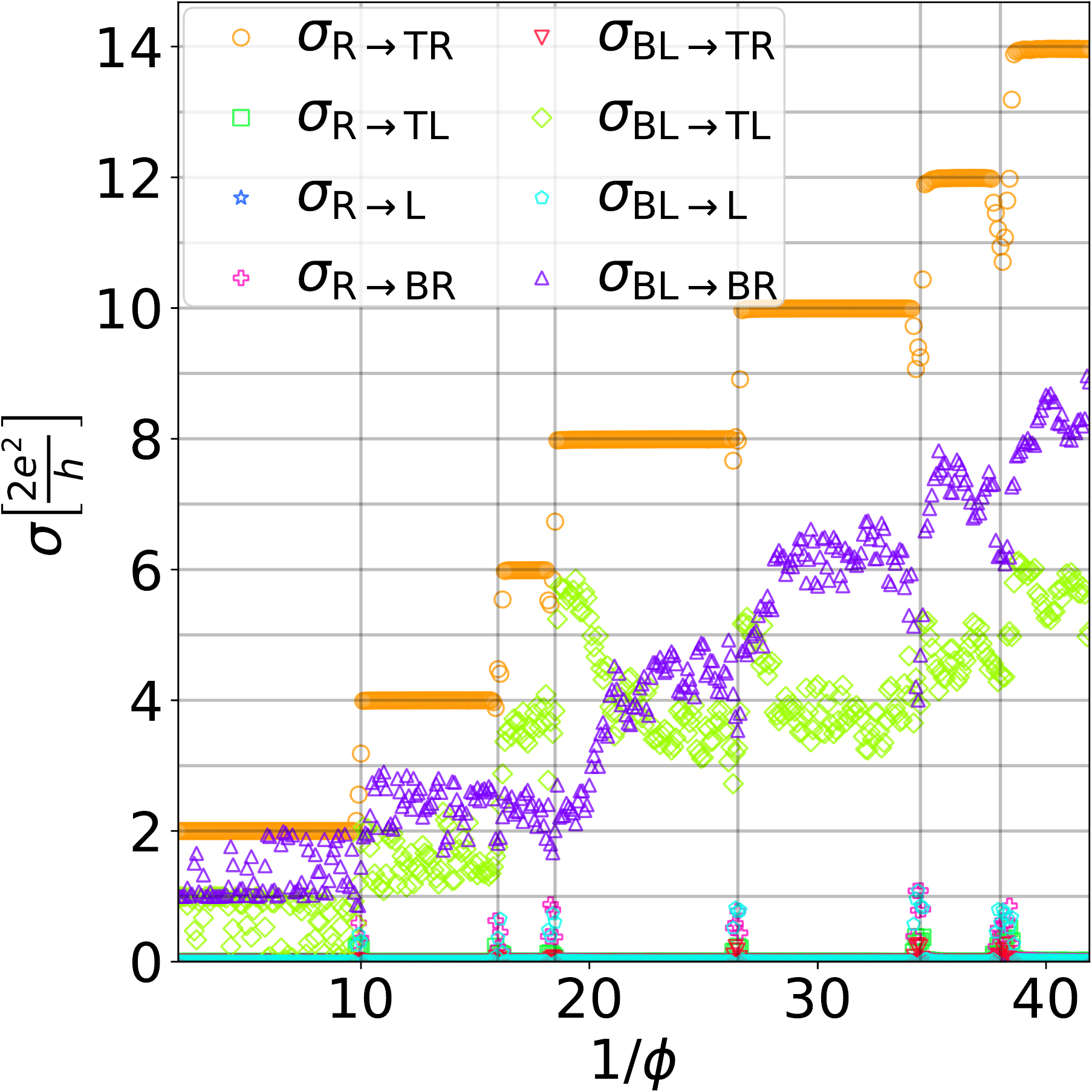}}\label{figure:finite_size_HLSW_right} }}%
  }%
  \caption{\finiteSizeCap{\gls{tlsw}}. \FarCutoff{}. \TensWidth{60}.}
  \label{figure:finite_size_TLSW}
}
 
A finite-size study for \gls{tlsw} systems is shown in \ref{figure:finite_size_TLSW}. Due to the strong fluctuations of $\SLSW{}$ and $\SPass{}$ in \gls{tlsw} systems, discussing finite-size effects is more difficult than for \gls{hlsw} and \gls{slsw} systems. As for the other system types, the conductance across the first four plateaus is similar for all system sizes. But even for large external magnetic fields, there are differences between \ref{figure:finite_size_HLSW_middle} and \ref{figure:finite_size_HLSW_right} close to plateau transitions. In particular the shape of $\SLSW{}$ and $\SPass{}$ differs near the third plateau. These changes however are fairly small in magnitude and given the satisfactory convergence of all other system types for these system sizes, using either the system size in \ref{figure:finite_size_HLSW_middle} or the one in \ref{figure:finite_size_HLSW_right} should be fine for $\phi^{-1}\in[0,40]$. The existence of finite-size effects for these systems however cannot be ruled out as confidently as for the other system types and finite-size effects should be kept in mind when discussing \gls{tlsw} systems.

Generically, systems in the regime of $\num{1e5}$ to $\num{2e5}$ sites are sufficient to discuss the first four plateaus of the magnetoconductance and systems of the order $\num{2e5}$ to $\num{3e5}$ are required to discuss the full range of magnetic field shown in the main body of the article. Even for larger plateaus, smaller system sizes should be sufficient to determine generic features of the conductivity shapes, but finite-size effects must be considered if such calculations are discussed.
\section{Numerical cutoff}\label{section:numerical_cutoff}
Another important parameter of the simulation is the cutoff distance $D_c$. For regular \gls{blg} and \gls{hlsw} systems we chose $D_c$, such that only nearest-neighbour hopping terms are nonzero. This should be adequate to discuss the \gls{iqhe}. For \gls{tblg} systems and systems with shear and tension however such a short cutoff might not model the physics properly. Thus convergence with respect to $D_c$ should be checked for \gls{slsw} and \gls{tlsw} systems.

\subsection{\slswhead}
\stfig{
  \includegraphics[width=\columnwidth]{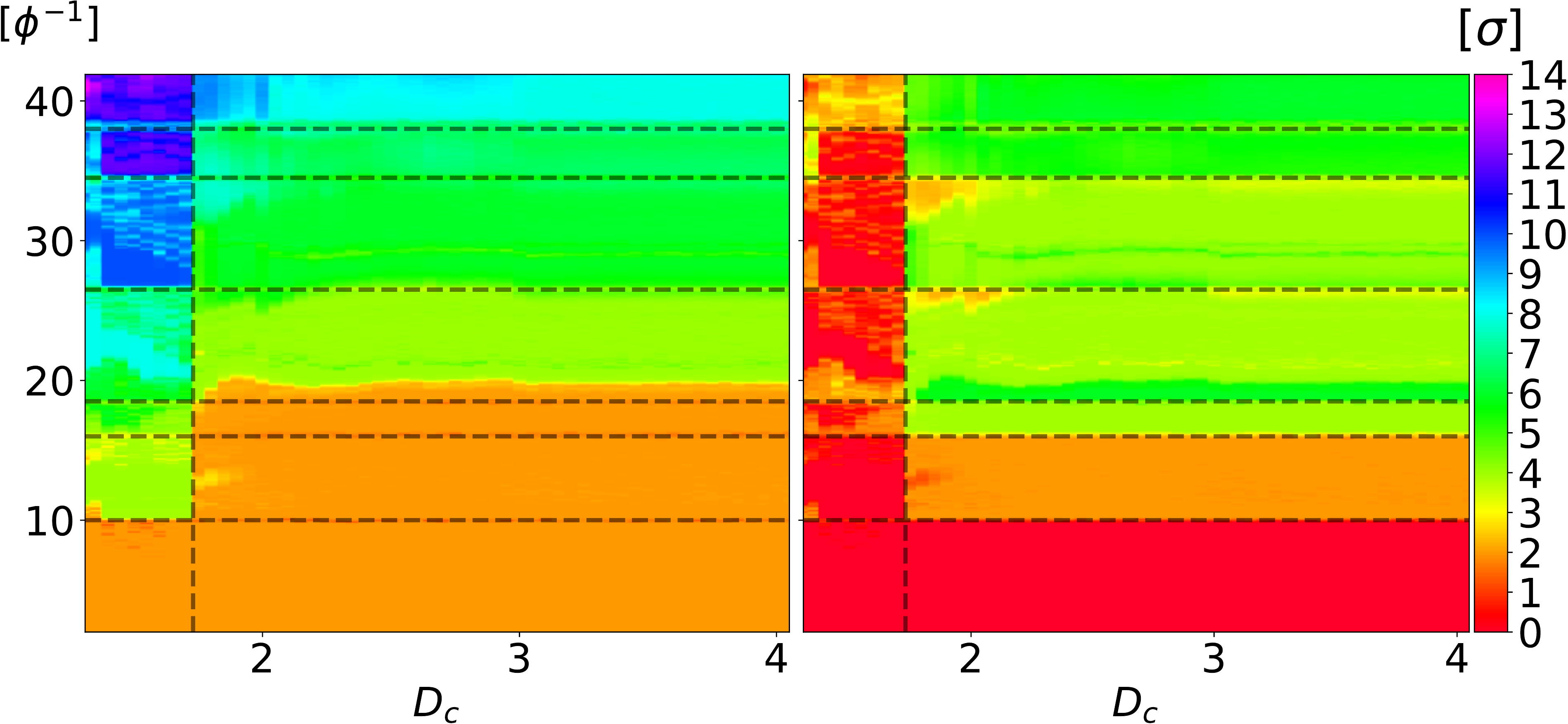}
  \caption{\parametercap{cutoff distance $D_c$}{a \gls{slsw} system}{\ShearLargeWidth}. \SkewWidth{44}.}\label{figure:interaction_cutoff_skew}%
}%
 
In \ref{figure:interaction_cutoff_skew} a calculation of different cutoff distances for \gls{slsw} systems is shown. The magnetoconductance shows a large jump at the value $D_c\approx1.73$ (position of vertical black line). In the interval $D_c\in\left[0, 1.73\right]$, the contribution to $\SPass{}$ on the right in \ref{figure:interaction_cutoff_skew} is much larger and the contribution to $\SLSW{}$ much smaller than to the right of the jump. The dependence on the cutoff distance to the right of the jump is strong and there is no obvious plateau structure for $\SPass{}$ in this regime. To the right of the jump, in the interval $D_c\in\left[1.73, 4.0\right]$, the magnetoconductance changes less rapidly with $D_c$ and there is a more apparent plateau structure when near the plateau centers of $\SRef{}$. The magnetoconductance functions show good convergence for values of $D_c>3.0$. Ultimately the structure of the calculation in \ref{figure:skew_hallbar} is well converged and a choice of $D_c=4.0$ is adequate for all presented results.

\subsection{\tlswhead}
\stfig{
  \includegraphics[width=\columnwidth]{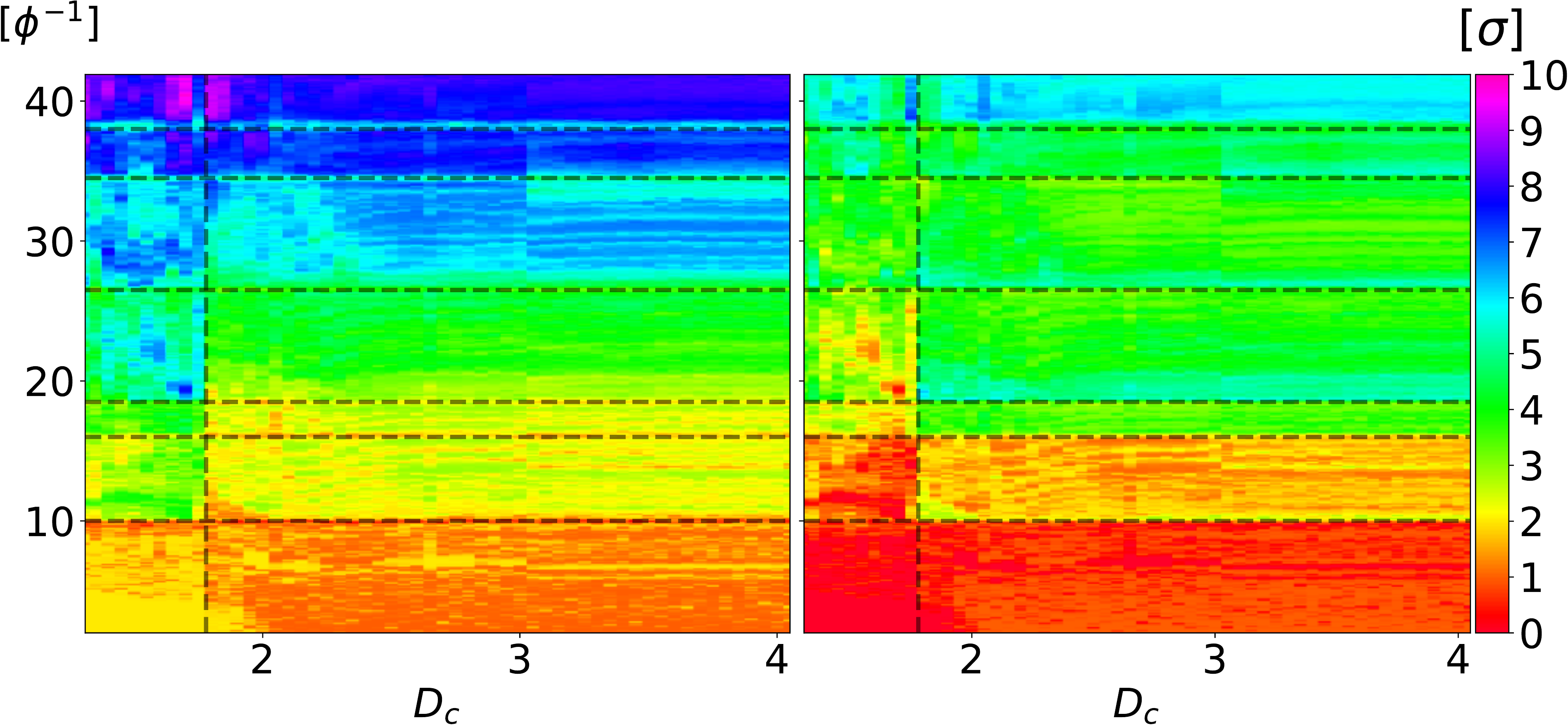}
  \caption{\parametercap{cutoff distance $D_c$}{a \gls{slsw} system}{\TensLargeWidth}. \TensWidth{60}.}\label{figure:interaction_cutoff_tension}%
}%
 
In \ref{figure:interaction_cutoff_tension} a calculation of different cutoff distances for \gls{tlsw} systems is shown. Just as for all other calculations, the magnetoconductance for this system type shows much stronger fluctuations than for the \gls{slsw} systems. Similarly to the \gls{slsw} case, there is a strong jump in the transmission function for a particular value of the cutoff distance. For the \gls{tlsw} system this value is $D_c\approx1.78$ (position of vertical black line). The change in transmission however is not as pronounced as in the \gls{slsw} case and the plateaus one to four are most affected, whereas the larger plateaus do not exhibit such a jump. There is a strong dependency on $D_c$ in both $\SPass{}$ and $\SLSW{}$ below $D_c\approx2.3$. After that, there is another minor change in conductance at $D_c=3.0$, where the decrease due to the fourth plateau transition in $\SRef{}$ becomes larger. For $D_c\leq 3.0$ however, the magnetoconductance is well converged keeping in mind the usual conductance fluctuations. As such the calculation for $D_c=4.0$ in \ref{figure:tensile_hallbar} and all other presented results in this paper should be appropriately converged with respect to the numerical hopping integral cutoff $D_c$.

The strong changes in the structure of the magnetoconductance calculations at particular values of $D_c$ make sense, since they correspond to the inclusion of particular hopping elements in the \gls{blg} lattice. In the case of the $D_c\approx1.7$ jump hoppings between sites of the same sublattice are added respectively.
\section{Other Fermi energies}\label{section:section_fermi}
As mentioned in the discussion of the parameters for all other calculations, the Fermi energy of all systems under consideration was chosen to be $\SI{0.95}{\electronvolt}\approx0.35t$. To identify energy dependencies a small study of the Fermi energy is performed for \gls{hlsw} with fully coupled upper sheet, \gls{slsw} and \gls{tlsw} systems. A small Fermi energy is difficult to investigate, since it requires smaller magnetic fields and thus larger length scales to suppress finite-size effects. Thus a large parameter study of very low-energy properties with a discrete model is currently not feasible due to computing time constraints.

\widefig{
  \adjustbox{width=\columnwidth}{%
    \includegraphics[width=\somesize]{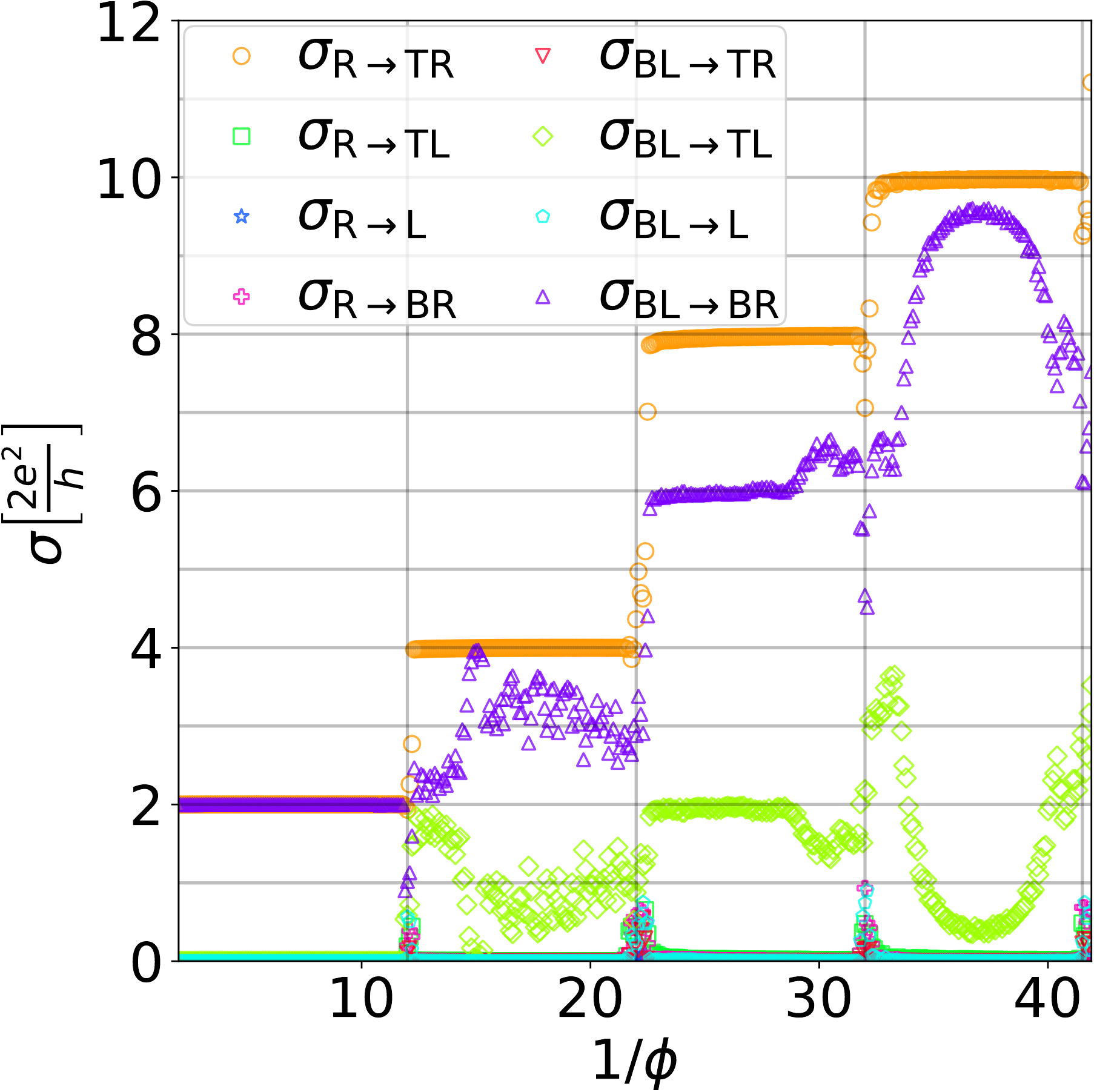}%
    \includegraphics[width=\somesize]{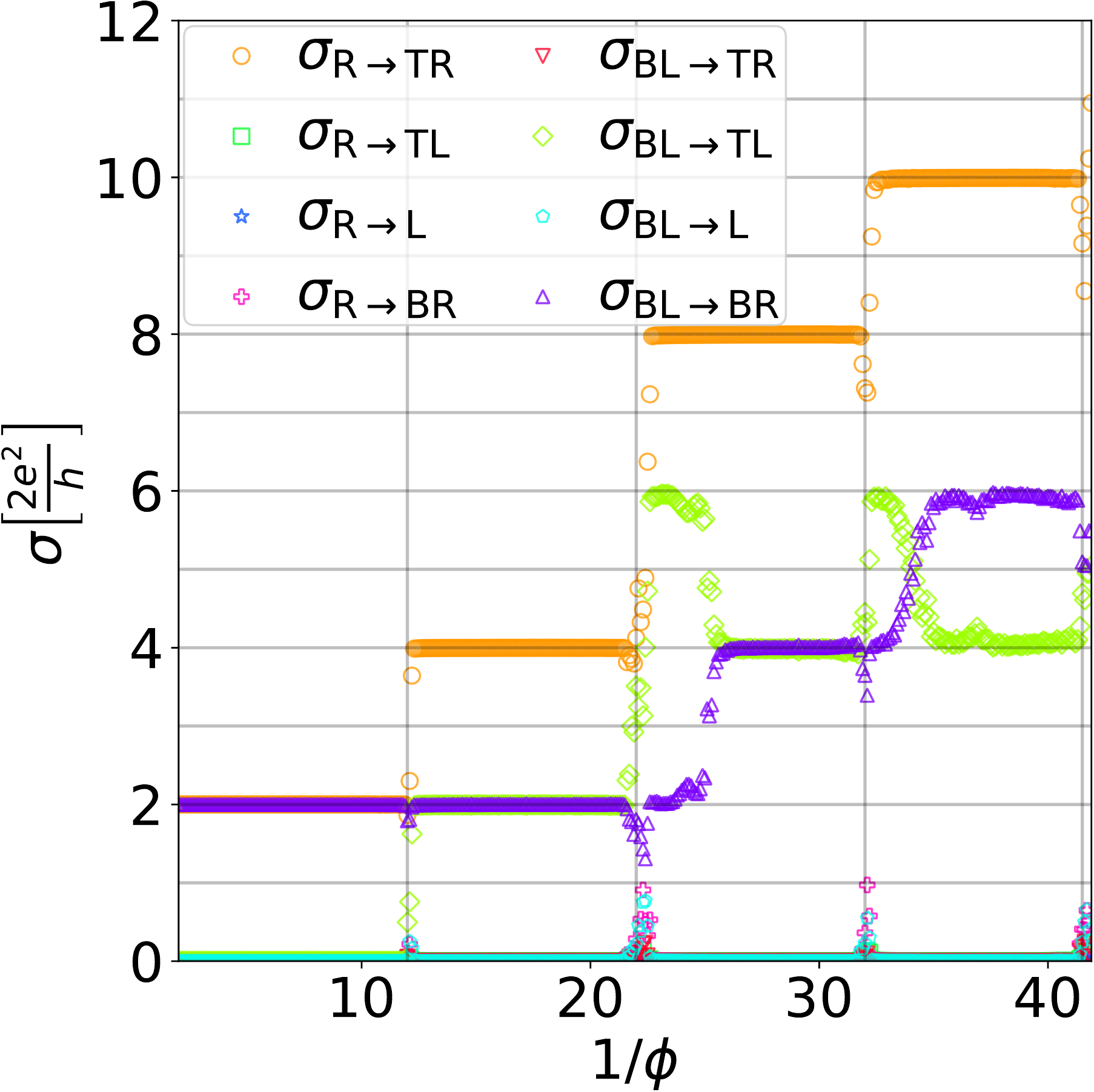}%
    \includegraphics[width=\somesize]{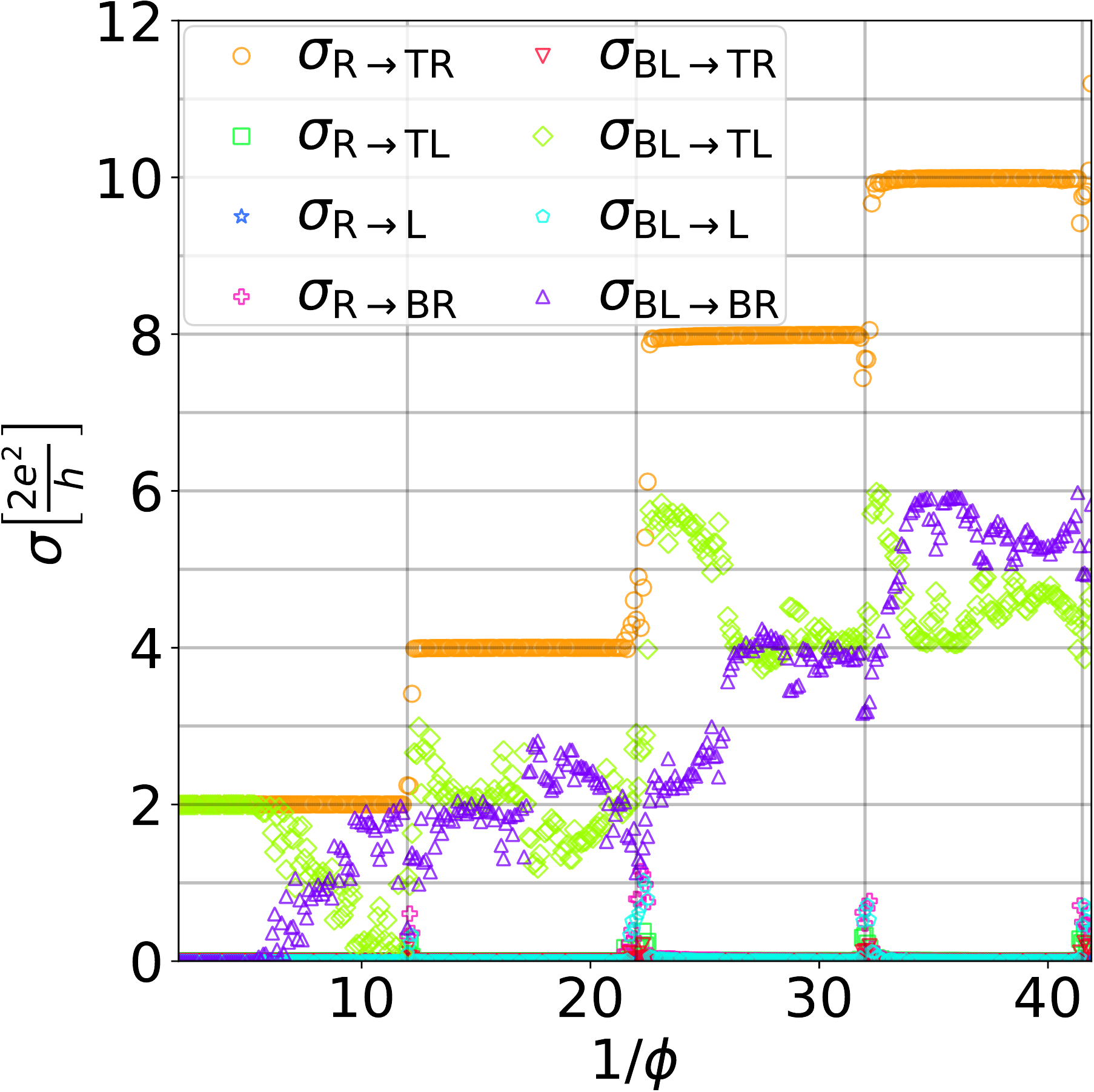}%
  }%

  \adjustbox{width=\columnwidth}{%
    \includegraphics[width=\somesize]{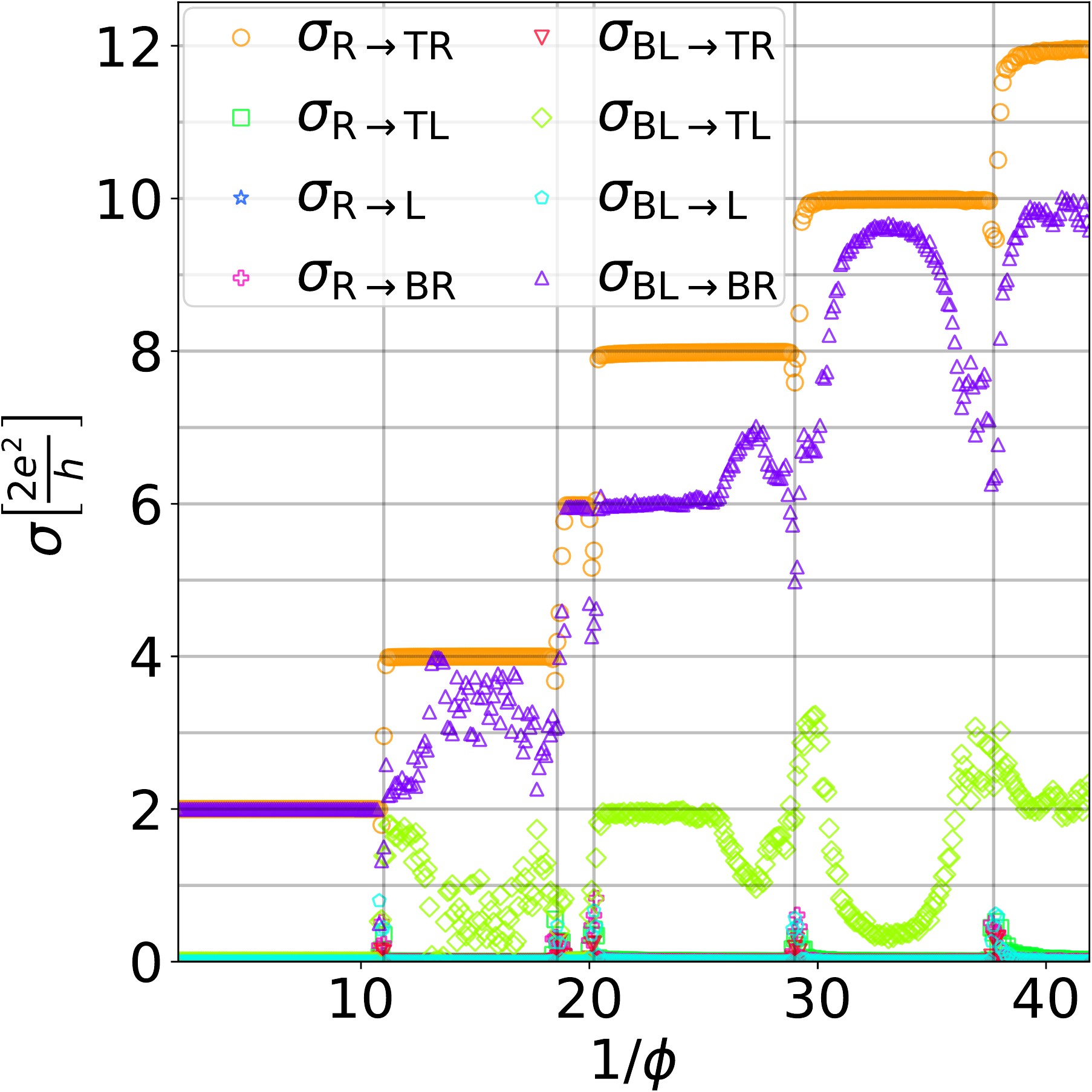}%
    \includegraphics[width=\somesize]{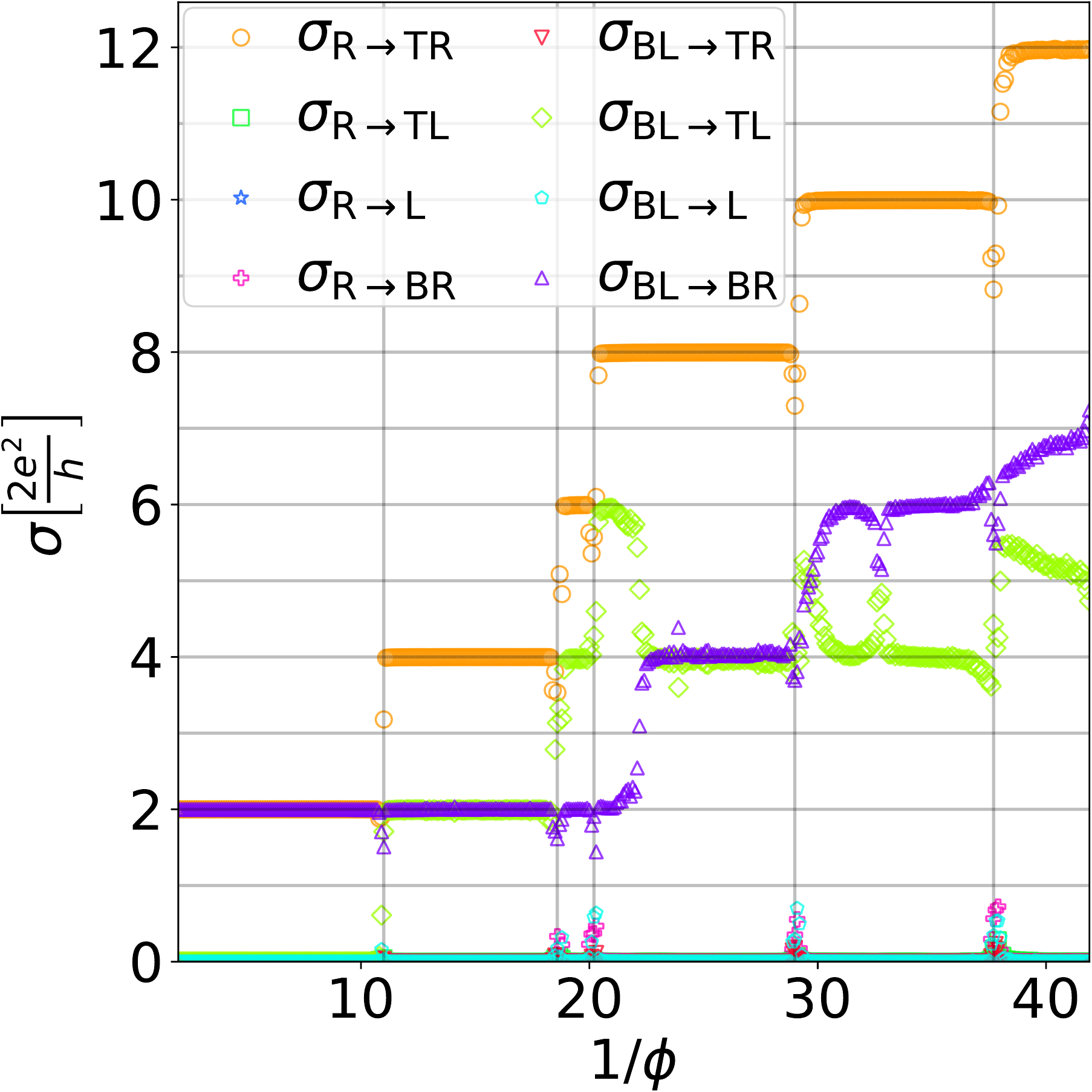}%
    \includegraphics[width=\somesize]{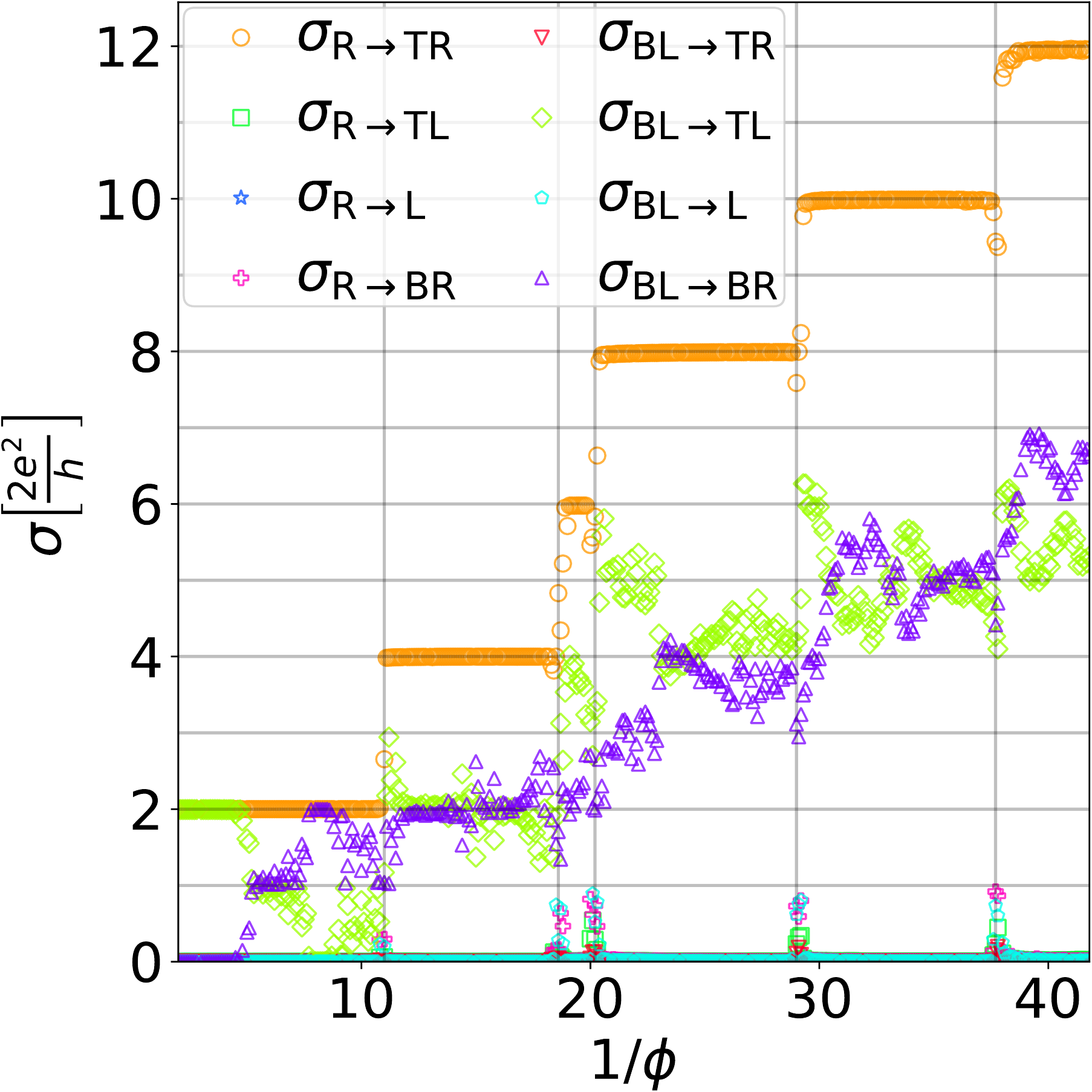}%
  }%

  \adjustbox{width=\columnwidth}{%
    \maybefbox{\includegraphics[width=\somesize]{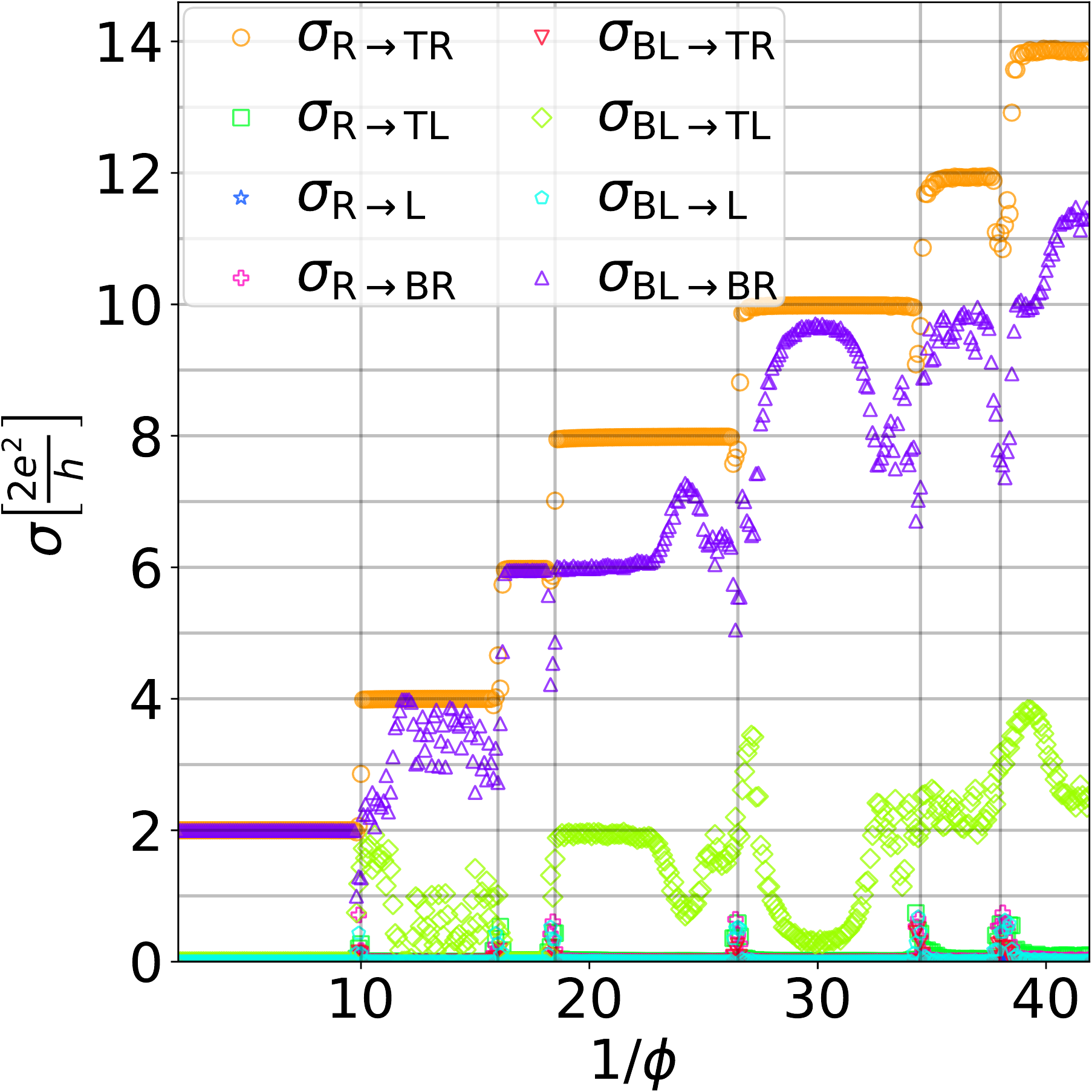}}%
    \maybefbox{\includegraphics[width=\somesize]{FIG_9_B.pdf}}%
    \maybefbox{\includegraphics[width=\somesize]{FIG_12_B.pdf}}%
  }%
  \caption{A study of magnetoconductance for three different Fermi energies. Transmission for \gls{hlsw} $N\approx\num{1.6e5}$ (left), \gls{slsw} $N\approx\num{1.7e5}$ (middle), \gls{tlsw} $N\approx\num{1.8e5}$ (right) systems. The energies are $E_{\text{F}}=\SI{0.85}{\electronvolt}$ (top), $E_{\text{F}}=\SI{0.9}{\electronvolt}$ (middle), $E_{\text{F}}=\SI{0.95}{\electronvolt}$ (bottom).}
  \label{figure:fermi_study}
}
 
The Fermi energy study is shown in \ref{figure:fermi_study}. The first obvious observation for all three system configurations is the difference in the Landau level filling for different Fermi energies. At lower energies and the same magnetic field, fewer bands will be filled and thus an analysis of the same region of magnetic field values at smaller Fermi energies will show a smaller number of plateaus. This is the essential reason why a large Fermi energy was used to perform calculations. To observe larger fillings for smaller energies, one would require smaller magnetic fields and thus larger systems to avoid finite-size effects.

There are changes to the plateau structure of $\SRef$ at different energies, but that is to be expected, since a different underlying band structure at a particular Fermi energy without magnetic field leads to different plateau transition positions. In particular, the width of the third plateau in $\SRef$ decreases for smaller Fermi energies.

For all three system configurations under consideration the general shapes of $\SPass$ and $\SLSW$ are remarkably similar for different energies supporting the generality of the discussed results. Although the generic shapes of $\SPass$ and $\SLSW$ are energy independent, there are also features, that warrant closer inspection. 

For the \gls{hlsw} systems on the left, the conductivities under consideration are largely the same except for two particular structures. The increase of $\SPass$ and decrease of $\SLSW$ at the transition from fourth to fifth plateau are attenuated for smaller energies. The transition from a smaller peak in $\SPass{}$ to a larger one from $E_{\text{F}}=\SI{0.85}{\electronvolt}$ to $E_{\text{F}}=\SI{0.9}{\electronvolt}$ and finally to $E_{\text{F}}=\SI{0.95}{\electronvolt}$ is apparent. There is also a decrease in $\SLSW$ at the transition from second to third plateau for $E_{\text{F}}=\SI{0.95}{\electronvolt}$, whose width decreases for $E_{\text{F}}=\SI{0.9}{\electronvolt}$ and has entirely vanished for $E_{\text{F}}=\SI{0.85}{\electronvolt}$

For the \gls{slsw} systems, the plateau structure is preserved in the range of interest. The conductance $\SPass{}$ is shifted to larger $\phi^{-1}$ for smaller Fermi energies. This causes structure in the conductance $\SLSW{}$, namely peaks around $\phi^{-1}=22$ and $\phi^{-1}=32$, due to the offset between plateau transitions in $\SRef{}$ and $\SPass{}$. The two conductances for the Fermi energy $E_{\text{F}}=\SI{0.95}{\electronvolt}$ merely seem to be well aligned such that $\SLSW{}$ has a plateau structure as well. Thus the monotonous plateau structure in $\SPass{}$ is generic for energies in the considered range, but the plateau structure of $\SLSW{}$ is not. The width change of the third plateau is also reflected in the width of the peak in $\SLSW{}$ near that transition.

Just as for the previous two system types, the general shape of the conductance calculations is similar for all three energies considered aside from the change in filling due to changed electron density at any particular value of $\phi^{-1}$. Like for the \gls{hlsw} calculations the change in width of the third plateau causes changes in the conductances $\SLSW{}$ and $\SPass{}$. Aside from that the general shape of the conductances is energy independent, but the particular fluctuations do change with energy. For example, the exact structures of $\SLSW{}$ and $\SPass{}$ around the fifth plateau clearly show a small energy dependence. However, the instability of this system type with respect to parameter changes has already been established and this behaviour is not surprising.

For all three system types, changes of features in the region between $\phi^{-1}=15$ and $\phi^{-1}=23$  for $E_{\text{F}}=\SI{0.95}{\electronvolt}$ coincide with the decrease in the width of the third plateau for smaller Fermi energies. It seems reasonable, that the magnetoconductivity for \gls{lsw} systems would reflect the changes in the homogeneous systems.

In conclusion, there are changes in the conductivities with Fermi energy. This behaviour however is not unexpected, since the homogeneous system conductance also changes with energy. These changes however do not fundamentally affect the previous discussions of these systems and this is a good indicator that the discussed properties like plateau formation are generic, at least in the energy range investigated.

\begin{thebibliography}{60}%
\makeatletter
\providecommand \@ifxundefined [1]{%
 \@ifx{#1\undefined}
}%
\providecommand \@ifnum [1]{%
 \ifnum #1\expandafter \@firstoftwo
 \else \expandafter \@secondoftwo
 \fi
}%
\providecommand \@ifx [1]{%
 \ifx #1\expandafter \@firstoftwo
 \else \expandafter \@secondoftwo
 \fi
}%
\providecommand \natexlab [1]{#1}%
\providecommand \enquote  [1]{``#1''}%
\providecommand \bibnamefont  [1]{#1}%
\providecommand \bibfnamefont [1]{#1}%
\providecommand \citenamefont [1]{#1}%
\providecommand \href@noop [0]{\@secondoftwo}%
\providecommand \href [0]{\begingroup \@sanitize@url \@href}%
\providecommand \@href[1]{\@@startlink{#1}\@@href}%
\providecommand \@@href[1]{\endgroup#1\@@endlink}%
\providecommand \@sanitize@url [0]{\catcode `\\12\catcode `\$12\catcode
  `\&12\catcode `\#12\catcode `\^12\catcode `\_12\catcode `\%12\relax}%
\providecommand \@@startlink[1]{}%
\providecommand \@@endlink[0]{}%
\providecommand \url  [0]{\begingroup\@sanitize@url \@url }%
\providecommand \@url [1]{\endgroup\@href {#1}{\urlprefix }}%
\providecommand \urlprefix  [0]{URL }%
\providecommand \Eprint [0]{\href }%
\providecommand \doibase [0]{http://dx.doi.org/}%
\providecommand \selectlanguage [0]{\@gobble}%
\providecommand \bibinfo  [0]{\@secondoftwo}%
\providecommand \bibfield  [0]{\@secondoftwo}%
\providecommand \translation [1]{[#1]}%
\providecommand \BibitemOpen [0]{}%
\providecommand \bibitemStop [0]{}%
\providecommand \bibitemNoStop [0]{.\EOS\space}%
\providecommand \EOS [0]{\spacefactor3000\relax}%
\providecommand \BibitemShut  [1]{\csname bibitem#1\endcsname}%
\let\auto@bib@innerbib\@empty
\justifying{}
\bibitem [{\citenamefont {Bonaccorso}\ \emph {et~al.}(2012)\citenamefont
  {Bonaccorso}, \citenamefont {Lombardo}, \citenamefont {Hasan}, \citenamefont
  {Sun}, \citenamefont {Colombo},\ and\ \citenamefont
  {Ferrari}}]{BONACCORSO2012564}%
  \BibitemOpen
  \bibfield  {author} {\bibinfo {author} {\bibfnamefont {F.}~\bibnamefont
  {Bonaccorso}}, \bibinfo {author} {\bibfnamefont {A.}~\bibnamefont
  {Lombardo}}, \bibinfo {author} {\bibfnamefont {T.}~\bibnamefont {Hasan}},
  \bibinfo {author} {\bibfnamefont {Z.}~\bibnamefont {Sun}}, \bibinfo {author}
  {\bibfnamefont {L.}~\bibnamefont {Colombo}}, \ and\ \bibinfo {author}
  {\bibfnamefont {A.~C.}\ \bibnamefont {Ferrari}},\ }\href {\doibase
  https://doi.org/10.1016/S1369-7021(13)70014-2} {\bibfield  {journal}
  {\bibinfo  {journal} {Materials Today}\ }\textbf {\bibinfo {volume} {15}},\
  \bibinfo {pages} {564 } (\bibinfo {year} {2012})}\BibitemShut {NoStop}%
\bibitem [{\citenamefont {Geim}\ and\ \citenamefont
  {Grigorieva}(2013)}]{Geim2013}%
  \BibitemOpen
  \bibfield  {author} {\bibinfo {author} {\bibfnamefont {A.~K.}\ \bibnamefont
  {Geim}}\ and\ \bibinfo {author} {\bibfnamefont {I.~V.}\ \bibnamefont
  {Grigorieva}},\ }\href {\doibase 10.1038/nature12385} {\bibfield  {journal}
  {\bibinfo  {journal} {Nature}\ }\textbf {\bibinfo {volume} {499}},\ \bibinfo
  {pages} {419} (\bibinfo {year} {2013})}\BibitemShut {NoStop}%
\bibitem [{\citenamefont {Qi}\ and\ \citenamefont {Zhang}(2011)}]{Xiao2011}%
  \BibitemOpen
  \bibfield  {author} {\bibinfo {author} {\bibfnamefont {X.-L.}\ \bibnamefont
  {Qi}}\ and\ \bibinfo {author} {\bibfnamefont {S.-C.}\ \bibnamefont {Zhang}},\
  }\href {\doibase 10.1103/RevModPhys.83.1057} {\bibfield  {journal} {\bibinfo
  {journal} {Rev. Mod. Phys.}\ }\textbf {\bibinfo {volume} {83}},\ \bibinfo
  {pages} {1057} (\bibinfo {year} {2011})}\BibitemShut {NoStop}%
\bibitem [{\citenamefont {Hasan}\ and\ \citenamefont {Kane}(2010)}]{Hasan2010}%
  \BibitemOpen
  \bibfield  {author} {\bibinfo {author} {\bibfnamefont {M.~Z.}\ \bibnamefont
  {Hasan}}\ and\ \bibinfo {author} {\bibfnamefont {C.~L.}\ \bibnamefont
  {Kane}},\ }\href {\doibase 10.1103/RevModPhys.82.3045} {\bibfield  {journal}
  {\bibinfo  {journal} {Rev. Mod. Phys.}\ }\textbf {\bibinfo {volume} {82}},\
  \bibinfo {pages} {3045} (\bibinfo {year} {2010})}\BibitemShut {NoStop}%
\bibitem [{\citenamefont {Schnyder}\ \emph {et~al.}(2009)\citenamefont
  {Schnyder}, \citenamefont {Ryu}, \citenamefont {Furusaki}, \citenamefont
  {Ludwig}, \citenamefont {Lebedev},\ and\ \citenamefont
  {Feigel’man}}]{schnyder2009}%
  \BibitemOpen
  \bibfield  {author} {\bibinfo {author} {\bibfnamefont {A.~P.}\ \bibnamefont
  {Schnyder}}, \bibinfo {author} {\bibfnamefont {S.}~\bibnamefont {Ryu}},
  \bibinfo {author} {\bibfnamefont {A.}~\bibnamefont {Furusaki}}, \bibinfo
  {author} {\bibfnamefont {A.~W.~W.}\ \bibnamefont {Ludwig}}, \bibinfo {author}
  {\bibfnamefont {V.}~\bibnamefont {Lebedev}}, \ and\ \bibinfo {author}
  {\bibfnamefont {M.}~\bibnamefont {Feigel’man}},\ }\href {\doibase
  10.1063/1.3149481} {\bibfield  {journal} {\bibinfo  {journal} {AIP Conference
  Proceedings}\ }\textbf {\bibinfo {volume} {1134}},\ \bibinfo {pages} {10}
  (\bibinfo {year} {2009})},\ \Eprint
  {http://arxiv.org/abs/https://aip.scitation.org/doi/pdf/10.1063/1.3149481}
  {https://aip.scitation.org/doi/pdf/10.1063/1.3149481} \BibitemShut {NoStop}%
\bibitem [{\citenamefont {Kitaev}(2001)}]{Kitaev2001}%
  \BibitemOpen
  \bibfield  {author} {\bibinfo {author} {\bibfnamefont {A.~Y.}\ \bibnamefont
  {Kitaev}},\ }\href {\doibase 10.1070/1063-7869/44/10s/s29} {\bibfield
  {journal} {\bibinfo  {journal} {Physics-Uspekhi}\ }\textbf {\bibinfo {volume}
  {44}},\ \bibinfo {pages} {131} (\bibinfo {year} {2001})}\BibitemShut
  {NoStop}%
\bibitem [{\citenamefont {Kitaev}\ \emph {et~al.}(2009)\citenamefont {Kitaev},
  \citenamefont {Lebedev},\ and\ \citenamefont {Feigel’man}}]{kitaev2009}%
  \BibitemOpen
  \bibfield  {author} {\bibinfo {author} {\bibfnamefont {A.}~\bibnamefont
  {Kitaev}}, \bibinfo {author} {\bibfnamefont {V.}~\bibnamefont {Lebedev}}, \
  and\ \bibinfo {author} {\bibfnamefont {M.}~\bibnamefont {Feigel’man}},\
  }\href {\doibase 10.1063/1.3149495} {\bibfield  {journal} {\bibinfo
  {journal} {AIP Conference Proceedings}\ }\textbf {\bibinfo {volume} {1134}},\
  \bibinfo {pages} {22} (\bibinfo {year} {2009})},\ \Eprint
  {http://arxiv.org/abs/https://aip.scitation.org/doi/pdf/10.1063/1.3149495}
  {https://aip.scitation.org/doi/pdf/10.1063/1.3149495} \BibitemShut {NoStop}%
\bibitem [{\citenamefont {Kitaev}(2003)}]{kitaev2003}%
  \BibitemOpen
  \bibfield  {author} {\bibinfo {author} {\bibfnamefont {A.}~\bibnamefont
  {Kitaev}},\ }\href {\doibase https://doi.org/10.1016/S0003-4916(02)00018-0}
  {\bibfield  {journal} {\bibinfo  {journal} {Annals of Physics}\ }\textbf
  {\bibinfo {volume} {303}},\ \bibinfo {pages} {2 } (\bibinfo {year}
  {2003})}\BibitemShut {NoStop}%
\bibitem [{\citenamefont {Freedman}\ \emph {et~al.}(2003)\citenamefont
  {Freedman}, \citenamefont {Kitaev}, \citenamefont {Larsen},\ and\
  \citenamefont {Wang}}]{freedman2003}%
  \BibitemOpen
  \bibfield  {author} {\bibinfo {author} {\bibfnamefont {M.~H.}\ \bibnamefont
  {Freedman}}, \bibinfo {author} {\bibfnamefont {A.}~\bibnamefont {Kitaev}},
  \bibinfo {author} {\bibfnamefont {M.~J.}\ \bibnamefont {Larsen}}, \ and\
  \bibinfo {author} {\bibfnamefont {Z.}~\bibnamefont {Wang}}\ }(\bibinfo {year}
  {2003})\ pp.\ \bibinfo {pages} {31--38},\ \bibinfo {note} {mathematical
  challenges of the 21st century (Los Angeles, CA, 2000)}\BibitemShut {NoStop}%
\bibitem [{\citenamefont {Laughlin}(1983)}]{laughlin1983}%
  \BibitemOpen
  \bibfield  {author} {\bibinfo {author} {\bibfnamefont {R.~B.}\ \bibnamefont
  {Laughlin}},\ }\href {\doibase 10.1103/PhysRevLett.50.1395} {\bibfield
  {journal} {\bibinfo  {journal} {Phys. Rev. Lett.}\ }\textbf {\bibinfo
  {volume} {50}},\ \bibinfo {pages} {1395} (\bibinfo {year}
  {1983})}\BibitemShut {NoStop}%
\bibitem [{\citenamefont {Tsui}\ \emph {et~al.}(1982)\citenamefont {Tsui},
  \citenamefont {Stormer},\ and\ \citenamefont {Gossard}}]{tsui1982}%
  \BibitemOpen
  \bibfield  {author} {\bibinfo {author} {\bibfnamefont {D.~C.}\ \bibnamefont
  {Tsui}}, \bibinfo {author} {\bibfnamefont {H.~L.}\ \bibnamefont {Stormer}}, \
  and\ \bibinfo {author} {\bibfnamefont {A.~C.}\ \bibnamefont {Gossard}},\
  }\href {\doibase 10.1103/PhysRevLett.48.1559} {\bibfield  {journal} {\bibinfo
   {journal} {Phys. Rev. Lett.}\ }\textbf {\bibinfo {volume} {48}},\ \bibinfo
  {pages} {1559} (\bibinfo {year} {1982})}\BibitemShut {NoStop}%
\bibitem [{\citenamefont {Balents}(2010)}]{Balents2010}%
  \BibitemOpen
  \bibfield  {author} {\bibinfo {author} {\bibfnamefont {L.}~\bibnamefont
  {Balents}},\ }\href {\doibase 10.1038/nature08917} {\bibfield  {journal}
  {\bibinfo  {journal} {Nature}\ }\textbf {\bibinfo {volume} {464}},\ \bibinfo
  {pages} {199} (\bibinfo {year} {2010})}\BibitemShut {NoStop}%
\bibitem [{\citenamefont {Kane}\ and\ \citenamefont
  {Mele}(2005)}]{kane2005quantum}%
  \BibitemOpen
  \bibfield  {author} {\bibinfo {author} {\bibfnamefont {C.~L.}\ \bibnamefont
  {Kane}}\ and\ \bibinfo {author} {\bibfnamefont {E.~J.}\ \bibnamefont
  {Mele}},\ }\href {\doibase 10.1103/PhysRevLett.95.226801} {\bibfield
  {journal} {\bibinfo  {journal} {Phys. Rev. Lett.}\ }\textbf {\bibinfo
  {volume} {95}},\ \bibinfo {pages} {226801} (\bibinfo {year}
  {2005})}\BibitemShut {NoStop}%
\bibitem [{\citenamefont {Frank}\ \emph {et~al.}(2007)\citenamefont {Frank},
  \citenamefont {Tanenbaum}, \citenamefont {van~der Zande},\ and\ \citenamefont
  {McEuen}}]{Frank2007}%
  \BibitemOpen
  \bibfield  {author} {\bibinfo {author} {\bibfnamefont {I.~W.}\ \bibnamefont
  {Frank}}, \bibinfo {author} {\bibfnamefont {D.~M.}\ \bibnamefont
  {Tanenbaum}}, \bibinfo {author} {\bibfnamefont {A.~M.}\ \bibnamefont {van~der
  Zande}}, \ and\ \bibinfo {author} {\bibfnamefont {P.~L.}\ \bibnamefont
  {McEuen}},\ }\href {\doibase 10.1116/1.2789446} {\bibfield  {journal}
  {\bibinfo  {journal} {Journal of Vacuum Science \& Technology B:
  Microelectronics and Nanometer Structures Processing, Measurement, and
  Phenomena}\ }\textbf {\bibinfo {volume} {25}},\ \bibinfo {pages} {2558}
  (\bibinfo {year} {2007})},\ \Eprint
  {http://arxiv.org/abs/https://avs.scitation.org/doi/pdf/10.1116/1.2789446}
  {https://avs.scitation.org/doi/pdf/10.1116/1.2789446} \BibitemShut {NoStop}%
\bibitem [{\citenamefont {Schwierz}(2010)}]{Schwierz2010}%
  \BibitemOpen
  \bibfield  {author} {\bibinfo {author} {\bibfnamefont {F.}~\bibnamefont
  {Schwierz}},\ }\href {\doibase 10.1038/nnano.2010.89} {\bibfield  {journal}
  {\bibinfo  {journal} {Nature Nanotechnology}\ }\textbf {\bibinfo {volume}
  {5}},\ \bibinfo {pages} {487} (\bibinfo {year} {2010})}\BibitemShut {NoStop}%
\bibitem [{Xia(2009)}]{Xia2009}%
  \BibitemOpen
  \href {\doibase 10.1038/nnano.2009.292} {\bibfield  {journal} {\bibinfo
  {journal} {Nature Nanotechnology}\ }\textbf {\bibinfo {volume} {4}},\
  \bibinfo {pages} {839} (\bibinfo {year} {2009})}\BibitemShut {NoStop}%
\bibitem [{\citenamefont {Ohta}\ \emph {et~al.}(2006)\citenamefont {Ohta},
  \citenamefont {Bostwick}, \citenamefont {Seyller}, \citenamefont {Horn},\
  and\ \citenamefont {Rotenberg}}]{Ohta951}%
  \BibitemOpen
  \bibfield  {author} {\bibinfo {author} {\bibfnamefont {T.}~\bibnamefont
  {Ohta}}, \bibinfo {author} {\bibfnamefont {A.}~\bibnamefont {Bostwick}},
  \bibinfo {author} {\bibfnamefont {T.}~\bibnamefont {Seyller}}, \bibinfo
  {author} {\bibfnamefont {K.}~\bibnamefont {Horn}}, \ and\ \bibinfo {author}
  {\bibfnamefont {E.}~\bibnamefont {Rotenberg}},\ }\href {\doibase
  10.1126/science.1130681} {\bibfield  {journal} {\bibinfo  {journal}
  {Science}\ }\textbf {\bibinfo {volume} {313}},\ \bibinfo {pages} {951}
  (\bibinfo {year} {2006})},\ \Eprint
  {http://arxiv.org/abs/https://science.sciencemag.org/content/313/5789/951.full.pdf}
  {https://science.sciencemag.org/content/313/5789/951.full.pdf} \BibitemShut
  {NoStop}%
\bibitem [{\citenamefont {Zhang}\ \emph {et~al.}(2009)\citenamefont {Zhang},
  \citenamefont {Tang}, \citenamefont {Girit}, \citenamefont {Hao},
  \citenamefont {Martin}, \citenamefont {Zettl}, \citenamefont {Crommie},
  \citenamefont {Shen},\ and\ \citenamefont {Wang}}]{Zhang2009}%
  \BibitemOpen
  \bibfield  {author} {\bibinfo {author} {\bibfnamefont {Y.}~\bibnamefont
  {Zhang}}, \bibinfo {author} {\bibfnamefont {T.-T.}\ \bibnamefont {Tang}},
  \bibinfo {author} {\bibfnamefont {C.}~\bibnamefont {Girit}}, \bibinfo
  {author} {\bibfnamefont {Z.}~\bibnamefont {Hao}}, \bibinfo {author}
  {\bibfnamefont {M.~C.}\ \bibnamefont {Martin}}, \bibinfo {author}
  {\bibfnamefont {A.}~\bibnamefont {Zettl}}, \bibinfo {author} {\bibfnamefont
  {M.~F.}\ \bibnamefont {Crommie}}, \bibinfo {author} {\bibfnamefont {Y.~R.}\
  \bibnamefont {Shen}}, \ and\ \bibinfo {author} {\bibfnamefont
  {F.}~\bibnamefont {Wang}},\ }\href {\doibase 10.1038/nature08105} {\bibfield
  {journal} {\bibinfo  {journal} {Nature}\ }\textbf {\bibinfo {volume} {459}},\
  \bibinfo {pages} {820} (\bibinfo {year} {2009})}\BibitemShut {NoStop}%
\bibitem [{\citenamefont {McCann}\ and\ \citenamefont
  {Koshino}(2013)}]{McCann_2013}%
  \BibitemOpen
  \bibfield  {author} {\bibinfo {author} {\bibfnamefont {E.}~\bibnamefont
  {McCann}}\ and\ \bibinfo {author} {\bibfnamefont {M.}~\bibnamefont
  {Koshino}},\ }\href {\doibase 10.1088/0034-4885/76/5/056503} {\bibfield
  {journal} {\bibinfo  {journal} {Reports on Progress in Physics}\ }\textbf
  {\bibinfo {volume} {76}},\ \bibinfo {pages} {056503} (\bibinfo {year}
  {2013})}\BibitemShut {NoStop}%
\bibitem [{\citenamefont {Su\'arez~Morell}\ \emph {et~al.}(2010)\citenamefont
  {Su\'arez~Morell}, \citenamefont {Correa}, \citenamefont {Vargas},
  \citenamefont {Pacheco},\ and\ \citenamefont {Barticevic}}]{Suarez2010}%
  \BibitemOpen
  \bibfield  {author} {\bibinfo {author} {\bibfnamefont {E.}~\bibnamefont
  {Su\'arez~Morell}}, \bibinfo {author} {\bibfnamefont {J.~D.}\ \bibnamefont
  {Correa}}, \bibinfo {author} {\bibfnamefont {P.}~\bibnamefont {Vargas}},
  \bibinfo {author} {\bibfnamefont {M.}~\bibnamefont {Pacheco}}, \ and\
  \bibinfo {author} {\bibfnamefont {Z.}~\bibnamefont {Barticevic}},\ }\href
  {\doibase 10.1103/PhysRevB.82.121407} {\bibfield  {journal} {\bibinfo
  {journal} {Phys. Rev. B}\ }\textbf {\bibinfo {volume} {82}},\ \bibinfo
  {pages} {121407} (\bibinfo {year} {2010})}\BibitemShut {NoStop}%
\bibitem [{\citenamefont {Huang}\ \emph {et~al.}(2018)\citenamefont {Huang},
  \citenamefont {Kim}, \citenamefont {Efimkin}, \citenamefont {Lovorn},
  \citenamefont {Taniguchi}, \citenamefont {Watanabe}, \citenamefont
  {MacDonald}, \citenamefont {Tutuc},\ and\ \citenamefont {LeRoy}}]{Huang2018}%
  \BibitemOpen
  \bibfield  {author} {\bibinfo {author} {\bibfnamefont {S.}~\bibnamefont
  {Huang}}, \bibinfo {author} {\bibfnamefont {K.}~\bibnamefont {Kim}}, \bibinfo
  {author} {\bibfnamefont {D.~K.}\ \bibnamefont {Efimkin}}, \bibinfo {author}
  {\bibfnamefont {T.}~\bibnamefont {Lovorn}}, \bibinfo {author} {\bibfnamefont
  {T.}~\bibnamefont {Taniguchi}}, \bibinfo {author} {\bibfnamefont
  {K.}~\bibnamefont {Watanabe}}, \bibinfo {author} {\bibfnamefont {A.~H.}\
  \bibnamefont {MacDonald}}, \bibinfo {author} {\bibfnamefont {E.}~\bibnamefont
  {Tutuc}}, \ and\ \bibinfo {author} {\bibfnamefont {B.~J.}\ \bibnamefont
  {LeRoy}},\ }\href {\doibase 10.1103/PhysRevLett.121.037702} {\bibfield
  {journal} {\bibinfo  {journal} {Phys. Rev. Lett.}\ }\textbf {\bibinfo
  {volume} {121}},\ \bibinfo {pages} {037702} (\bibinfo {year}
  {2018})}\BibitemShut {NoStop}%
\bibitem [{\citenamefont {Yankowitz}\ \emph {et~al.}(2019)\citenamefont
  {Yankowitz}, \citenamefont {Chen}, \citenamefont {Polshyn}, \citenamefont
  {Zhang}, \citenamefont {Watanabe}, \citenamefont {Taniguchi}, \citenamefont
  {Graf}, \citenamefont {Young},\ and\ \citenamefont {Dean}}]{Yankowitz1059}%
  \BibitemOpen
  \bibfield  {author} {\bibinfo {author} {\bibfnamefont {M.}~\bibnamefont
  {Yankowitz}}, \bibinfo {author} {\bibfnamefont {S.}~\bibnamefont {Chen}},
  \bibinfo {author} {\bibfnamefont {H.}~\bibnamefont {Polshyn}}, \bibinfo
  {author} {\bibfnamefont {Y.}~\bibnamefont {Zhang}}, \bibinfo {author}
  {\bibfnamefont {K.}~\bibnamefont {Watanabe}}, \bibinfo {author}
  {\bibfnamefont {T.}~\bibnamefont {Taniguchi}}, \bibinfo {author}
  {\bibfnamefont {D.}~\bibnamefont {Graf}}, \bibinfo {author} {\bibfnamefont
  {A.~F.}\ \bibnamefont {Young}}, \ and\ \bibinfo {author} {\bibfnamefont
  {C.~R.}\ \bibnamefont {Dean}},\ }\href {\doibase 10.1126/science.aav1910}
  {\bibfield  {journal} {\bibinfo  {journal} {Science}\ }\textbf {\bibinfo
  {volume} {363}},\ \bibinfo {pages} {1059} (\bibinfo {year} {2019})},\ \Eprint
  {http://arxiv.org/abs/https://science.sciencemag.org/content/363/6431/1059.full.pdf}
  {https://science.sciencemag.org/content/363/6431/1059.full.pdf} \BibitemShut
  {NoStop}%
\bibitem [{\citenamefont {Cao}\ \emph {et~al.}(2018{\natexlab{a}})\citenamefont
  {Cao}, \citenamefont {Fatemi}, \citenamefont {Demir}, \citenamefont {Fang},
  \citenamefont {Tomarken}, \citenamefont {Luo}, \citenamefont
  {Sanchez-Yamagishi}, \citenamefont {Watanabe}, \citenamefont {Taniguchi},
  \citenamefont {Kaxiras}, \citenamefont {Ashoori},\ and\ \citenamefont
  {Jarillo-Herrero}}]{Cao2018Insulator}%
  \BibitemOpen
  \bibfield  {author} {\bibinfo {author} {\bibfnamefont {Y.}~\bibnamefont
  {Cao}}, \bibinfo {author} {\bibfnamefont {V.}~\bibnamefont {Fatemi}},
  \bibinfo {author} {\bibfnamefont {A.}~\bibnamefont {Demir}}, \bibinfo
  {author} {\bibfnamefont {S.}~\bibnamefont {Fang}}, \bibinfo {author}
  {\bibfnamefont {S.~L.}\ \bibnamefont {Tomarken}}, \bibinfo {author}
  {\bibfnamefont {J.~Y.}\ \bibnamefont {Luo}}, \bibinfo {author} {\bibfnamefont
  {J.~D.}\ \bibnamefont {Sanchez-Yamagishi}}, \bibinfo {author} {\bibfnamefont
  {K.}~\bibnamefont {Watanabe}}, \bibinfo {author} {\bibfnamefont
  {T.}~\bibnamefont {Taniguchi}}, \bibinfo {author} {\bibfnamefont
  {E.}~\bibnamefont {Kaxiras}}, \bibinfo {author} {\bibfnamefont {R.~C.}\
  \bibnamefont {Ashoori}}, \ and\ \bibinfo {author} {\bibfnamefont
  {P.}~\bibnamefont {Jarillo-Herrero}},\ }\href {\doibase 10.1038/nature26154}
  {\bibfield  {journal} {\bibinfo  {journal} {Nature}\ }\textbf {\bibinfo
  {volume} {556}},\ \bibinfo {pages} {80} (\bibinfo {year}
  {2018}{\natexlab{a}})}\BibitemShut {NoStop}%
\bibitem [{\citenamefont {Cao}\ \emph {et~al.}(2018{\natexlab{b}})\citenamefont
  {Cao}, \citenamefont {Fatemi}, \citenamefont {Fang}, \citenamefont
  {Watanabe}, \citenamefont {Taniguchi}, \citenamefont {Kaxiras},\ and\
  \citenamefont {Jarillo-Herrero}}]{Cao2018}%
  \BibitemOpen
  \bibfield  {author} {\bibinfo {author} {\bibfnamefont {Y.}~\bibnamefont
  {Cao}}, \bibinfo {author} {\bibfnamefont {V.}~\bibnamefont {Fatemi}},
  \bibinfo {author} {\bibfnamefont {S.}~\bibnamefont {Fang}}, \bibinfo {author}
  {\bibfnamefont {K.}~\bibnamefont {Watanabe}}, \bibinfo {author}
  {\bibfnamefont {T.}~\bibnamefont {Taniguchi}}, \bibinfo {author}
  {\bibfnamefont {E.}~\bibnamefont {Kaxiras}}, \ and\ \bibinfo {author}
  {\bibfnamefont {P.}~\bibnamefont {Jarillo-Herrero}},\ }\href {\doibase
  10.1038/nature26160} {\bibfield  {journal} {\bibinfo  {journal} {Nature}\
  }\textbf {\bibinfo {volume} {556}},\ \bibinfo {pages} {43} (\bibinfo {year}
  {2018}{\natexlab{b}})}\BibitemShut {NoStop}%
\bibitem [{\citenamefont {Castro~Neto}\ \emph {et~al.}(2009)\citenamefont
  {Castro~Neto}, \citenamefont {Guinea}, \citenamefont {Peres}, \citenamefont
  {Novoselov},\ and\ \citenamefont {Geim}}]{Castro2009}%
  \BibitemOpen
  \bibfield  {author} {\bibinfo {author} {\bibfnamefont {A.~H.}\ \bibnamefont
  {Castro~Neto}}, \bibinfo {author} {\bibfnamefont {F.}~\bibnamefont {Guinea}},
  \bibinfo {author} {\bibfnamefont {N.~M.~R.}\ \bibnamefont {Peres}}, \bibinfo
  {author} {\bibfnamefont {K.~S.}\ \bibnamefont {Novoselov}}, \ and\ \bibinfo
  {author} {\bibfnamefont {A.~K.}\ \bibnamefont {Geim}},\ }\href {\doibase
  10.1103/RevModPhys.81.109} {\bibfield  {journal} {\bibinfo  {journal} {Rev.
  Mod. Phys.}\ }\textbf {\bibinfo {volume} {81}},\ \bibinfo {pages} {109}
  (\bibinfo {year} {2009})}\BibitemShut {NoStop}%
\bibitem [{\citenamefont {Gusynin}\ and\ \citenamefont
  {Sharapov}(2005)}]{Gusynin2005}%
  \BibitemOpen
  \bibfield  {author} {\bibinfo {author} {\bibfnamefont {V.~P.}\ \bibnamefont
  {Gusynin}}\ and\ \bibinfo {author} {\bibfnamefont {S.~G.}\ \bibnamefont
  {Sharapov}},\ }\href {\doibase 10.1103/PhysRevLett.95.146801} {\bibfield
  {journal} {\bibinfo  {journal} {Phys. Rev. Lett.}\ }\textbf {\bibinfo
  {volume} {95}},\ \bibinfo {pages} {146801} (\bibinfo {year}
  {2005})}\BibitemShut {NoStop}%
\bibitem [{\citenamefont {Novoselov}\ \emph {et~al.}(2006)\citenamefont
  {Novoselov}, \citenamefont {McCann}, \citenamefont {Morozov}, \citenamefont
  {Fal'ko}, \citenamefont {Katsnelson}, \citenamefont {Zeitler}, \citenamefont
  {Jiang}, \citenamefont {Schedin},\ and\ \citenamefont
  {Geim}}]{novoselov2006unconventional}%
  \BibitemOpen
  \bibfield  {author} {\bibinfo {author} {\bibfnamefont {K.~S.}\ \bibnamefont
  {Novoselov}}, \bibinfo {author} {\bibfnamefont {E.}~\bibnamefont {McCann}},
  \bibinfo {author} {\bibfnamefont {S.~V.}\ \bibnamefont {Morozov}}, \bibinfo
  {author} {\bibfnamefont {V.~I.}\ \bibnamefont {Fal'ko}}, \bibinfo {author}
  {\bibfnamefont {M.~I.}\ \bibnamefont {Katsnelson}}, \bibinfo {author}
  {\bibfnamefont {U.}~\bibnamefont {Zeitler}}, \bibinfo {author} {\bibfnamefont
  {D.}~\bibnamefont {Jiang}}, \bibinfo {author} {\bibfnamefont
  {F.}~\bibnamefont {Schedin}}, \ and\ \bibinfo {author} {\bibfnamefont
  {A.~K.}\ \bibnamefont {Geim}},\ }\href {\doibase 10.1038/nphys245} {\bibfield
   {journal} {\bibinfo  {journal} {Nature Physics}\ }\textbf {\bibinfo {volume}
  {2}},\ \bibinfo {pages} {177} (\bibinfo {year} {2006})}\BibitemShut {NoStop}%
\bibitem [{\citenamefont {Yan}\ \emph {et~al.}(2011)\citenamefont {Yan},
  \citenamefont {Peng}, \citenamefont {Zhou}, \citenamefont {Li},\ and\
  \citenamefont {Liu}}]{Yan2011}%
  \BibitemOpen
  \bibfield  {author} {\bibinfo {author} {\bibfnamefont {K.}~\bibnamefont
  {Yan}}, \bibinfo {author} {\bibfnamefont {H.}~\bibnamefont {Peng}}, \bibinfo
  {author} {\bibfnamefont {Y.}~\bibnamefont {Zhou}}, \bibinfo {author}
  {\bibfnamefont {H.}~\bibnamefont {Li}}, \ and\ \bibinfo {author}
  {\bibfnamefont {Z.}~\bibnamefont {Liu}},\ }\href {\doibase 10.1021/nl104000b}
  {\bibfield  {journal} {\bibinfo  {journal} {Nano Letters}\ }\textbf {\bibinfo
  {volume} {11}},\ \bibinfo {pages} {1106} (\bibinfo {year} {2011})},\ \bibinfo
  {note} {pMID: 21322597}\BibitemShut {NoStop}%
\bibitem [{\citenamefont {Riedl}\ \emph {et~al.}(2009)\citenamefont {Riedl},
  \citenamefont {Coletti}, \citenamefont {Iwasaki}, \citenamefont {Zakharov},\
  and\ \citenamefont {Starke}}]{riedl2009quasi}%
  \BibitemOpen
  \bibfield  {author} {\bibinfo {author} {\bibfnamefont {C.}~\bibnamefont
  {Riedl}}, \bibinfo {author} {\bibfnamefont {C.}~\bibnamefont {Coletti}},
  \bibinfo {author} {\bibfnamefont {T.}~\bibnamefont {Iwasaki}}, \bibinfo
  {author} {\bibfnamefont {A.~A.}\ \bibnamefont {Zakharov}}, \ and\ \bibinfo
  {author} {\bibfnamefont {U.}~\bibnamefont {Starke}},\ }\href {\doibase
  10.1103/PhysRevLett.103.246804} {\bibfield  {journal} {\bibinfo  {journal}
  {Phys. Rev. Lett.}\ }\textbf {\bibinfo {volume} {103}},\ \bibinfo {pages}
  {246804} (\bibinfo {year} {2009})}\BibitemShut {NoStop}%
\bibitem [{\citenamefont {Speck}\ \emph {et~al.}(2010)\citenamefont {Speck},
  \citenamefont {Ostler}, \citenamefont {R{\"{o}}hrl}, \citenamefont {Jobst},
  \citenamefont {Waldmann}, \citenamefont {Hundhausen}, \citenamefont {Ley},
  \citenamefont {Weber},\ and\ \citenamefont {Seyller}}]{speck2010quasi}%
  \BibitemOpen
  \bibfield  {author} {\bibinfo {author} {\bibfnamefont {F.}~\bibnamefont
  {Speck}}, \bibinfo {author} {\bibfnamefont {M.}~\bibnamefont {Ostler}},
  \bibinfo {author} {\bibfnamefont {J.}~\bibnamefont {R{\"{o}}hrl}}, \bibinfo
  {author} {\bibfnamefont {J.}~\bibnamefont {Jobst}}, \bibinfo {author}
  {\bibfnamefont {D.}~\bibnamefont {Waldmann}}, \bibinfo {author}
  {\bibfnamefont {M.}~\bibnamefont {Hundhausen}}, \bibinfo {author}
  {\bibfnamefont {L.}~\bibnamefont {Ley}}, \bibinfo {author} {\bibfnamefont
  {H.~B.}\ \bibnamefont {Weber}}, \ and\ \bibinfo {author} {\bibfnamefont
  {T.}~\bibnamefont {Seyller}},\ }in\ \href {\doibase
  10.4028/www.scientific.net/MSF.645-648.629} {\emph {\bibinfo {booktitle}
  {Silicon Carbide and Related Materials 2009}}},\ \bibinfo {series} {Materials
  Science Forum}, Vol.\ \bibinfo {volume} {645}\ (\bibinfo  {publisher} {Trans
  Tech Publications Ltd},\ \bibinfo {year} {2010})\ pp.\ \bibinfo {pages}
  {629--632}\BibitemShut {NoStop}%
\bibitem [{\citenamefont {Alden}\ \emph {et~al.}(2013)\citenamefont {Alden},
  \citenamefont {Tsen}, \citenamefont {Huang}, \citenamefont {Hovden},
  \citenamefont {Brown}, \citenamefont {Park}, \citenamefont {Muller},\ and\
  \citenamefont {McEuen}}]{Alden11256}%
  \BibitemOpen
  \bibfield  {author} {\bibinfo {author} {\bibfnamefont {J.~S.}\ \bibnamefont
  {Alden}}, \bibinfo {author} {\bibfnamefont {A.~W.}\ \bibnamefont {Tsen}},
  \bibinfo {author} {\bibfnamefont {P.~Y.}\ \bibnamefont {Huang}}, \bibinfo
  {author} {\bibfnamefont {R.}~\bibnamefont {Hovden}}, \bibinfo {author}
  {\bibfnamefont {L.}~\bibnamefont {Brown}}, \bibinfo {author} {\bibfnamefont
  {J.}~\bibnamefont {Park}}, \bibinfo {author} {\bibfnamefont {D.~A.}\
  \bibnamefont {Muller}}, \ and\ \bibinfo {author} {\bibfnamefont {P.~L.}\
  \bibnamefont {McEuen}},\ }\href {\doibase 10.1073/pnas.1309394110} {\bibfield
   {journal} {\bibinfo  {journal} {Proceedings of the National Academy of
  Sciences}\ }\textbf {\bibinfo {volume} {110}},\ \bibinfo {pages} {11256}
  (\bibinfo {year} {2013})},\ \Eprint
  {http://arxiv.org/abs/https://www.pnas.org/content/110/28/11256.full.pdf}
  {https://www.pnas.org/content/110/28/11256.full.pdf} \BibitemShut {NoStop}%
\bibitem [{\citenamefont {Butz}\ \emph {et~al.}(2014)\citenamefont {Butz},
  \citenamefont {Dolle}, \citenamefont {Niekiel}, \citenamefont {Weber},
  \citenamefont {Waldmann}, \citenamefont {Weber}, \citenamefont {Meyer},\ and\
  \citenamefont {Spiecker}}]{Butz2014}%
  \BibitemOpen
  \bibfield  {author} {\bibinfo {author} {\bibfnamefont {B.}~\bibnamefont
  {Butz}}, \bibinfo {author} {\bibfnamefont {C.}~\bibnamefont {Dolle}},
  \bibinfo {author} {\bibfnamefont {F.}~\bibnamefont {Niekiel}}, \bibinfo
  {author} {\bibfnamefont {K.}~\bibnamefont {Weber}}, \bibinfo {author}
  {\bibfnamefont {D.}~\bibnamefont {Waldmann}}, \bibinfo {author}
  {\bibfnamefont {H.~B.}\ \bibnamefont {Weber}}, \bibinfo {author}
  {\bibfnamefont {B.}~\bibnamefont {Meyer}}, \ and\ \bibinfo {author}
  {\bibfnamefont {E.}~\bibnamefont {Spiecker}},\ }\href {\doibase
  10.1038/nature12780} {\bibfield  {journal} {\bibinfo  {journal} {Nature}\
  }\textbf {\bibinfo {volume} {505}},\ \bibinfo {pages} {533} (\bibinfo {year}
  {2014})}\BibitemShut {NoStop}%
\bibitem [{\citenamefont {Gong}\ and\ \citenamefont {Mele}(2014)}]{Gong2014}%
  \BibitemOpen
  \bibfield  {author} {\bibinfo {author} {\bibfnamefont {X.}~\bibnamefont
  {Gong}}\ and\ \bibinfo {author} {\bibfnamefont {E.~J.}\ \bibnamefont
  {Mele}},\ }\href {\doibase 10.1103/PhysRevB.89.121415} {\bibfield  {journal}
  {\bibinfo  {journal} {Phys. Rev. B}\ }\textbf {\bibinfo {volume} {89}},\
  \bibinfo {pages} {121415} (\bibinfo {year} {2014})}\BibitemShut {NoStop}%
\bibitem [{\citenamefont {Li}\ \emph {et~al.}(2016)\citenamefont {Li},
  \citenamefont {Wang}, \citenamefont {McFaul}, \citenamefont {Zern},
  \citenamefont {Ren}, \citenamefont {Watanabe}, \citenamefont {Taniguchi},
  \citenamefont {Qiao},\ and\ \citenamefont {Zhu}}]{Li2016}%
  \BibitemOpen
  \bibfield  {author} {\bibinfo {author} {\bibfnamefont {J.}~\bibnamefont
  {Li}}, \bibinfo {author} {\bibfnamefont {K.}~\bibnamefont {Wang}}, \bibinfo
  {author} {\bibfnamefont {K.~J.}\ \bibnamefont {McFaul}}, \bibinfo {author}
  {\bibfnamefont {Z.}~\bibnamefont {Zern}}, \bibinfo {author} {\bibfnamefont
  {Y.}~\bibnamefont {Ren}}, \bibinfo {author} {\bibfnamefont {K.}~\bibnamefont
  {Watanabe}}, \bibinfo {author} {\bibfnamefont {T.}~\bibnamefont {Taniguchi}},
  \bibinfo {author} {\bibfnamefont {Z.}~\bibnamefont {Qiao}}, \ and\ \bibinfo
  {author} {\bibfnamefont {J.}~\bibnamefont {Zhu}},\ }\href {\doibase
  10.1038/nnano.2016.158} {\bibfield  {journal} {\bibinfo  {journal} {Nature
  Nanotechnology}\ }\textbf {\bibinfo {volume} {11}},\ \bibinfo {pages} {1060}
  (\bibinfo {year} {2016})}\BibitemShut {NoStop}%
\bibitem [{\citenamefont {Ju}\ \emph {et~al.}(2015)\citenamefont {Ju},
  \citenamefont {Shi}, \citenamefont {Nair}, \citenamefont {Lv}, \citenamefont
  {Jin}, \citenamefont {Velasco}, \citenamefont {Ojeda-Aristizabal},
  \citenamefont {Bechtel}, \citenamefont {Martin}, \citenamefont {Zettl},
  \citenamefont {Analytis},\ and\ \citenamefont {Wang}}]{Ju2015}%
  \BibitemOpen
  \bibfield  {author} {\bibinfo {author} {\bibfnamefont {L.}~\bibnamefont
  {Ju}}, \bibinfo {author} {\bibfnamefont {Z.}~\bibnamefont {Shi}}, \bibinfo
  {author} {\bibfnamefont {N.}~\bibnamefont {Nair}}, \bibinfo {author}
  {\bibfnamefont {Y.}~\bibnamefont {Lv}}, \bibinfo {author} {\bibfnamefont
  {C.}~\bibnamefont {Jin}}, \bibinfo {author} {\bibfnamefont {J.}~\bibnamefont
  {Velasco}}, \bibinfo {author} {\bibfnamefont {C.}~\bibnamefont
  {Ojeda-Aristizabal}}, \bibinfo {author} {\bibfnamefont {H.~A.}\ \bibnamefont
  {Bechtel}}, \bibinfo {author} {\bibfnamefont {M.~C.}\ \bibnamefont {Martin}},
  \bibinfo {author} {\bibfnamefont {A.}~\bibnamefont {Zettl}}, \bibinfo
  {author} {\bibfnamefont {J.}~\bibnamefont {Analytis}}, \ and\ \bibinfo
  {author} {\bibfnamefont {F.}~\bibnamefont {Wang}},\ }\href {\doibase
  10.1038/nature14364} {\bibfield  {journal} {\bibinfo  {journal} {Nature}\
  }\textbf {\bibinfo {volume} {520}},\ \bibinfo {pages} {650} (\bibinfo {year}
  {2015})}\BibitemShut {NoStop}%
\bibitem [{\citenamefont {Rickhaus}\ \emph {et~al.}(2018)\citenamefont
  {Rickhaus}, \citenamefont {Wallbank}, \citenamefont {Slizovskiy},
  \citenamefont {Pisoni}, \citenamefont {Overweg}, \citenamefont {Lee},
  \citenamefont {Eich}, \citenamefont {Liu}, \citenamefont {Watanabe},
  \citenamefont {Taniguchi}, \citenamefont {Ihn},\ and\ \citenamefont
  {Ensslin}}]{rickhaus2018transport}%
  \BibitemOpen
  \bibfield  {author} {\bibinfo {author} {\bibfnamefont {P.}~\bibnamefont
  {Rickhaus}}, \bibinfo {author} {\bibfnamefont {J.}~\bibnamefont {Wallbank}},
  \bibinfo {author} {\bibfnamefont {S.}~\bibnamefont {Slizovskiy}}, \bibinfo
  {author} {\bibfnamefont {R.}~\bibnamefont {Pisoni}}, \bibinfo {author}
  {\bibfnamefont {H.}~\bibnamefont {Overweg}}, \bibinfo {author} {\bibfnamefont
  {Y.}~\bibnamefont {Lee}}, \bibinfo {author} {\bibfnamefont {M.}~\bibnamefont
  {Eich}}, \bibinfo {author} {\bibfnamefont {M.-H.}\ \bibnamefont {Liu}},
  \bibinfo {author} {\bibfnamefont {K.}~\bibnamefont {Watanabe}}, \bibinfo
  {author} {\bibfnamefont {T.}~\bibnamefont {Taniguchi}}, \bibinfo {author}
  {\bibfnamefont {T.}~\bibnamefont {Ihn}}, \ and\ \bibinfo {author}
  {\bibfnamefont {K.}~\bibnamefont {Ensslin}},\ }\href {\doibase
  10.1021/acs.nanolett.8b02387} {\bibfield  {journal} {\bibinfo  {journal}
  {Nano Letters}\ }\textbf {\bibinfo {volume} {18}},\ \bibinfo {pages} {6725}
  (\bibinfo {year} {2018})}\BibitemShut {NoStop}%
\bibitem [{\citenamefont {Yin}\ \emph {et~al.}(2016)\citenamefont {Yin},
  \citenamefont {Jiang}, \citenamefont {Qiao},\ and\ \citenamefont
  {He}}]{Yin2016}%
  \BibitemOpen
  \bibfield  {author} {\bibinfo {author} {\bibfnamefont {L.-J.}\ \bibnamefont
  {Yin}}, \bibinfo {author} {\bibfnamefont {H.}~\bibnamefont {Jiang}}, \bibinfo
  {author} {\bibfnamefont {J.-B.}\ \bibnamefont {Qiao}}, \ and\ \bibinfo
  {author} {\bibfnamefont {L.}~\bibnamefont {He}},\ }\href {\doibase
  10.1038/ncomms11760} {\bibfield  {journal} {\bibinfo  {journal} {Nature
  Communications}\ }\textbf {\bibinfo {volume} {7}},\ \bibinfo {pages} {11760}
  (\bibinfo {year} {2016})}\BibitemShut {NoStop}%
\bibitem [{\citenamefont {Kisslinger}\ \emph {et~al.}(2015)\citenamefont
  {Kisslinger}, \citenamefont {Ott}, \citenamefont {Heide}, \citenamefont
  {Kampert}, \citenamefont {Butz}, \citenamefont {Spiecker}, \citenamefont
  {Shallcross},\ and\ \citenamefont {Weber}}]{kisslinger2015linear}%
  \BibitemOpen
  \bibfield  {author} {\bibinfo {author} {\bibfnamefont {F.}~\bibnamefont
  {Kisslinger}}, \bibinfo {author} {\bibfnamefont {C.}~\bibnamefont {Ott}},
  \bibinfo {author} {\bibfnamefont {C.}~\bibnamefont {Heide}}, \bibinfo
  {author} {\bibfnamefont {E.}~\bibnamefont {Kampert}}, \bibinfo {author}
  {\bibfnamefont {B.}~\bibnamefont {Butz}}, \bibinfo {author} {\bibfnamefont
  {E.}~\bibnamefont {Spiecker}}, \bibinfo {author} {\bibfnamefont
  {S.}~\bibnamefont {Shallcross}}, \ and\ \bibinfo {author} {\bibfnamefont
  {H.~B.}\ \bibnamefont {Weber}},\ }\href {\doibase 10.1038/nphys3368}
  {\bibfield  {journal} {\bibinfo  {journal} {Nature Physics}\ }\textbf
  {\bibinfo {volume} {11}},\ \bibinfo {pages} {650} (\bibinfo {year}
  {2015})}\BibitemShut {NoStop}%
\bibitem [{\citenamefont {San-Jose}\ \emph {et~al.}(2014)\citenamefont
  {San-Jose}, \citenamefont {Gorbachev}, \citenamefont {Geim}, \citenamefont
  {Novoselov},\ and\ \citenamefont {Guinea}}]{SanJose2014}%
  \BibitemOpen
  \bibfield  {author} {\bibinfo {author} {\bibfnamefont {P.}~\bibnamefont
  {San-Jose}}, \bibinfo {author} {\bibfnamefont {R.~V.}\ \bibnamefont
  {Gorbachev}}, \bibinfo {author} {\bibfnamefont {A.~K.}\ \bibnamefont {Geim}},
  \bibinfo {author} {\bibfnamefont {K.~S.}\ \bibnamefont {Novoselov}}, \ and\
  \bibinfo {author} {\bibfnamefont {F.}~\bibnamefont {Guinea}},\ }\href
  {\doibase 10.1021/nl500230a} {\bibfield  {journal} {\bibinfo  {journal} {Nano
  Letters}\ }\textbf {\bibinfo {volume} {14}},\ \bibinfo {pages} {2052}
  (\bibinfo {year} {2014})},\ \bibinfo {note} {pMID: 24605877},\ \Eprint
  {http://arxiv.org/abs/https://doi.org/10.1021/nl500230a}
  {https://doi.org/10.1021/nl500230a} \BibitemShut {NoStop}%
\bibitem [{\citenamefont {Shallcross}\ \emph {et~al.}(2017)\citenamefont
  {Shallcross}, \citenamefont {Sharma},\ and\ \citenamefont
  {Weber}}]{Shallcross2017}%
  \BibitemOpen
  \bibfield  {author} {\bibinfo {author} {\bibfnamefont {S.}~\bibnamefont
  {Shallcross}}, \bibinfo {author} {\bibfnamefont {S.}~\bibnamefont {Sharma}},
  \ and\ \bibinfo {author} {\bibfnamefont {H.~B.}\ \bibnamefont {Weber}},\
  }\href {\doibase 10.1038/s41467-017-00397-8} {\bibfield  {journal} {\bibinfo
  {journal} {Nature Communications}\ }\textbf {\bibinfo {volume} {8}},\
  \bibinfo {pages} {342} (\bibinfo {year} {2017})}\BibitemShut {NoStop}%
\bibitem [{\citenamefont {Weckbecker}\ \emph {et~al.}(2019)\citenamefont
  {Weckbecker}, \citenamefont {Gupta}, \citenamefont {Rost}, \citenamefont
  {Sharma},\ and\ \citenamefont {Shallcross}}]{Weckbecker2019}%
  \BibitemOpen
  \bibfield  {author} {\bibinfo {author} {\bibfnamefont {D.}~\bibnamefont
  {Weckbecker}}, \bibinfo {author} {\bibfnamefont {R.}~\bibnamefont {Gupta}},
  \bibinfo {author} {\bibfnamefont {F.}~\bibnamefont {Rost}}, \bibinfo {author}
  {\bibfnamefont {S.}~\bibnamefont {Sharma}}, \ and\ \bibinfo {author}
  {\bibfnamefont {S.}~\bibnamefont {Shallcross}},\ }\href {\doibase
  10.1103/PhysRevB.99.195405} {\bibfield  {journal} {\bibinfo  {journal} {Phys.
  Rev. B}\ }\textbf {\bibinfo {volume} {99}},\ \bibinfo {pages} {195405}
  (\bibinfo {year} {2019})}\BibitemShut {NoStop}%
\bibitem [{\citenamefont {Maher}\ \emph {et~al.}(2014)\citenamefont {Maher},
  \citenamefont {Wang}, \citenamefont {Gao}, \citenamefont {Forsythe},
  \citenamefont {Taniguchi}, \citenamefont {Watanabe}, \citenamefont {Abanin},
  \citenamefont {Papi{\'c}}, \citenamefont {Cadden-Zimansky}, \citenamefont
  {Hone}, \citenamefont {Kim},\ and\ \citenamefont {Dean}}]{Maher61}%
  \BibitemOpen
  \bibfield  {author} {\bibinfo {author} {\bibfnamefont {P.}~\bibnamefont
  {Maher}}, \bibinfo {author} {\bibfnamefont {L.}~\bibnamefont {Wang}},
  \bibinfo {author} {\bibfnamefont {Y.}~\bibnamefont {Gao}}, \bibinfo {author}
  {\bibfnamefont {C.}~\bibnamefont {Forsythe}}, \bibinfo {author}
  {\bibfnamefont {T.}~\bibnamefont {Taniguchi}}, \bibinfo {author}
  {\bibfnamefont {K.}~\bibnamefont {Watanabe}}, \bibinfo {author}
  {\bibfnamefont {D.}~\bibnamefont {Abanin}}, \bibinfo {author} {\bibfnamefont
  {Z.}~\bibnamefont {Papi{\'c}}}, \bibinfo {author} {\bibfnamefont
  {P.}~\bibnamefont {Cadden-Zimansky}}, \bibinfo {author} {\bibfnamefont
  {J.}~\bibnamefont {Hone}}, \bibinfo {author} {\bibfnamefont {P.}~\bibnamefont
  {Kim}}, \ and\ \bibinfo {author} {\bibfnamefont {C.~R.}\ \bibnamefont
  {Dean}},\ }\href {\doibase 10.1126/science.1252875} {\bibfield  {journal}
  {\bibinfo  {journal} {Science}\ }\textbf {\bibinfo {volume} {345}},\ \bibinfo
  {pages} {61} (\bibinfo {year} {2014})},\ \Eprint
  {http://arxiv.org/abs/https://science.sciencemag.org/content/345/6192/61.full.pdf}
  {https://science.sciencemag.org/content/345/6192/61.full.pdf} \BibitemShut
  {NoStop}%
\bibitem [{\citenamefont {Kou}\ \emph {et~al.}(2014)\citenamefont {Kou},
  \citenamefont {Feldman}, \citenamefont {Levin}, \citenamefont {Halperin},
  \citenamefont {Watanabe}, \citenamefont {Taniguchi},\ and\ \citenamefont
  {Yacoby}}]{Kou2014}%
  \BibitemOpen
  \bibfield  {author} {\bibinfo {author} {\bibfnamefont {A.}~\bibnamefont
  {Kou}}, \bibinfo {author} {\bibfnamefont {B.~E.}\ \bibnamefont {Feldman}},
  \bibinfo {author} {\bibfnamefont {A.~J.}\ \bibnamefont {Levin}}, \bibinfo
  {author} {\bibfnamefont {B.~I.}\ \bibnamefont {Halperin}}, \bibinfo {author}
  {\bibfnamefont {K.}~\bibnamefont {Watanabe}}, \bibinfo {author}
  {\bibfnamefont {T.}~\bibnamefont {Taniguchi}}, \ and\ \bibinfo {author}
  {\bibfnamefont {A.}~\bibnamefont {Yacoby}},\ }\href {\doibase
  10.1126/science.1250270} {\bibfield  {journal} {\bibinfo  {journal}
  {Science}\ }\textbf {\bibinfo {volume} {345}},\ \bibinfo {pages} {55}
  (\bibinfo {year} {2014})}\BibitemShut {NoStop}%
\bibitem [{\citenamefont {Barlas}\ \emph {et~al.}(2008)\citenamefont {Barlas},
  \citenamefont {C\^ot\'e}, \citenamefont {Nomura},\ and\ \citenamefont
  {MacDonald}}]{Barlas2008}%
  \BibitemOpen
  \bibfield  {author} {\bibinfo {author} {\bibfnamefont {Y.}~\bibnamefont
  {Barlas}}, \bibinfo {author} {\bibfnamefont {R.}~\bibnamefont {C\^ot\'e}},
  \bibinfo {author} {\bibfnamefont {K.}~\bibnamefont {Nomura}}, \ and\ \bibinfo
  {author} {\bibfnamefont {A.~H.}\ \bibnamefont {MacDonald}},\ }\href {\doibase
  10.1103/PhysRevLett.101.097601} {\bibfield  {journal} {\bibinfo  {journal}
  {Phys. Rev. Lett.}\ }\textbf {\bibinfo {volume} {101}},\ \bibinfo {pages}
  {097601} (\bibinfo {year} {2008})}\BibitemShut {NoStop}%
\bibitem [{\citenamefont {Kisslinger}\ \emph {et~al.}(2019)\citenamefont
  {Kisslinger}, \citenamefont {Rienm\"{u}ller}, \citenamefont {Ott},
  \citenamefont {Kampert},\ and\ \citenamefont {Weber}}]{Kisslinger2019}%
  \BibitemOpen
  \bibfield  {author} {\bibinfo {author} {\bibfnamefont {F.}~\bibnamefont
  {Kisslinger}}, \bibinfo {author} {\bibfnamefont {D.}~\bibnamefont
  {Rienm\"{u}ller}}, \bibinfo {author} {\bibfnamefont {C.}~\bibnamefont {Ott}},
  \bibinfo {author} {\bibfnamefont {E.}~\bibnamefont {Kampert}}, \ and\
  \bibinfo {author} {\bibfnamefont {H.~B.}\ \bibnamefont {Weber}},\ }\href
  {\doibase 10.1002/andp.201800188} {\bibfield  {journal} {\bibinfo  {journal}
  {Annalen der Physik}\ }\textbf {\bibinfo {volume} {531}},\ \bibinfo {pages}
  {1800188} (\bibinfo {year} {2019})}\BibitemShut {NoStop}%
\bibitem [{\citenamefont {Thouless}\ and\ \citenamefont
  {Kirkpatrick}(1981)}]{Thouless_1981}%
  \BibitemOpen
  \bibfield  {author} {\bibinfo {author} {\bibfnamefont {D.~J.}\ \bibnamefont
  {Thouless}}\ and\ \bibinfo {author} {\bibfnamefont {S.}~\bibnamefont
  {Kirkpatrick}},\ }\href {\doibase 10.1088/0022-3719/14/3/007} {\bibfield
  {journal} {\bibinfo  {journal} {Journal of Physics C: Solid State Physics}\
  }\textbf {\bibinfo {volume} {14}},\ \bibinfo {pages} {235} (\bibinfo {year}
  {1981})}\BibitemShut {NoStop}%
\bibitem [{\citenamefont {Lewenkopf}\ and\ \citenamefont
  {Mucciolo}(2013)}]{Lewenkopf2013}%
  \BibitemOpen
  \bibfield  {author} {\bibinfo {author} {\bibfnamefont {C.~H.}\ \bibnamefont
  {Lewenkopf}}\ and\ \bibinfo {author} {\bibfnamefont {E.~R.}\ \bibnamefont
  {Mucciolo}},\ }\href {\doibase 10.1007/s10825-013-0458-7} {\bibfield
  {journal} {\bibinfo  {journal} {Journal of Computational Electronics}\
  }\textbf {\bibinfo {volume} {12}},\ \bibinfo {pages} {203} (\bibinfo {year}
  {2013})}\BibitemShut {NoStop}%
\bibitem [{\citenamefont {Petersen}\ \emph {et~al.}(2008)\citenamefont
  {Petersen}, \citenamefont {Sørensen}, \citenamefont {Hansen}, \citenamefont
  {Skelboe},\ and\ \citenamefont {Stokbro}}]{petersen2008block}%
  \BibitemOpen
  \bibfield  {author} {\bibinfo {author} {\bibfnamefont {D.~E.}\ \bibnamefont
  {Petersen}}, \bibinfo {author} {\bibfnamefont {H.~H.~B.}\ \bibnamefont
  {Sørensen}}, \bibinfo {author} {\bibfnamefont {P.~C.}\ \bibnamefont
  {Hansen}}, \bibinfo {author} {\bibfnamefont {S.}~\bibnamefont {Skelboe}}, \
  and\ \bibinfo {author} {\bibfnamefont {K.}~\bibnamefont {Stokbro}},\ }\href
  {\doibase https://doi.org/10.1016/j.jcp.2007.11.035} {\bibfield  {journal}
  {\bibinfo  {journal} {Journal of Computational Physics}\ }\textbf {\bibinfo
  {volume} {227}},\ \bibinfo {pages} {3174 } (\bibinfo {year}
  {2008})}\BibitemShut {NoStop}%
\bibitem [{\citenamefont {Khomyakov}\ and\ \citenamefont
  {Brocks}(2004)}]{khomyakov2004real}%
  \BibitemOpen
  \bibfield  {author} {\bibinfo {author} {\bibfnamefont {P.~A.}\ \bibnamefont
  {Khomyakov}}\ and\ \bibinfo {author} {\bibfnamefont {G.}~\bibnamefont
  {Brocks}},\ }\href {\doibase 10.1103/PhysRevB.70.195402} {\bibfield
  {journal} {\bibinfo  {journal} {Phys. Rev. B}\ }\textbf {\bibinfo {volume}
  {70}},\ \bibinfo {pages} {195402} (\bibinfo {year} {2004})}\BibitemShut
  {NoStop}%
\bibitem [{\citenamefont {Khomyakov}\ \emph {et~al.}(2005)\citenamefont
  {Khomyakov}, \citenamefont {Brocks}, \citenamefont {Karpan}, \citenamefont
  {Zwierzycki},\ and\ \citenamefont {Kelly}}]{khomyakov2005conductance}%
  \BibitemOpen
  \bibfield  {author} {\bibinfo {author} {\bibfnamefont {P.~A.}\ \bibnamefont
  {Khomyakov}}, \bibinfo {author} {\bibfnamefont {G.}~\bibnamefont {Brocks}},
  \bibinfo {author} {\bibfnamefont {V.}~\bibnamefont {Karpan}}, \bibinfo
  {author} {\bibfnamefont {M.}~\bibnamefont {Zwierzycki}}, \ and\ \bibinfo
  {author} {\bibfnamefont {P.~J.}\ \bibnamefont {Kelly}},\ }\href {\doibase
  10.1103/PhysRevB.72.035450} {\bibfield  {journal} {\bibinfo  {journal} {Phys.
  Rev. B}\ }\textbf {\bibinfo {volume} {72}},\ \bibinfo {pages} {035450}
  (\bibinfo {year} {2005})}\BibitemShut {NoStop}%
\bibitem [{\citenamefont {Rammer}\ and\ \citenamefont
  {Smith}(1986)}]{rammer1986}%
  \BibitemOpen
  \bibfield  {author} {\bibinfo {author} {\bibfnamefont {J.}~\bibnamefont
  {Rammer}}\ and\ \bibinfo {author} {\bibfnamefont {H.}~\bibnamefont {Smith}},\
  }\href {\doibase 10.1103/RevModPhys.58.323} {\bibfield  {journal} {\bibinfo
  {journal} {Rev. Mod. Phys.}\ }\textbf {\bibinfo {volume} {58}},\ \bibinfo
  {pages} {323} (\bibinfo {year} {1986})}\BibitemShut {NoStop}%
\bibitem [{\citenamefont {Cresti}\ \emph {et~al.}(2003)\citenamefont {Cresti},
  \citenamefont {Farchioni}, \citenamefont {Grosso},\ and\ \citenamefont
  {Parravicini}}]{Cresti2003}%
  \BibitemOpen
  \bibfield  {author} {\bibinfo {author} {\bibfnamefont {A.}~\bibnamefont
  {Cresti}}, \bibinfo {author} {\bibfnamefont {R.}~\bibnamefont {Farchioni}},
  \bibinfo {author} {\bibfnamefont {G.}~\bibnamefont {Grosso}}, \ and\ \bibinfo
  {author} {\bibfnamefont {G.~P.}\ \bibnamefont {Parravicini}},\ }\href
  {\doibase 10.1103/PhysRevB.68.075306} {\bibfield  {journal} {\bibinfo
  {journal} {Phys. Rev. B}\ }\textbf {\bibinfo {volume} {68}},\ \bibinfo
  {pages} {075306} (\bibinfo {year} {2003})}\BibitemShut {NoStop}%
\bibitem [{\citenamefont {Moon}\ and\ \citenamefont
  {Koshino}(2012)}]{tb_param_2012}%
  \BibitemOpen
  \bibfield  {author} {\bibinfo {author} {\bibfnamefont {P.}~\bibnamefont
  {Moon}}\ and\ \bibinfo {author} {\bibfnamefont {M.}~\bibnamefont {Koshino}},\
  }\href {\doibase 10.1103/PhysRevB.85.195458} {\bibfield  {journal} {\bibinfo
  {journal} {Phys. Rev. B}\ }\textbf {\bibinfo {volume} {85}},\ \bibinfo
  {pages} {195458} (\bibinfo {year} {2012})}\BibitemShut {NoStop}%
\bibitem [{\citenamefont {Zhang}\ \emph {et~al.}(2013)\citenamefont {Zhang},
  \citenamefont {MacDonald},\ and\ \citenamefont {Mele}}]{zhang2013valley}%
  \BibitemOpen
  \bibfield  {author} {\bibinfo {author} {\bibfnamefont {F.}~\bibnamefont
  {Zhang}}, \bibinfo {author} {\bibfnamefont {A.~H.}\ \bibnamefont
  {MacDonald}}, \ and\ \bibinfo {author} {\bibfnamefont {E.~J.}\ \bibnamefont
  {Mele}},\ }\href {\doibase 10.1073/pnas.1308853110} {\bibfield  {journal}
  {\bibinfo  {journal} {Proceedings of the National Academy of Sciences}\
  }\textbf {\bibinfo {volume} {110}},\ \bibinfo {pages} {10546} (\bibinfo
  {year} {2013})},\ \Eprint
  {http://arxiv.org/abs/https://www.pnas.org/content/110/26/10546.full.pdf}
  {https://www.pnas.org/content/110/26/10546.full.pdf} \BibitemShut {NoStop}%
\bibitem [{\citenamefont {Groth}\ \emph {et~al.}(2014)\citenamefont {Groth},
  \citenamefont {Wimmer}, \citenamefont {Akhmerov},\ and\ \citenamefont
  {Waintal}}]{Groth_2014}%
  \BibitemOpen
  \bibfield  {author} {\bibinfo {author} {\bibfnamefont {C.~W.}\ \bibnamefont
  {Groth}}, \bibinfo {author} {\bibfnamefont {M.}~\bibnamefont {Wimmer}},
  \bibinfo {author} {\bibfnamefont {A.~R.}\ \bibnamefont {Akhmerov}}, \ and\
  \bibinfo {author} {\bibfnamefont {X.}~\bibnamefont {Waintal}},\ }\href
  {\doibase 10.1088/1367-2630/16/6/063065} {\bibfield  {journal} {\bibinfo
  {journal} {New Journal of Physics}\ }\textbf {\bibinfo {volume} {16}},\
  \bibinfo {pages} {063065} (\bibinfo {year} {2014})}\BibitemShut {NoStop}%
\bibitem [{\citenamefont {Kazymyrenko}\ and\ \citenamefont
  {Waintal}(2008)}]{Kazymyrenko2008}%
  \BibitemOpen
  \bibfield  {author} {\bibinfo {author} {\bibfnamefont {K.}~\bibnamefont
  {Kazymyrenko}}\ and\ \bibinfo {author} {\bibfnamefont {X.}~\bibnamefont
  {Waintal}},\ }\href {\doibase 10.1103/PhysRevB.77.115119} {\bibfield
  {journal} {\bibinfo  {journal} {Phys. Rev. B}\ }\textbf {\bibinfo {volume}
  {77}},\ \bibinfo {pages} {115119} (\bibinfo {year} {2008})}\BibitemShut
  {NoStop}%
\bibitem [{\citenamefont {Lee}\ and\ \citenamefont
  {Joannopoulos}(1981)}]{Lee1981}%
  \BibitemOpen
  \bibfield  {author} {\bibinfo {author} {\bibfnamefont {D.~H.}\ \bibnamefont
  {Lee}}\ and\ \bibinfo {author} {\bibfnamefont {J.~D.}\ \bibnamefont
  {Joannopoulos}},\ }\href {\doibase 10.1103/PhysRevB.23.4997} {\bibfield
  {journal} {\bibinfo  {journal} {Phys. Rev. B}\ }\textbf {\bibinfo {volume}
  {23}},\ \bibinfo {pages} {4997} (\bibinfo {year} {1981})}\BibitemShut
  {NoStop}%
\bibitem [{\citenamefont {Ando}(1991)}]{Ando1991}%
  \BibitemOpen
  \bibfield  {author} {\bibinfo {author} {\bibfnamefont {T.}~\bibnamefont
  {Ando}},\ }\href {\doibase 10.1103/PhysRevB.44.8017} {\bibfield  {journal}
  {\bibinfo  {journal} {Phys. Rev. B}\ }\textbf {\bibinfo {volume} {44}},\
  \bibinfo {pages} {8017} (\bibinfo {year} {1991})}\BibitemShut {NoStop}%
\bibitem [{\citenamefont {S\o{}rensen}\ \emph {et~al.}(2008)\citenamefont
  {S\o{}rensen}, \citenamefont {Hansen}, \citenamefont {Petersen},
  \citenamefont {Skelboe},\ and\ \citenamefont {Stokbro}}]{sorensen2008krylov}%
  \BibitemOpen
  \bibfield  {author} {\bibinfo {author} {\bibfnamefont {H.~H.~B.}\
  \bibnamefont {S\o{}rensen}}, \bibinfo {author} {\bibfnamefont {P.~C.}\
  \bibnamefont {Hansen}}, \bibinfo {author} {\bibfnamefont {D.~E.}\
  \bibnamefont {Petersen}}, \bibinfo {author} {\bibfnamefont {S.}~\bibnamefont
  {Skelboe}}, \ and\ \bibinfo {author} {\bibfnamefont {K.}~\bibnamefont
  {Stokbro}},\ }\href {\doibase 10.1103/PhysRevB.77.155301} {\bibfield
  {journal} {\bibinfo  {journal} {Phys. Rev. B}\ }\textbf {\bibinfo {volume}
  {77}},\ \bibinfo {pages} {155301} (\bibinfo {year} {2008})}\BibitemShut
  {NoStop}%
\bibitem [{\citenamefont {Sancho}\ \emph {et~al.}(1985)\citenamefont {Sancho},
  \citenamefont {Sancho}, \citenamefont {Sancho},\ and\ \citenamefont
  {Rubio}}]{sancho1985highly}%
  \BibitemOpen
  \bibfield  {author} {\bibinfo {author} {\bibfnamefont {M.~P.~L.}\
  \bibnamefont {Sancho}}, \bibinfo {author} {\bibfnamefont {J.~M.~L.}\
  \bibnamefont {Sancho}}, \bibinfo {author} {\bibfnamefont {J.~M.~L.}\
  \bibnamefont {Sancho}}, \ and\ \bibinfo {author} {\bibfnamefont
  {J.}~\bibnamefont {Rubio}},\ }\href {\doibase 10.1088/0305-4608/15/4/009}
  {\bibfield  {journal} {\bibinfo  {journal} {Journal of Physics F: Metal
  Physics}\ }\textbf {\bibinfo {volume} {15}},\ \bibinfo {pages} {851}
  (\bibinfo {year} {1985})}\BibitemShut {NoStop}%
\end{thebibliography}
\end{document}